\documentclass[journal]{IEEEtran}
\usepackage[table,xcdraw]{xcolor}

\usepackage[sort&compress, square, numbers]{natbib}

\usepackage{graphicx}
\usepackage{lmodern,babel,adjustbox,booktabs}
\usepackage{todonotes}
\usepackage{longtable}
\usepackage{hyperref}
\usepackage{amsfonts}
\usepackage{authblk}

\usepackage{tabularx}
\usepackage{multirow}
\newcolumntype{C}{>{\centering\arraybackslash}X}
\newcolumntype{L}{>{\raggedright\arraybackslash}X}
\newcolumntype{P}[1]{>{\raggedright\arraybackslash}p{#1}}

\newcommand{\etal}{\textit{et al}.~}
\newcommand{\ie}{\textit{i}.\textit{e}.~}
\newcommand{\eg}{\textit{e}.\textit{g}.~}

\usepackage{booktabs}
\usepackage[capitalise]{cleveref}

\usepackage[percent]{overpic}

\begin{document}

\title{Domain Generalization in Computational Pathology: Survey and Guidelines}

\author{Mostafa~Jahanifar\textsuperscript{1},
Manahil Raza\textsuperscript{1},
Kesi Xu\textsuperscript{1},
Trinh Vuong\textsuperscript{2},
Rob Jewsbury\textsuperscript{1},
Adam Shephard\textsuperscript{1},
Neda Zamanitajeddin\textsuperscript{1},
Jin Tae Kwak\textsuperscript{2},
Shan E Ahmed Raza\textsuperscript{1},
Fayyaz Minhas\textsuperscript{1},
Nasir Rajpoot\textsuperscript{1}
\thanks{Corresponding author: Mostafa Jahanifar (e-mail: mostafa.jahanifar@warwick.ac.uk)}
}

\affil{\textsuperscript{1}Tissue Image Analytics center, Department of Computer Science, University of Warwick, UK}
\affil{\textsuperscript{2}School of Electrical Engineering, Korea University, Korea}

\markboth{Domain generalization in CPath}%
{M. Jahanifar \MakeLowercase{etal}: Domain generalization in CPath}

\maketitle

\begin{abstract}
Deep learning models have exhibited exceptional effectiveness in Computational Pathology (CPath) by tackling intricate tasks across an array of histology image analysis applications. Nevertheless, the presence of out-of-distribution data (stemming from a multitude of sources such as disparate imaging devices and diverse tissue preparation methods) can cause \emph{domain shift} (DS). DS decreases the generalization of trained models to unseen datasets with slightly different data distributions, prompting the need for innovative \emph{domain generalization} (DG) solutions.
Recognizing the potential of DG methods to significantly influence diagnostic and prognostic models in cancer studies and clinical practice, we present this survey along with guidelines on achieving DG in CPath.
We rigorously define various DS types, systematically review and categorize existing DG approaches and resources in CPath, and provide insights into their advantages, limitations, and applicability. We also conduct thorough benchmarking experiments with 28 cutting-edge DG algorithms to address a complex DG problem.
Our findings suggest that careful experiment design and CPath-specific Stain Augmentation technique can be very effective. However, there is no one-size-fits-all solution for DG in CPath. Therefore, we establish clear guidelines for detecting and managing DS depending on different scenarios.
While most of the concepts, guidelines, and recommendations are given for applications in CPath, they are applicable to most medical image analysis tasks as well.

\end{abstract}


\IEEEpeerreviewmaketitle

\section{Introduction}

Image analysis and machine learning (ML) are important parts of Computational Pathology (CPath) to analyze and extract meaningful information from various types of pathology-related data which often includes whole slide images (WSI) of tissue samples. The nuanced characteristics of large histopathological images and the inherent diversity of clinical data make it hard to solve intricate problems in CPath using classical image processing or ML algorithms. Fortunately, with the advent of deep learning (DL) algorithms \cite{lecun2015deep,goodfellow2016deep} and improvement of convolutional neural networks (CNNs) \cite{li2021survey} over the recent decades, CPath has also experienced remarkable success, yielding state-of-the-art (SOTA) results in various diagnostic and prognostic tasks across distinct datasets \cite{verghese2023computational,song2023artificial}. Sophisticated DL-driven artificial intelligence (AI) systems have expanded the range of automatically solvable problems \cite{cui2021artificial,srinidhi2021deep}, such as various classification \cite{liu2019artificial,abdollahi2022detection}, detection \cite{koohababni2018nuclei,cai2019efficient,krithiga2021breast,42_javed2021cpath}, regression \cite{janowczyk2016deep, dawood2021albrt,song2022deep,laves2020well}, segmentation \cite{ronneberger2015u,graham2019hover,hollandi2022nucleus,saha2018her2net, shephard2021simultaneous}, and survival prediction \cite{173_kather2019cpath,lu2022slidegraph+,li2018graph,bychkov2018deep,luo2017comprehensive,rathore2022survival} tasks.

\begin{figure*}
    \centering
    \includegraphics[width=\textwidth]{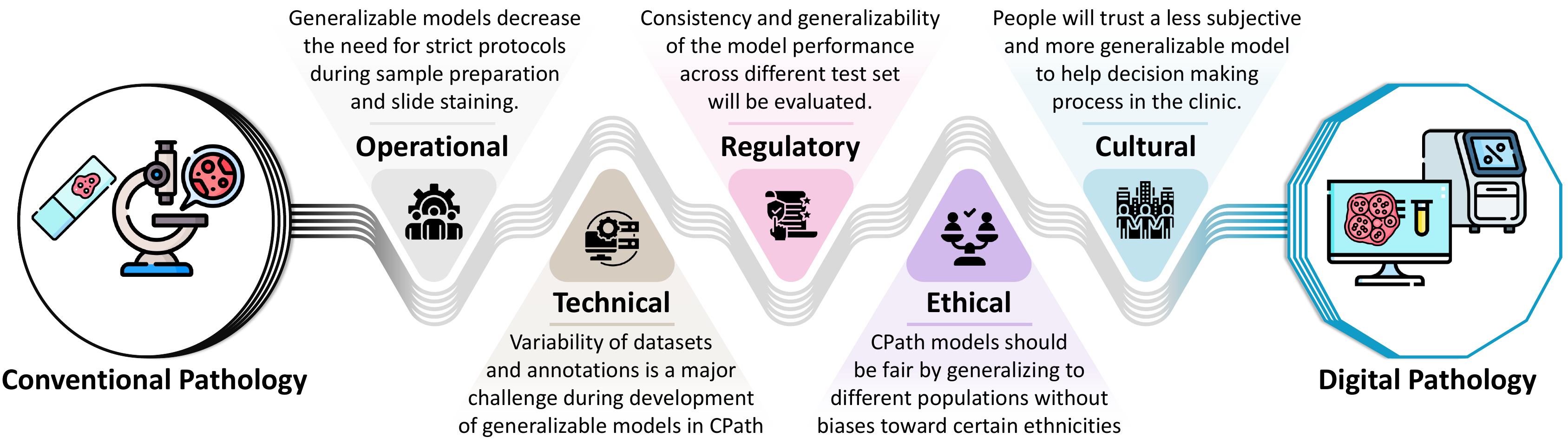}
    \caption{Challenges of transforming conventional pathology into digital pathology in the clinic \cite{verghese2023computational}. In each transition challenge, domain generalization of CPath models plays a significant role.}
    \label{fig:motivation}
\end{figure*}

With all the recent advances of DL in CPath, the question of the practicality of these models in the clinical setting has also been investigated \cite{asif2023unleashing,asif2021towards,94_kleppe2021cpath,tizhoosh2018artificial,van2021deep}. 
One of the main effectiveness of DL models hinges on their ability to generalize well beyond the training data and handle the inherent variability in histology images. Verghese \etal \cite{verghese2023computational} outlined multiple challenges in transiting from conventional pathology to digital pathology in practice, categorized under operational, technical, regulatory, ethical, and cultural challenges. We believe domain generalizability of DL models plays an important role in resolving these hurdles in the path, as exemplified in \cref{fig:motivation}. Achieving true practicality necessitates addressing critical questions related to the generalizability of DL models: How can models trained on one dataset seamlessly adapt to new datasets from diverse sources? Can a DL model trained on small images perform well on WSIs? What about generalizability to the datasets that have been annotated differently? Can a DL model trained for a specific task perform well on samples from different stages of cancer, ethnicity, and gender? 

These questions invariably lead us to reconsider the Identically and Independently Distributed (\emph{i.i.d}) assumption, a cornerstone of traditional ML that assumes data in source and target domains are coming from the same distribution. The presence of Out-of-Distribution (OOD) data in CPath (arising from variations in imaging devices, tissue preparation techniques, staining procedures, and labeling protocols \cite{13_stacke2021cpath,17_nisar2022cpath,149_khan2022cpath,fuchs2011computational}) voids the \emph{i.i.d} assumption and poses a considerable challenge to generalize to unseen domains. This challenge is characterized as \emph{domain shift} (DS), wherein the source and target domains exhibit differences in data distribution that hinder the direct application of trained models to unseen datasets  \cite{ben2010theory,moreno2012unifying,hu2020domain,hendrycks2018benchmarking}. DL models in CPath have also demonstrated a notable vulnerability to DS, as well as common corruptions and perturbations \cite{85_zhang2022cpath,82_vu2022cpath,29_pohjonen2022cpath,163_faryna2021cpath}.

One obvious solution to improve DL against DS is to collect more variation of data distribution in the training domain. Nevertheless, in practical terms, it proves exceedingly challenging to comprehensively assemble all conceivable data distributions, particularly within real-world contexts. This challenge is particularly pronounced in medical image analysis, given the substantial cost associated with data curation, including factors such as time constraints, the necessity for skilled annotators, and stringent patient privacy considerations, among other formidable hurdles \cite{van2019quality,varoquaux2022machine}. Therefore, addressing domain shift requires innovative solutions, leading us to the realm of \emph{domain generalization} (DG). Unlike domain adaptation that seeks to align source and target domains by collecting some data from the target domain \cite{122_farahani2021general,12_guan2022general}, domain generalization aims to create models robust enough to generalize across diverse domains without direct exposure to target domain data \cite{blanchard2011generalizing,muandet2013domain}.

The history of generalization in machine learning can be traced back to foundational works like that of Vapnik \cite{vapnik1999erm}, who provided a theoretical overview of generalization in neural networks. As data streaming gained prominence, Kifer \etal \cite{kifer2004detecting} delved into detecting shifts in these streams using non-parametric tests. A few years later, the challenge of differing training and test data distributions, known as `covariate shift', was addressed by Bickel \etal \cite{bickel2007discriminative}.
The concept of domain generalization later emerged, with its formal introduction by Blanchard \etal \cite{blanchard2011generalizing} as a distinct ML problem, and was subsequently termed by Muandet \etal \cite{muandet2013domain}. This problem was motivated by practical challenges, such as automating the cell classification process in medical applications where classifiers trained on data from previous patients struggled to extend their performance to new patients \cite{blanchard2011generalizing}. Since then, many methods have been proposed to deal with DS in the ML community that approached DG from different angles \cite{110_zhou2023general,112_wang2021general,111_shen2021general,55_csurka2017general,sheth2022domain}.

As we delve into the historical trajectory of DG, its emergence becomes evident in the context of CPath, where DS is a common occurrence due to the inherent variations in histopathological data. Attempts to improve generalizability through stain variation can be dated back to 2001 when the first stain normalization methods were proposed \cite{198_ruifrok2001cpath,200_reinhard2001cpath}. However, the DS problem in CPath and DG solutions for DL models were brought to the foreground with the `MITOS \& ATYPIA' challenge \cite{126_roux2014_cpath} in 2014 where contestants had to detect mitosis and evaluate the nuclear atypia in histology images coming from different scanners. Datasets and challenges like `Camelyon' \cite{74_ehteshamibejnordi2017cpath} for lymph node metastases detection in breast cancer (BC) further paved the way for DG research in the CPath field. In recent years, mitosis domain generalization (MIDOG) challenges/datasets \cite{64_aubreville2023cpath,65_aubreville2022cpath,166_aubreville2023cpath} have considerably fueled DG research in CPath. To name a few examples of DG method in CPath, we can mention data augmentation techniques to synthetically generate samples from unseen domains \cite{21_tellez2019cpath,101_chen2022cpath,63_jarkman2022cpath,26_chen2022cpath,100_xue2021cpath,92_yamashita2021cpath}, methods that try to learn aligned features representations across different domains \cite{189_lafarge2019cpath,54_shin2021cpath,78_ke2021cpath,109_wang2021cpath,118_gadermayr2019cpath}, pretraining the model with a large amount of unlabeled data to learn generalizable representations \cite{1_marini2021cpath,89_koohbanani2021cpath,144_fan2021cpath,2_fashi2022cpath,5_yang2022cpath,159_kang2023cpath}, and histology-specific model designs that are tailored to the task at hand \cite{38_lafarge2021cpath,40_zhang2022cpath,131_yaar2020cpath,119_graham2023cpath,130_yu2023cpath}.

There have been various methods developed for DG over the past years in ML and computer vision (CV) communities and comprehensive review papers have also been published to summarize those efforts and shed some light on future perspectives of the research field \cite{110_zhou2023general,112_wang2021general,111_shen2021general,55_csurka2017general,sheth2022domain}. However, despite the ongoing interest in DG in CPath and medical image analysis communities, no comprehensive survey on the topic exists hitherto to review the DG methodologies employed and their efficacy in the context of CPath.

This paper fills this critical gap and elucidates the DG landscape by offering a comprehensive and systematic review of DG methods and resources tailored for CPath, along with providing guidance for future DG research to bolster the robustness and applicability of DL models in CPath.
We begin by providing clear definitions of DS and its various types, supported by concise mathematical formulations and practical examples. We then introduce the concept of DG and differentiate it from related concepts within ML (\cref{sec:defs}).
Moving forward, we conduct an exhaustive examination of existing DG approaches in CPath, analyzing and categorizing them to shed light on their strengths, limitations, and suitability for addressing diverse DS challenges specific to CPath (\cref{sec:dg_survey}). Recognizing the pivotal role of resources in DG methodology development, we devote a section to reviewing available resources, including toolboxes and datasets, to facilitate DG research (\cref{sec:resources}).
To bridge the gap between recent DG advancements in CV and the CPath community, and to underscore the potential of the reviewed resources, we present a DG algorithm benchmark in \cref{sec:benchmark}. Our aim is to transcend mere methodological review and to inspire future DG-related research in CPath while offering practical guidance for researchers seeking to design more generalizable DL models for their specific applications. Therefore, drawing from the definitions and categorization of the DS problem in \cref{sec:defs}, we distill clear and concise guidelines for DG in \cref{sec:guidelines}. These guidelines provide researchers with a pragmatic framework for detecting and addressing DS challenges.
Lastly, in \cref{sec:discussion}, we emphasize the implications of our review study, identify potential future research directions for DG in CPath, explore emerging technologies applicable to DG, and acknowledge the limitations of our work.

\begin{figure*}[!ht]
    \centering
    \includegraphics[width=0.9\textwidth]{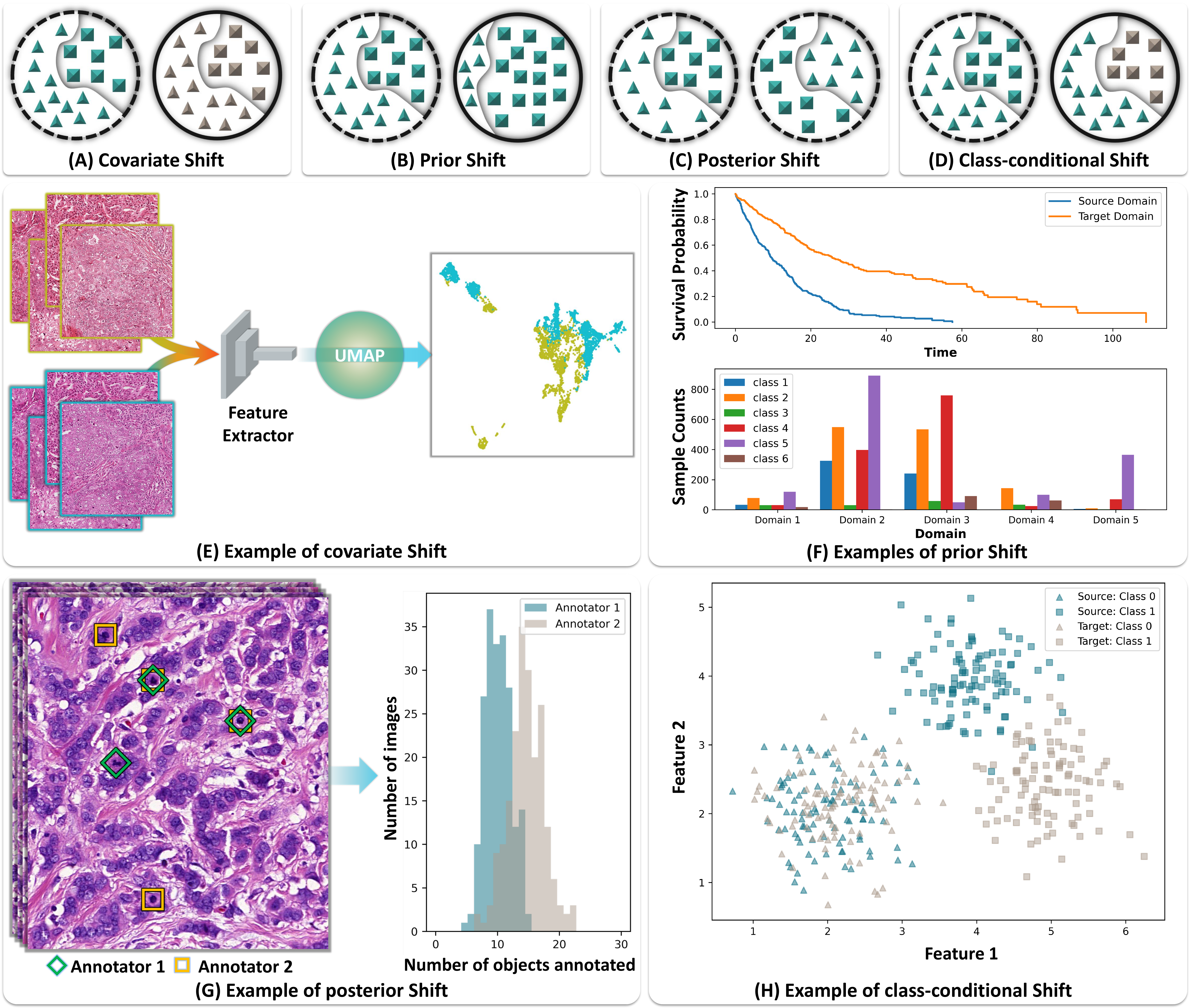}
    \caption{Schematics illustrating four distinct types of domain shifts (A-D) and their corresponding examples observed in CPath (E-H). In all panels, shapes (triangles and rectangles) and colors represent classes and domains, respectively. (A;E): Covariate Shift; variations in scanning technologies lead to differences in image appearance and feature space. (B;F): Prior Shift; changes in label distribution between source and target domains, exemplified in a survival analysis study and a 6-class classification problem. (C;G): Posterior Shift; inter-observer variability in mitosis annotation results in distribution shifts when annotated by two different annotators (two domains). (D;H): Class-conditional Shift; shifts in the distributions of two morphological features within the same cell label (class 1) between early-stage and late-stage patient domains.}
    \label{fig:shifts}
\end{figure*}

\section{Definitions}
\label{sec:defs}
A domain refers to the specific distribution underlying a dataset or data source, characterized by unique attributes such as image resolution, staining methods, patient demographics, and so forth. Mathematically, a domain in this context signifies the joint distribution spanning the input (or feature) space ($X$) and the label space ($Y$), represented as $D=P_{XY}$. Distinct joint distributions for source ($s$) and target ($t$) domains can be denoted as $D^s = P^s_{XY}$ and $D^t = P^t_{XY}$, respectively. It is important to note that there can be multiple source domains, each potentially encompassing data sourced from different centers, scanners, and datasets, among other variables. For the sake of clarity in our discourse, the `source domains' strictly refer to all the data pools available during model training or method design, and the `target domain' designates the pool of unseen test set data on which we intend to evaluate the trained model.


\subsection{Domain Shift (DS)}
\label{sec:ds}
Domain shift arises when there exists a discrepancy in the joint distribution between the source and target domains, \ie, $P^s_{XY} \neq P^t_{XY}$. Leveraging Bayes' theorem, we can reconstruct the joint distribution as follows:
\begin{equation}
{P_{XY}} = {P_{X|Y}}{P_Y} = {P_{Y|X}}{P_X},
\end{equation}
where the terms $P_{X|Y}$, $P_Y$, $P_{Y|X}$, and $P_X$ represent class-conditional, posterior, prior, and covariate distributions, respectively. This allows us to further categorize DS into four types, depending on whether the observed shift in joint distributions across the source and target domains is precipitated by shifts in one of the four constituent distributions. Schematics of these four types of DS are illustrated in \cref{fig:shifts}A-D where each source domain is surrounded with dashed or solid circles and the samples from different classes have different shapes (triangles or rectangles).

It is worth noting that this is a theoretical abstraction of the domain shift problem. In real-world scenarios, such an abstraction may not always be feasible. Nonetheless, this categorization lends us deeper insights into the problem and facilitates the design of effective strategies to mitigate domain shift and enhance domain generalization.

\subsubsection{Covariate Shift}
\label{sec:ds1}
Covariate shift is seen when the distributions of source and target covariates or features are different, \ie, $P^s_X \neq P^t_X$. In \cref{fig:shifts}A, covariate shift is shown when the appearance of samples (presented by their color) from both classes changes over domains but the label distributions (indicated by sectioning shadow line in the circles) align.
In CPath, this type of shift is quite prevalent and constitutes the primary focus of numerous research studies. Covariate shifts in CPath can be attributed to a plethora of factors including, but not limited to, different scanning technologies, variations in staining protocols and sample preparation methods, or tissue samples originating from different cancers or even different species. It is prudent to anticipate a covariate shift in any practical CPath problem. \cref{fig:shifts}E shows an example of covariate shift where samples of the same tissue slides are scanned with two different scanners, resulting in visually distinct color and feature space representations. 

\subsubsection{Prior Shift}
\label{sec:ds2}
The prior shift is characterized by differences in the distribution of priors (labels) between the source and target domains, \ie, $P^s_Y \neq P^t_Y$. A schematic of the prior shift is depicted in \cref{fig:shifts}B where the appearance of samples is the same over two domains but their label distributions vary significantly.
This kind of domain shift is common in CPath when the source and target domains are pulled from different datasets. For instance, when the proportion of classes differs between domains (bottom plot in \cref{fig:shifts}F) or in survival analysis where the number of events (or probability of survival of individuals) in the source and target domains starkly contrast (top plot in \cref{fig:shifts}F). The extension of model application from ROIs to WSIs can also instigate a prior shift because models are often trained on specific ROIs and encounter a very different distribution of labels in WSIs. A case in point is a mitosis detection algorithm, trained on image tiles abundant with mitosis, which may falter when applied to normal WSIs where mitosis is rare.

\subsubsection{Posterior Shift}
\label{sec:ds3}
The posterior shift or `concept shift' is characterized by a discrepancy in the conditional label distribution across source and target domains, specifically when $P^s_{Y|X} \neq P^t_{Y|X}$. In essence, posterior shifts represent variations in labels for the same data as illustrated in \cref{fig:shifts}C where samples in the same source domain (both circles' boundaries are dashed lines)  are labeled completely differently by two different annotators although their respective label distributions are aligned (this is an extreme case where the label of all examples are swapped). This phenomenon is typically encountered in subjective labeling tasks in CPath, such as mitosis annotation (as shown in the example of \cref{fig:shifts}G) \cite{veta2016mitosis,saldanha2020global}, Gleason grading \cite{kweldam2016gleason,175_bulten2022cpath}, nuclear pleomorphism assessment \cite{chowdhury2006interobserver,robbins1995histological}, etc.
However, posterior shifts are not solely due to subjective labeling discrepancies. For example, in survival analysis, a posterior shift can occur when treatment disparities between source and target populations lead to divergent survival outcomes (which are not subjective labels), despite identical initial covariates. 
Although posterior shift is a prevalent issue in many CPath applications, it is often under-addressed and necessitates tailored strategies.

\subsubsection{Class-conditional Shift}
\label{sec:ds4}
Class-conditional shift refers to the scenario where the conditional distributions of covariates 
for a specific class or sets of classes are different across source and target domains, \ie, $P^s_{X|Y=y} \neq P^t_{X|Y=y}$. Essentially, this happens when the same label is associated with different data characteristics in the source and target domains. This shift is modeled in \cref{fig:shifts}D where the appearance of only one class (rectangle) has changed across two domains while the distributions of other classes are aligned. In CPath, this kind of shift could arise when the characteristics of a certain pathology class vary between different populations, different disease stages, or different treatment responses. For instance, the morphological features of cancer cells from early-stage patients may differ significantly from those in late-stage patients (shown in \cref{fig:shifts}H). Although these cells belong to the same class 1 (tumor epithelial for example), the features associated with them change across different stages of the disease (source and target domains) while features for class 0 cells (connective cells for example) do not change, illustrating a class-conditional shift. Furthermore, this type of shift can also be caused by interobserver variability in pathological grading where one observer might label certain morphological changes in tissue differently from another, leading to the same class label being associated with different image features. While there is some overlap with covariate shift (DS Type 1) in terms of feature variation, class-conditional shift specifically focuses on within-class variations, and addressing it does not inherently resolve covariate shifts which involve broader, class-agnostic variations in the data distribution. This type of domain shift, while less frequently addressed, presents a significant challenge in CPath and requires the development of robust models that can generalize well across these shifts.

\subsection{Domain generalization}
The fundamental notion of `generalization' implies the ability of a predictive function $f:X\rightarrow Y$ to perform well on unseen data. Specifically, when a model is trained on data from source domain $(x,y) \sim P_{XY}^{s}$, it should generate accurate predictions in target domain as well, $E[f(x)|x\sim P_X^t]<\epsilon$, under the usual \emph{i.i.d} assumption, $P^s_{XY} = P^t_{XY}$.

Domain generalization aims to build a model using data from the source domains $D^s$ that perform reasonably well on the unseen target domains $D^t$, even when the joint distributions of the source and target domains do not match, $P^s_{XY} \neq P^t_{XY}$ (see \cref{fig:related}A). In other words, DG is a strategy to learn from the source domain in a way that generalizes to novel target domains despite any distributional shifts. The domains are typically similar but distinct, with each associated with a different joint distribution.

DG can be studied under two different settings: multi-source and single-source. Multi-source DG assumes that data from multiple distinct but related domains are available, and the model aims to learn patterns that generalize across these domains \cite{blanchard2011generalizing}. Single-source DG, on the other hand, assumes that training data comes from a single domain, a setup that is closely related to the problem of OOD robustness \cite{salehi2021unified,bulusu2020anomalous,tang2021selfnorm}, especially in the context of CPath \cite{208_foote2022cpath,foote2021now,ghaffari2022adversarial}.

\subsection{Related Concepts}
\begin{figure*}
    \centering
    \includegraphics[width=0.7\textwidth]{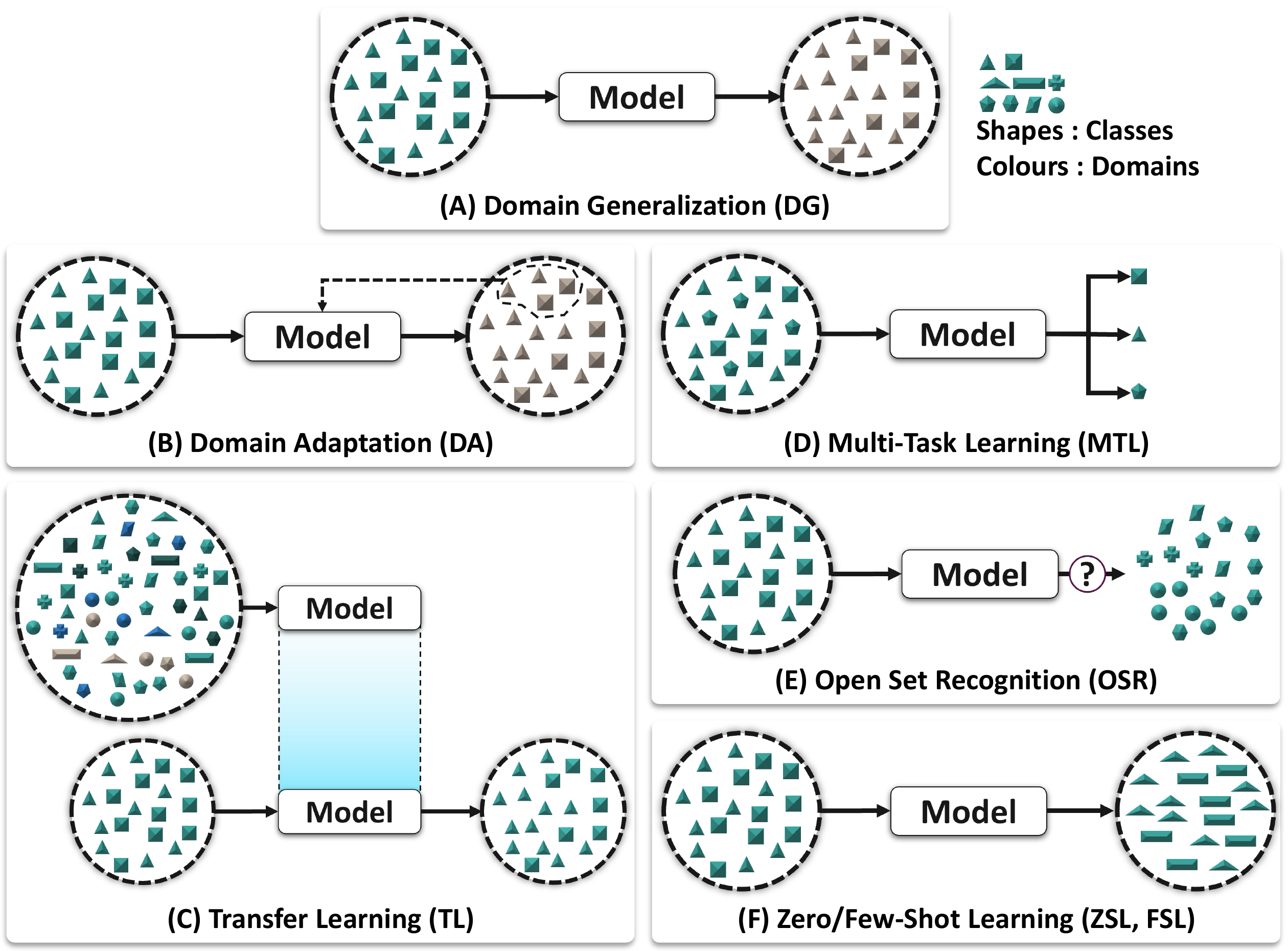}
    \caption{Schematic representation of Domain generalization and other related topics where the shape and color of objects represent their classes and domains, respectively. (A): Domain generalization (DG) trains a model that performs well on data from the unseen domain (characterised by different colors), (B): Domain Adaptation (DA) uses data from the unseen domain during fine-tuning to adapt, (C): Transfer Learning (TL) robust models trained on large-scale datasets are fine-tuned on new tasks, (D): Multi-Task Learning (MTL) simultaneously learns multiple related tasks using a shared model, (E): Open Set Recognition tries to identify and reject unknown classes, and (F): Zero/Few-Shot Learning (ZSL, FSL) uses no or few samples from new task to deal with changes in the labels.}
    \label{fig:related}
\end{figure*}

While there are other topics in machine learning that may appear similar to DG, they differ in several important ways. Some of the most important concepts and their differences from DG are explained below \cite{110_zhou2023general}:

\paragraph{Domain Adaptation (DA)} perhaps the most closely related topic to DG, deals with a scenario where the source and target domains have different data distributions for the same task.
The only difference with DG is that DA assumes that some form of target domain data (either labeled or unlabeled) is available during training (see \cref{fig:related}B). 

\paragraph{Transfer Learning (TL)} involves learning representations from a source domain, and transferring these learned representations to enhance learning in a related but different target domain \cite{weiss2016survey}. The most common practice in TL is pre-training a model on a large-scale dataset (such as ImageNet \cite{deng2009imagenet}) and then fine-tuning the model on a smaller, specific target task (see \cref{fig:related}C). Unlike DG, TL typically assumes the availability of target domain data (often in large amounts) during fine-tuning.

\paragraph{Multi-Task Learning (MTL)} aims to simultaneously learn multiple related tasks using a shared model, leveraging the shared information and interdependencies among the tasks to improve the overall performance. This is typically achieved by designing a shared representation with task-specific branches or layers in the model \cite{zhang2021survey} (see \cref{fig:related}D). However, unlike DG, MTL typically assumes that the distributions of the source and target tasks are identical, and it does not explicitly handle distribution shifts.

\paragraph{Open Set Recognition (OSR)} refers to the situation where some test samples may come from classes that have not been seen during training \ie, $\mathcal{Y}^s \neq \mathcal{Y}^t$. It assumes that the classes encountered during training (the closed set) do not cover all possible classes that may appear at test time, and thus it needs to discriminate between known and unknown classes \cite{scheirer2012toward,bendale2015towards} (see \cref{fig:related}E). Unlike DG, the main challenge in OSR lies in identifying and correctly rejecting unknown classes, rather than performing well on new domains with known classes.

\paragraph{Zero/Few-Shot Learning (ZSL)} focuses on predicting classes that were not observed during training. ZSL often involves learning a semantic embedding space, where the learned features of an instance and the semantic representation of its class are close to each other (see \cref{fig:related}F). At test time, the class with the closest semantic representation to the instance's features in the embedding space is predicted \cite{wang2019survey}. While there may be similarities with DG, ZSL mainly deals with changes in the labels, whereas DG handles shifts in both the feature and label distributions.

It is essential to consider these nuances to distinguish DG from these related concepts. Understanding the particularities of each setting helps to identify the most suitable approach for different real-world tasks. For more information on these comparisons refer to \cite{110_zhou2023general,112_wang2021general,122_farahani2021general,55_csurka2017general}. Although we have introduced the differences between these concepts and DG, it is important to note that some techniques from MTL and ZSL can also be utilized for DG, as we will see in the next section.

\section{Domain generalization methods: a survey}
\label{sec:dg_survey}
This section offers a comprehensive overview of domain generalization methods that have been proposed or applied in the context of CPath. Following the established approach in the literature \cite{110_zhou2023general,112_wang2021general}, and based on our analysis of DG methodologies, we have classified the reviewed papers into eight main groups: domain alignment, data augmentation, domain separation, meta-learning, ensemble learning, model design, pretraining, and regularization strategies. The subsequent sections will delve into the working principles of these categories and present exemplary CPath papers that fall within each group. The schematic overviews of all method categories are depicted in \cref{fig:methods}, enabling the reader to compare their respective working principles. Additionally, we have compiled a summary of the reviewed papers in \cref{tab:methods}, along with their respective advantages and disadvantages.

\subsection{Domain Alignment}
\label{sec:alignment}
Domain alignment-based approaches aim to bridge the gap between diverse source domains to capture feature representations that are invariant to domain variations. This alignment can be achieved through modifications in the data (image), feature, and/or classifier spaces, as illustrated in \cref{fig:methods}A and further elaborated in subsequent sections. In the CV community, numerous methods have been proposed to align the distribution of feature representations across different domains. This is commonly achieved through domain adversarial learning \cite{ganin2016dann,135_li2018general,goodfellow2020generative} or by minimizing distributions moment distances \cite{erfani2016robust,muandet2013domain}, contrastive losses \cite{yoon2019generalizable,motiian2017unified,kim2021selfreg}, Maximum Mean Discrepancy (MMD) distance \cite{gretton2012kernel,wang2021rethinking}, or KL Divergence \cite{wang2021respecting,134_liu2021general}. However, in the context of CPath, various research efforts have been made to align input data, such as applying stain normalization techniques, in order to obtain comparable feature representations from diverse domains.

\subsubsection{Domain alignment by stain normalization}
Stain normalization (SN) is a preprocessing step that counters discrepancies in the color of histology images due to varied staining procedures and scanner variations \cite{197_bejnordi2014cpath}. Core to this method is the utilization of a target image with a desired stain distribution. The objective is to modify source images such that their color distributions align with that of the target. Techniques in SN span from basic linear scaling and histogram matching \cite{gonzalez2009digital} to advanced methods represented by the likes of Ruifrok \cite{198_ruifrok2001cpath}, Macenko \cite{199_macenko2009cpath}, Reinhard \cite{200_reinhard2001cpath}, Vahadane \cite{201_vahadane2016cpath}, and Khan \cite{202_khan2014cpath}. These approaches exploit the stain matrix (representing stain colors) and the stain concentration (indicating pixel-wise stain quantity).  These methods, in essence, adjust the color of the source image without altering its structural details by keeping its stain concentrations intact while replacing its stain matrix with that of the target image.

This process has been essential in CPath studies, enhancing the performance in H\&E slide analysis \cite{79_kather2019cpath,80_sebai2020cpath,196_jahanifar2022cpath,21_tellez2019cpath,bilal2021development,171_bueno2020cpath,184_razavi2021cpath}. Notably, some works focused on extracting the Hematoxylin component (H-Channel) using SN algorithms, which minimized domain variability given that the H-channel across different centers exhibits less variability \cite{90_zhang2022cpath,23_anand2020cpath}. Accessibility to such methods through toolboxes like TIAtoolbox has increased their uptake \cite{pocock2022tiatoolbox}. Moreover, contemporary research aims to devise reliable SN techniques for WSIs \cite{203_zheng2019cpath, 204_hoque2021cpath, 205_vicory2015cpath, 206_tosta2019cpath}.

However, the effectiveness of SN in CPath is not without challenges. Studies have spotlighted the instability of modern techniques \cite{82_vu2022cpath} and how performance variability is affected \cite{202_khan2014cpath,149_khan2022cpath}. Furthermore, it is noted that SN introduces computational overhead during inference \cite{21_tellez2019cpath}. A recent exploration indicated that conventional SN methods cannot entirely erase site-specific data from WSIs, thus, not ensuring domain/site agnostic feature extraction \cite{207_dawood2023cpath}.

However, the value of SN remains undeniable, especially when it comes to achieving accurate quantitative measurements and comparisons in slides stained with specific immunohistochemical markers \cite{chung2007quantitative,van2016image,tam2016method,199_macenko2009cpath}.

\subsubsection{Domain alignment using generative models}
The application of generative adversarial neural networks (GANs) \cite{goodfellow2020generative,creswell2018generative} and style transfer \cite{gatys2015neural,jing2019neural} has ushered a series of advancements in CPath with a specific emphasis on the alignment of pathology images from diverse domains. A central theme in this endeavor has been the use of CycleGAN-based models \cite{zhu2017unpaired} for domain alignment, stain normalization, and stain translation. CycleGANs have been observed to be a better choice than traditional stain normalization methods, particularly when the domain gap is significant \cite{21_tellez2019cpath,31_boyd2022cpath,103_shaban2019cpath,118_gadermayr2019cpath}. Pioneering works include the study by Moyes \etal \cite{27_moyes2023cpath} that introduced a multi-channel autoencoder for domain mapping, and Shin \etal's \cite{54_shin2021cpath} approach of style transferring training images to the style of test images. 

Using this technology, a few unique contributions have been made to SN. Cong \etal \cite{81_cong2022cpath} introduced the color Adaptive Generative Network (CAGAN) to ensure consistency in outputs while addressing stain color variations. Xing \etal \cite{106_xing2022cpath} combined GANs with a nuclei detection model, emphasizing consistent predictions between translated and original images. In contrast, the workflow proposed by Jia \etal \cite{84_jia2022cpath} operated in the feature space, relying on adversarial training for texture feature encoding. This idea of feature space alignment was further explored by Ke \etal \cite{78_ke2021cpath} and Cong \etal \cite{99_cong2021cpath}, the latter eliminating the need for paired ground truth data from source and target domains. Zhao \etal \cite{83_zhao2022cpath} also reformulated SN as a self-supervised re-staining process, setting a new benchmark over traditional GAN-based approaches.
Additionally, some studies extended GAN-based models to tackle unique challenges in CPath. For instance, Geng \etal \cite{15_geng2022cpath} addressed defocus blur in WSIs by generating focused images from unfocused counterparts. On the other hand, Wagner \etal \cite{113_wagner2022cpath} introduced BottleGAN for federated learning in CPath, aiming to normalize local data distributions across laboratories, thus improving generalization across lab datasets.
In general, GANs have shown significant promise in CPath for domain alignment and SN, often surpassing traditional techniques. However, a noteworthy limitation is their sensitivity to architectural changes, which might inadvertently alter diagnostic markers, potentially limiting their reliability in certain applications \cite{121_vasiljević2022cpath}.

\subsubsection{Feature space alignment}
In CPath, the arena of feature space alignment introduces several pivotal methodologies, primarily centered around adversarial training, KL divergence techniques, and other ancillary approaches.
The popularity of domain adversarial training \cite{ganin2016dann} is witnessed by its deployment in numerous studies. Quiros et al. \cite{143_quiros2021cpath} pioneered a framework via unsupervised learning and GANs, targeting phenotype representations anchored on tissue and cellular attributes. This venture significantly demystified morphological nuances across cancer variations. Analogously, domain adversarial training \cite{ganin2016dann} found applications in molecular subtyping from pathology imagery by Sirinukunwattana et al. \cite{183_kirsuk2020cpath}. The challenge of ensuring robustness against variability from multiple imaging sources, such as different scanners, was addressed by similar approaches in \cite{188_wilm2021cpath,189_lafarge2019cpath}. Wang et al. \cite{109_wang2021cpath} then brought forward a nuclei detection framework, encapsulating both image and instance-level alignment orchestrated through adversarial learning. Inspired by HoVer-Net \cite{graham2019hover}, Li et al. \cite{3_li2022cpath} introduced a self-supervised domain adaptation strategy for nuclei segmentation, encompassing class-level feature alignment for domain gap minimization and a pseudo-labeling mechanism bolstered by nuclei-level prototyping. The outcomes manifested the viability of class-aware adaptation, accentuating the strength of self-supervised learning in domain adversarial training.

Sikaroudi et al. \cite{28_sikaroudi2022cpath} unveiled a representation learning approach that remained undeterred by hospital-specific variances. The approach married KL divergence and Triplet loss functions, ensuring cohesion and separation of instances, irrespective of the domain. Salehi et al. \cite{147_salehi2022cpath} implemented a domain generalization mechanism for classifying Hematological malignancies, using a Mask R-CNN-based strategy to capture white blood cell attributes. The captured attributes, when compressed in a latent space, were subjected to domain adaptation through group normalization. The ensemble of KL divergence, feature similarity, and cluster-based loss was further reinforced by Raipuria et al. \cite{14_raipuria2022cpath} to achieve stain invariance. By juxtaposing raw and stain-altered images in a dual-strategy, their model surpassed counterparts leveraging traditional stain normalization. Additionally, Sharma et al. \cite{19_sharma2022cpath} applied Jensen-Shannon Divergence-based mutual information loss  \cite{englesson2021generalized}, reflecting a broader palette of segmentation methods for nuclei.
As versatile feature space alignment methods are, most of them require domain labels which are not always available. Furthermore, to align the distributions of different domains effectively, some methods require large amounts of data to accurately estimate the underlying distributions.

\subsection{Data augmentation}
\label{sec:augment}
Data augmentation enhances model generalization by introducing training data variations \cite{chlap2021review,shorten2019survey}. This is achieved through computational transformations of existing datasets or by collecting new data, as in Akram \etal\cite{58_akram2018cpath}, where a pretrained model mined mitotic candidates from unlabeled WSIs. Although data augmentation has been explored in both image and feature spaces \cite{devries2017dataset,li2021simple,liu2020deep,ko2020embedding}, in CPath, most techniques focus on image transformations or generative neural networks.

\begin{figure*}[!ht]
    \centering
    \includegraphics[width=\textwidth]{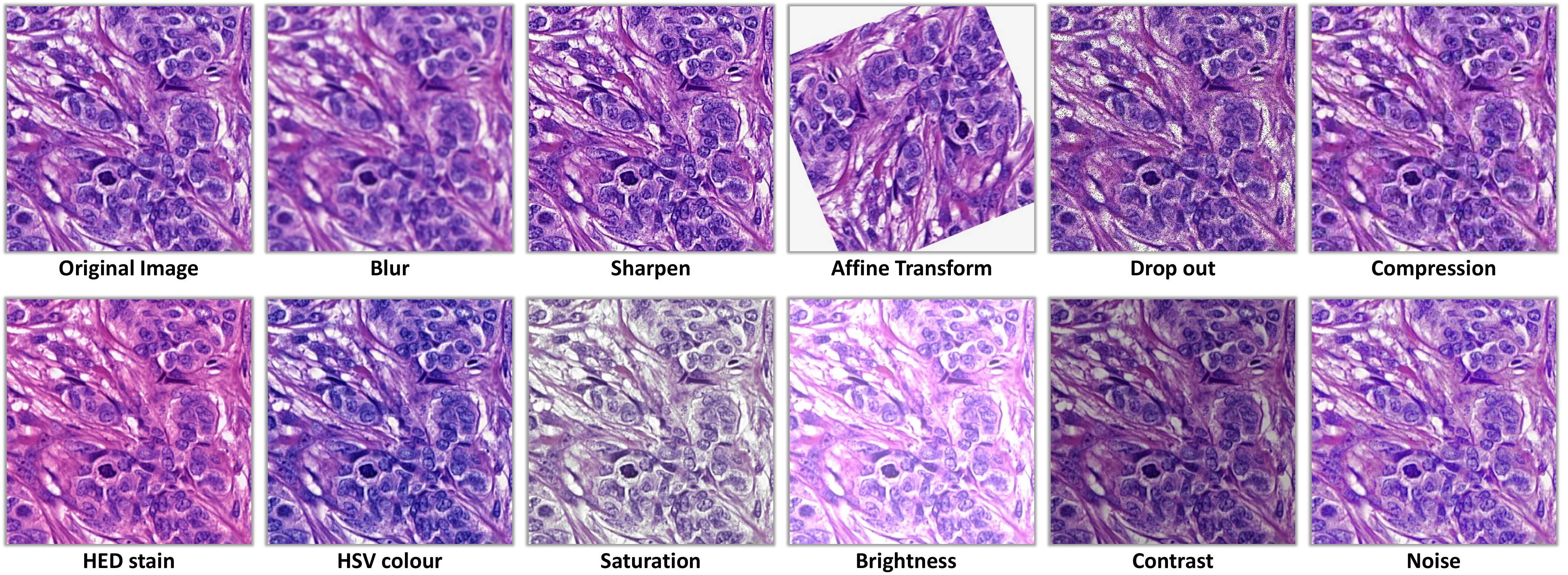}
    \caption{Examples of commonly used image augmentation techniques for CPath.}
    \label{fig:augs}
\end{figure*}

\subsubsection{Image transformation}
Enhancing domain generalizability of models in CPath is frequently achieved via image transformation techniques. Central to these strategies are geometric augmentations, brightness and contrast adjustments, and HSV color jittering, among others. These transformations, when applied in CPath studies \cite{59_alemi-koohbanani2020cpath,NuClick_SPIE,63_jarkman2022cpath,85_zhang2022cpath,93_bándi2023cpath,97_zhu2021cpath,150_holland2020cpath,163_faryna2021cpath,29_pohjonen2022cpath,94_kleppe2021cpath,184_razavi2021cpath,185_kondo2021cpath}, have shown promising improvements in model generalizability against prevalent variances in histology datasets, such as color variations and compression artifacts \cite{82_vu2022cpath,21_tellez2019cpath,29_pohjonen2022cpath}.

Unique to histology is stain augmentation (SA), notably the Hematoxylin-Eosin-DAB (HED) stain augmentation. This technique decomposes RGB histology images into stain components, making random perturbations and then recomposing the images. The effectiveness of HED augmentation has been corroborated by its contribution to classifier training \cite{21_tellez2019cpath}, mitosis detection algorithms \cite{196_jahanifar2022cpath,209_jahanifar2022cpath,186_nateghi2021cpath}, and as a Test-Time Augmentation (TTA) technique \cite{114_xu2022cpath}. Notably, methods have been developed to counter the generation of unrealistic color samples with SA. Chanh \etal \cite{56_chang2021cpath} proposed the Stain-MixUp method, and Marini \etal \cite{148_marini2023cpath} introduced a data-driven color augmentation method, with both focusing on enhancing the diversity of color appearance. Moreover, Shen \etal \cite{57_shen2022cpath} combined stain normalization and augmentation to generate biologically realistic samples, while Faryna \etal \cite{163_faryna2021cpath} automated data augmentation policy selection by improving RandAugment framework \cite{cubuk2020randaugment}, tailor-made for histology-specific augmentations.

A novel angle to data augmentation involves Fourier-based approaches \cite{xu2023fourier}. Through the Fourier transformation of images, various studies \cite{32_li2022cpath,61_wang2023cpath,162_yang2021cpath} manipulated the amplitude spectrum, maintaining that domain-invariant details predominantly reside in the phase spectrum post-transformation.

While traditional augmentation techniques remain invaluable for training neural networks in CPath \cite{perez2017effectiveness,29_pohjonen2022cpath, 21_tellez2019cpath, 140_tellez2018cpath,163_faryna2021cpath}, it is imperative to find a balance to prevent label shift and guarantee the generation of biologically plausible images \cite{shorten2019survey,148_marini2023cpath,56_chang2021cpath,57_shen2022cpath,163_faryna2021cpath}. Strong augmentations, as suggested by Pohjonen \etal \cite{29_pohjonen2022cpath}, can be advantageous in bolstering the consistency of CNNs in some classification application but may prove destructive in other sensitive applications.

\subsubsection{Generative neural networks}
Generative neural networks have surfaced as pivotal tools in computational pathology (CPath) to confront issues related to limited data availability and dataset size \cite{goodfellow2020generative, antoniou2017data}. These networks, leveraging their capability to craft new data samples while maintaining inherent semantic content, allow for inventive data variations surpassing rudimentary geometric and color modifications \cite{calimeri2017biomedical,celard2023survey}.

Central to the augmentation efforts, various techniques have been introduced. The neural style transfer was used by Yamashita \etal \cite{92_yamashita2021cpath} to modify image styles, capitalizing on both medically relevant and irrelevant style transformations. Echoing this, Chung \etal \cite{187_chung2021cpath} manipulated images between labeled and unlabeled domains, while Wu \etal \cite{25_wu2019cpath} innovatively simulated stain effects for Glomerulus classification. Techniques like SHI-GAN by Falahkheirkhah \etal \cite{98_falahkheirkhah2023cpath} and GAN-based test-time augmentation by Scalbert \etal \cite{102_scalbert2022cpath} emphasized the generative networks' versatility, while Zaffar \etal's EmbAugmenter \cite{161_zaffar2023cpath} showcased robustness improvements in the embedding space.

Tackling staining challenges, Vasiljević \etal \cite{35_vasiljević2021cpath} and Haan \etal \cite{193_haan2021cpath} employed CycleGAN and StarGAN, focusing on synthetic generation across diverse staining domains. Lin \etal's InsMix technique \cite{62_lin2022cpath}, through nuclear instance augmentation, and Roy \etal's residual cycle-GAN \cite{104_roy2021cpath} underlined the significance of cross-domain transformations. In an intriguing exploration, Tsirikoglou \etal \cite{146_tsirikoglou2021cpath} bridged domain gaps by augmenting tumor images from various cancers, emphasizing the value of style transferred out-of-domain samples.

Related to conditional GANs \cite{mirza2014conditional} for augmentation, HistoGAN by Xue \etal \cite{100_xue2021cpath} generated realistic breast cancer histopathology patches, while Chen \etal \cite{101_chen2022cpath} targeted the renal cell carcinoma subtypes, addressing limited subtype-specific data challenges. Fan \etal \cite{141_fan2022cpath}, on the other hand, transitioned WSI styles, demonstrating the potential of generative models to counteract class-conditional shifts due to varying sample preparation methods.

However, as promising as these methodologies are for CPath, they come with inherent challenges. Training complexities, significant data requirements, and the potential risk of generating misleading data, especially in stain transfer applications, demand caution. Visual assessments might be insufficient for the apt training of such models, emphasizing the need for rigorous verification \cite{35_vasiljević2021cpath}.

\subsection{Domain Separation}
\label{sec:separation}
\begin{figure*}
    \centering
    \includegraphics[width=\textwidth]{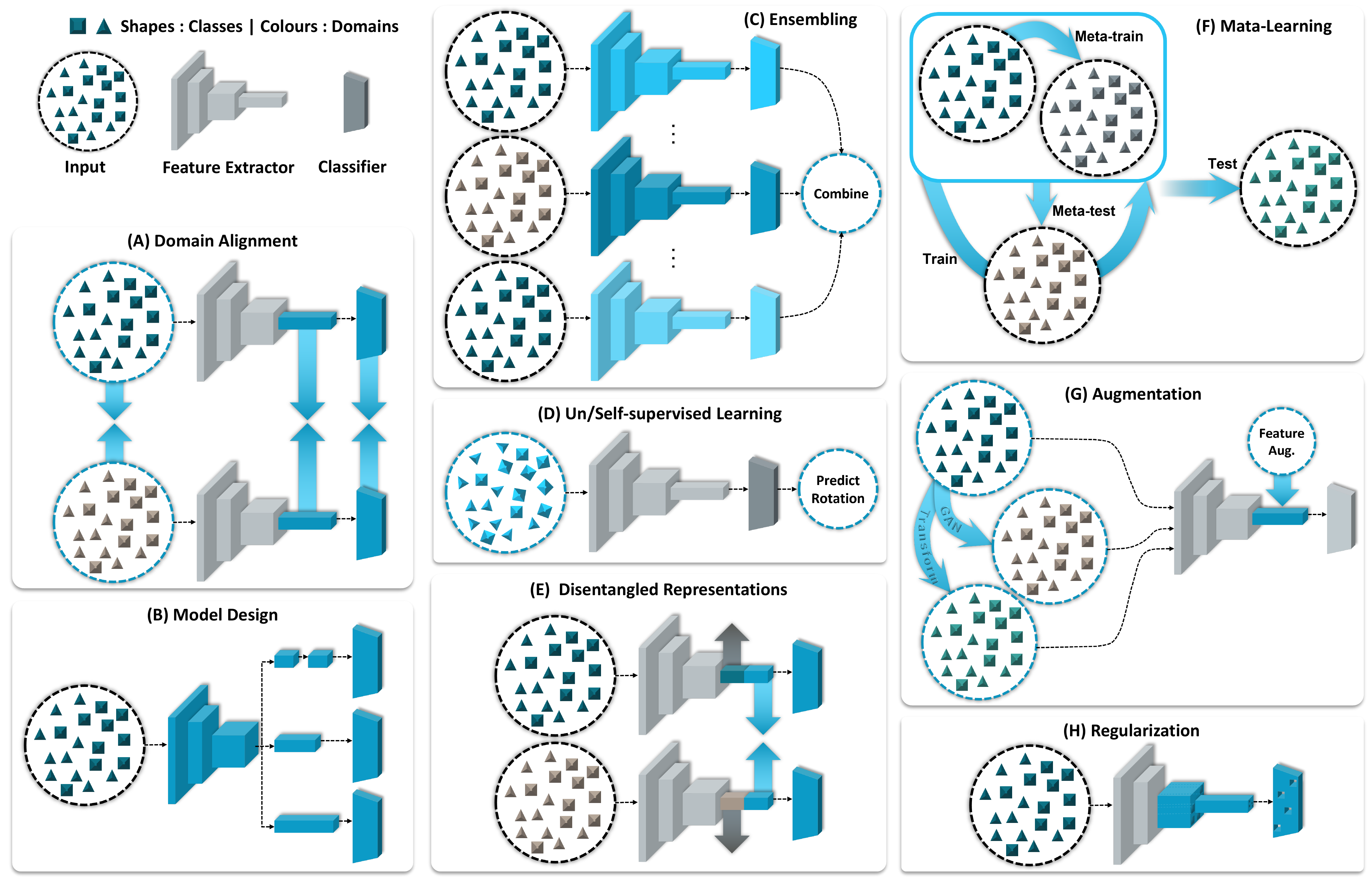}
    \caption{Different categories of domain generalization methods illustrated where samples come from different classes (shape: rectangle or triangle) and domains (colors). Depending on the category and the algorithm, DG methods can operate on any of the `input data', `feature extractor', or `classifier' sections of a DL approach, which are highlighted with blue color for different categories.}
    \label{fig:methods}
\end{figure*}

Unlike domain alignment methods, domain separation methods aim to learn disentangled representations for images originating from different domains. This is typically accomplished by segregating the domain-specific and domain-agnostic feature maps \cite{li2017deeper} or using generative models to learn independent feature spaces for different domains \cite{ilse2020diva}. Once latent subspaces are separated for different domains, single or multiple classifiers can be trained on the domain-agnostic or domain-specific feature representations \cite{chattopadhyay2020learning} (as depicted in \cref{fig:methods}E).

However, this category has attracted less attention in both general CV and medical image analysis communities. 
In the limited research conducted, Chikontwe \etal \cite{107_chikontwe2022cpath} proposed a method, called FRMIL, which improved model performance in MIL tasks (such as breast cancer metastasis detection or MSI detection in colorectal cancer) by adjusting the feature distributions to enhance separability. The method was primarily focused on re-calibrating instance features within the same domain. 
Using generative models, Wagner \etal \cite{49_wagner2021cpath}  proposed a new color transfer model that can be used as a stain augmentation technique. Essentially a GAN tailored for CPath that ``disentangles" the content of the image, \ie, the morphological tissue structure, from the stain color attributes, and thus can preserve the structure while altering the color.
Additionally, some research studies in CPath focused on isolating only the Hematoxylin component (H-channel) of histology images to design domain-independent models \cite{90_zhang2022cpath,23_anand2020cpath}.

Domain separation in CPath enhances generalization through distinct domain-specific and agnostic features, facilitating easier adaptation to new domains and clearer model interpretation. However, true feature disentanglement is challenging, requiring diverse multi-domain data, with any imperfections risking reduced performance.

\subsection{Meta-learning}
\label{sec:meta}

Meta-learning, or ``learning to learn," has emerged as a versatile technique within domain generalization (DG), demonstrating potential in tackling its challenges \cite{hospedales2021meta}. It equips models to swiftly adapt to unfamiliar domains or tasks \cite{li2017meta,li2018mldg,finn2017maml,lee2019meta}. Notably, the Model-Agnostic Meta-Learning (MAML) algorithm captures the essence of this approach. It trains on a variety of tasks, facilitating knowledge transfer to new tasks while being exposed to domain shifts \cite{finn2017maml}. This capability is enhanced when coupled with other DG techniques like adversarial training and contrastive losses \cite{hospedales2021meta}.

Within the CPath landscape, meta-learning has been applied in several innovative ways. For classification, Fagerblom \etal \cite{48_fagerblom2021cpath} employed MAML+, an advanced version of MAML with cosine annealing meta-optimizer learning rate scheduler, to tackle tasks like Her2 status classification, with the technique outperforming its standard counterpart \cite{finn2017maml}. Sikaroudi \etal introduced methods for embedding across magnification levels and training hospital-independent classifiers \cite{16_sikaroudi2021cpath,28_sikaroudi2022cpath}. Liu and team \cite{44_liu2021cpath} enhanced the generalization of patient-independent MSI classifiers through meta-contrastive learning, while Li's work \cite{129_li2022multi} on a multi-modalities study demonstrated superiority in classifying histology images with varying preparation procedures.

For segmentation, Han and colleagues \cite{46_han2022cpath} proposed a fusion of MAML and multi-task learning, concentrating on nuclear segmentation. Yuan \etal \cite{45_yuan2021cpath} introduced MetaHistoSeg, a comprehensive framework for histopathology image segmentation. Lastly, Shakeri \etal \cite{22_shakeri2022cpath} conducted benchmarks on several meta-learning approaches, extending their research into few-shot learning algorithms on histology image classification tasks.

Meta-learning presents both notable strengths and challenges in computational applications. Its adaptability allows integration with any model architecture and efficient scaling to handle vast datasets. Moreover, it stands out in swiftly adapting to new domains, particularly under significant domain shifts. However, meta-learning's computational intensity and dependence on multiple hyperparameters complicate its deployment \cite{hospedales2021meta}. The need for diverse source domains during training further restricts its universal applicability. While promising, its efficacy is determined by the balance between its advantages and inherent complexities.

\subsection{Ensemble Learning}
\label{sec:ensemble}
Ensemble learning improves accuracy by combining predictions from diverse base models, overcoming individual limitations \cite{yang2023survey}. In DG, this diversity is harnessed through methods like aggregation, bagging, boosting, or stacking \cite{dietterich2000ensemble}. 
In CPath, various approaches have been employed to enhance performance using ensembling techniques \cite{pimkin2018ensembling,rathore2022survival,abdollahi2022detection,nguyen2021classification,fuchs2009randomized,139_linmans2023cpath}. Traditional aggregation methods involved training the same model on different subsamples of the data or training different models on the same data and then taking the arithmetic average of the test set predictions \cite{dietterich2000ensemble,191_liang2021cpath,192_kotte2022cpath}. Linman \etal \cite{139_linmans2023cpath}, found such ensembles of deep CNNs to work well in detecting far out-of-distribution data. However, its limitation lies in the fact that the algorithm may not adequately represent the required data, especially in complex and diverse domains.

To overcome this limitation, several studies in CPath have explored an ensemble of distinct model architectures trained on consistent data. For instance, Yengec-Tasdemir \etal \cite{52_yengec-tasdemir2023cpath} integrated predictions from different model variants, whereas Luz \etal \cite{53_luz2021cpath} applied a single architecture across varied image processing methods. Both demonstrated superior results with such ensemble methods, emphasizing their utility in tackling domain-specific challenges, such as those arising from scanner or center variations.

A distinct ensemble approach, frequently employed in challenges where test data remains elusive during training, capitalizes on cross-validation \cite{eisenmann2023winner}. Here, different models are trained across varied data folds, with their collective predictions then integrated to enhance the ensemble's robustness \cite{209_jahanifar2022cpath, 196_jahanifar2022cpath, 210_shephard2022cpath, 51_hägele2022cpath}. Model stacking integrates predictions from various base models using a meta-model. Techniques like linear/logistic regressions or neural networks help in this combination. Sohail \textit{\etal} \cite{50_sohail2021cpath} utilized predictions from diverse base models as inputs for a multi-layer perceptron, aiming to classify patches for mitosis presence. This method surpassed individual approaches, emphasizing the value of understanding intricate inter-model relationships. 



By leveraging model diversity and combining predictions from multiple models, these approaches have shown improved results compared to individual models. This technique is very flexible (can be used in conjunction with other DG methods) and applicable to almost any model design and domain shift. However, one major downside of the ensembling method is the extra computational cost that using different models brings during training and inference.

\subsection{Tailored model design strategies}
\label{sec:design}
Within the sphere of CPath, various tailored model design strategies have been explored to address domain-specific challenges, ensuring robustness and improved generalizability. These design strategies hinge on the idiosyncrasies of histopathology images, along with their diverse sources.

\subsubsection{Problem-specific designs}
Numerous research papers have put forth distinctive model designs and loss functions addressing CPath challenges and achieving better generalizability. Chen et al.\cite{151_chen2021cpath} developed an approach for WSI classification targeting thyroid, colon, and cervical samples, utilizing unit-level CNNs fused with attention mechanisms. Similarly, Li et al.\cite{155_li2023cpath} introduced the versatile embedded fusion mutual learning model for tissue classification in various cancers, emphasizing its mutual learning and feature fusion capabilities. Notably, Tang et al. \cite{145_tang2021cpath} adopted neural architecture search (NAS) to find the most generalizable classifier architecture for CPath applications.

For nuclei detection, Javed et al.\cite{42_javed2021cpath} incorporated spatial nucleus constraints and contextual information to improve generalizability. Rojas-Moraleda et al.\cite{115_rojas-moraleda2017cpath} proposed a multi-phase segmentation method, emphasizing cell nuclei properties through a series of image processing steps that work across domains. Xie et al.\cite{41_xie2018cpath} showcased a structured regression model, excelling in nuclei detection across different modalities. Razavi et al. \cite{39_razavi2022cpath} leveraged conditional GANs for simultaneous mitosis and nuclei segmentation, outperforming standard UNet for generalization to new domains.

Silva et al.\cite{36_silva2022cpath} addressed class-conditional shifts with a novel network architecture coupled with attention mechanisms that focused on glomeruli segmentation task across different staining domains. 
Highlighting the significance of rotational symmetry in histology images, Graham et al.\cite{37_graham2020cpath} presented dense steerable filter CNNs, leading to enhanced feature discernment robust against rotation. Similarly, Lafarge et al.\cite{38_lafarge2021cpath} proposed a rotational-equivariant representation, bolstering model efficiency. Anand et al. \cite{23_anand2020cpath} proposed a `switching loss' technique to adaptively adjust for the contribution of imbalanced background and foreground pixels in the segmentation task, alleviating prior shift during training. 

\subsubsection{Multi-Task Learning (MTL)}
MTL has emerged as a promising method in DG across various tasks in CPath. By training synchronously on multiple tasks, models glean more pertinent and resilient features. Notable implementations include Dabass et al.'s\cite{120_dabass2022cpath} multi-task UNet for gland segmentation and tumor classification in colorectal cancer and Wang et al.'s\cite{43_wang2021cpath} three-headed model for segmenting and classifying tumor tissues in Hepatocellular carcinoma region segmentation. Graham et al.\cite{119_graham2023cpath} further proposed the Cerberus MTL approach, leveraging data from multiple independent sources to concurrently segment glands, lumens, and nuclei in colon tissue. Furthermore, Vuong et al.\cite{182_vuong2021cpath} combined categorical and ordinal learning via MTL for cancer grading.

Tailored model designs in CPath offer several advantages, such as enhancing performance through domain-specific designs and shared representation. MTL is efficient, conserving computational resources, especially for WSI processing, and also serves as a protective measure against overfitting. However, these strategies come with their own challenges. Sophisticated designs introduce increased complexity. Additionally, MTL can be difficult to balance across tasks and may pose scalability challenges, potentially amplifying both model size and complexity.

\subsection{Pretraining strategies}
\label{sec:pretrain}
The abundance of unlabeled data in CPath can pave the way for enhanced feature encoding through pretraining strategies. These strategies not only bolster the generalizability across novel domains but also facilitate integration into comprehensive frameworks \cite{schmarje2021survey,142_tran2022cpath}. Pretraining algorithms notably encompass self-supervised \cite{jing2020self,zhang2023dive,159_kang2023cpath,154_sikaroudi2020cpath}, unsupervised \cite{dike2018unsupervised,hu2018unsupervised}, and semi-supervised \cite{van2020survey,1_marini2021cpath,0_neto2022cpath} learning methods. These can be singularly used or merged with other frameworks \cite{schmarje2021survey}. A notable combination of these methods is seen in \cite{142_tran2022cpath} where hierarchical contrastive learning amalgamated fully-supervised, self-supervised, and semi-supervised learning strategies to classify colon tissue and leukemia single cells.

\subsubsection{Self-supervised learning}
Self-supervised learning (SSL) emerges as a dominant pretraining approach where models harness their generated labels or annotations to derive valuable insights. This is mainly achieved either through pretext tasks or contrastive learning methods, each contributing distinctively to the CPath context \cite{jaiswal2020survey,123_chikontwe2022cpath}.

Pretext learning involves pretraining on auxiliary tasks (such as rotation prediction \cite{yamaguchi2021image} or solving a Jigsaw puzzle \cite{noroozi2016unsupervised}) using self-generated labels for capturing better representations. In the domain of pretext tasks, a myriad of studies have showcased innovative methods. For instance, techniques have been employed to mask out bands in hyperspectral images and regress them, yielding enhanced classification outcomes for pancreatic and gastric cancer-related conditions \cite{88_liu2022cpath,7_xie2022micro}. Other contributions in this realm involve encoding WSIs into discrete representations for slide retrieval \cite{194_chen2022cpath}, self-supervised segmentation nuclei \cite{116_haq2022cpath}, and pretraining Visual-Image-Transformers (ViT) \cite{dosovitskiy2020image}  using a specialized self-supervised technique called iBOT  \cite{zhou2021ibot} in \cite{195_filiot2023cpath}. Pathology-specific self-supervision tasks are explored in \cite{89_koohbanani2021cpath}, while a colorization task is investigated in \cite{144_fan2021cpath}. 

Self Supervised Contrastive Learning (SSCL) leverages differences between similar and dissimilar samples for pretraining \cite{jaiswal2020survey,he2020momentum}. Demonstrating its efficacy, Ciga \etal's model pretrains on an extensive dataset from 57 histopathology sources, enhancing performance in various tasks \cite{164_ciga2022cpath}. Beyond this, a fusion of contrastive learning with supervised loss has been explored for cancer sub-typing and WSI retrieval \cite{2_fashi2022cpath}. The field has also witnessed advanced techniques like a multi-channel autoencoder that refines image-patch representations \cite{27_moyes2023cpath}, lesion-aware contrastive learning \cite{6_li2022cpath} within the CLAM \cite{lu2021data} and TransMIL \cite{shao2021transmil} frameworks, and the IMPaSh \cite{180_vuong2023cpath} framework's integration of patch-shuffling with MoCo \cite{he2020momentum} for colorectal image classification. Moreover, a conditional variational autoencoder generator utilized contrastive loss to achieve commendable results in zero-shot learning, nearly matching fully supervised approaches \cite{8_mahapatra2021multi}. Expanding horizons feature Abbet \etal's design \cite{91_abbet2022cpath} targeted domain adaptation using self-supervised learning and contrastive loss, while challenges got innovative solutions from frameworks applying contrastive loss directly at the WSI level \cite{178_lazard2023cpath}, or from Sikaroudi \etal's triplet network which deepened histopathology insights \cite{154_sikaroudi2020cpath}. The field continues to flourish with methods like multi-view contrastive learning \cite{159_kang2023cpath} and varied SSCL approaches in CPath, highlighting stain-based, color perturbation, and spatially guided techniques \cite{4_qin2022cpath, 90_zhang2022cpath, 91_abbet2022cpath, 156_huang2021cpath, 157_li2021cpath}, collectively accentuating the evolving potential of contrastive learning in CPath.

Ultimately, a synergistic approach involving both pretext tasks and contrastive learning can prove highly effective for pretraining, as demonstrated by the work of Yang \etal \cite{5_yang2022cpath}. In their study, they employed a combination of a cross-stain prediction pretext task and a contrastive learning loss, showcasing the potential of merging these strategies.

\subsubsection{Semi-supervised learning}
Semi-supervised learning uses both labeled and unlabeled data to enhance deep learning models in situations with scarce training data. Initially trained on a limited labeled dataset, the model then assigns pseudo-labels to unlabeled data, which further refines the model. By incorporating more data, this approach addresses the challenges of limited labeled resources, increasing data diversity and enabling the model to capture diverse patterns and improve generalization \cite{van2020survey,schmarje2021survey}.

Diverse studies have fortified this approach in CPath. For instance, a semi-supervised framework for prostate image classification employed a teacher model to generate pseudo-labels, paving the way for weakly supervised student models \cite{1_marini2021cpath}. Another technique augmented images from a source domain and utilized pseudo-labels for predictions, shedding light on the method's versatility \cite{87_bozorgtabar2021cpath}. Furthermore, Neto \etal's approach leaned on a vast quantum of weakly labeled data, yielding superior model generalization across disparate centers \cite{0_neto2022cpath}. Lastly, findings from a study by Sikaroudi \etal accentuated the significance of semi-supervised models in CPath, although domain-specific datasets occasionally showcased marginally superior performance \cite{177_sikaroudi2023cpath}.

While pretraining strategies effectively harness the rich unlabeled data in CPath, enhance feature extraction, and demonstrate robustness to certain "nuisance factors", they also present challenges. The features extracted might not always be universally suitable, limited gains are observed with medium-sized labeled datasets, and the computational cost for training remains high. Nonetheless, the evolutionary trajectory of these methods underscores their significant potential in the continuously evolving landscape of CPath.

\subsection{Regularization strategies}
\label{sec:regularize}

Regularization techniques have always held prominence in machine learning to avoid overfitting and ensure model generalizability \cite{tian2022comprehensive}. They primarily work by introducing constraints or penalties that govern model complexity, driving the identification of more general patterns \cite{moradi2020survey,tian2022comprehensive,huang2020rsc}.
One prominent approach is the modulation of the loss function. L1 and L2 regularizations stand out here, influencing model weights to achieve a balance between interpretability, sparsity, and robustness.

An application in CPath by Minhas \etal \cite{10_minhas2021cpath} incorporated an L1 regularization term for a survival prediction model, enhancing its risk assessment capacity for distant metastasis. Furthermore, Liang \etal \cite{153_liang2023cpath} leveraged flooding, a regularization term, to set a minimum threshold on training loss, ensuring improved performance on unseen datasets.
Dropout \cite{srivastava2014dropout}, another regularization technique, randomly nullifies a fraction of input units or weights during training to boost model robustness. Jiang \textit{\etal} \cite{9_jiang2023cpath} employed feature-level dropout in a multi-head attention network for survival prediction, targeting reduced overfitting and better generalization.
Data augmentation, especially in image classification, offers another perspective on regularization. Su \textit{\etal} \cite{95_su2023cpath} proposed the Semi-LAC method, which stresses the learning of consistent features, regardless of the image being original or augmented, using a directional consistency loss. A modality-specific pruning technique was proposed by Cai \etal \cite{20_cai2021cpath} which revolved around using a domain classifier to ascertain image sources and subsequently pruning unrelated model parameters. Tailored for multi-source Ki67 images, this method, apart from exhibiting superior performance, ensured a compact model structure, beneficial for real-world deployments.

Overall, regularization strategies, including modification of the loss function and dropout, are flexible techniques that can reduce the chances of overfitting and complexity of the model \cite{moradi2020survey}, therefore improving model generalization. However, such methods can be sensitive to hyper-parameters and drastic usage may lead to under-fitting \cite{tian2022comprehensive}.

\begin{table*}[]
\label{tab:methods}
\centering
\caption{Summary of domain generalization method categories with their advantages and disadvantages.}
\begin{tabular}{@{}p{2cm}p{2cm}p{6.4cm}p{6.4cm}@{}}
\toprule\hline
\multicolumn{1}{c}{\textbf{Category}}        & \multicolumn{1}{c}{\textbf{Refrences}} & \multicolumn{1}{c}{\textbf{Advantages}}                                                              & \multicolumn{1}{c}{\textbf{Disadvantages}}                                                    \\ \midrule
\multirow{3}{2cm}{Domain alignment (\cref{sec:alignment})}          & \multirow{3}{2cm}{\cite{3_li2022cpath,14_raipuria2022cpath,15_geng2022cpath,19_sharma2022cpath,27_moyes2023cpath,28_sikaroudi2022cpath,31_boyd2022cpath,54_shin2021cpath,78_ke2021cpath,79_kather2019cpath,80_sebai2020cpath,81_cong2022cpath,82_vu2022cpath,83_zhao2022cpath,84_jia2022cpath,90_zhang2022cpath,99_cong2021cpath,103_shaban2019cpath,106_xing2022cpath,109_wang2021cpath,113_wagner2022cpath,118_gadermayr2019cpath,121_vasiljević2022cpath,143_quiros2021cpath,147_salehi2022cpath,149_khan2022cpath,171_bueno2020cpath,183_kirsuk2020cpath,184_razavi2021cpath,188_wilm2021cpath,189_lafarge2019cpath,197_bejnordi2014cpath,198_ruifrok2001cpath,199_macenko2009cpath,200_reinhard2001cpath,201_vahadane2016cpath,202_khan2014cpath,203_zheng2019cpath,204_hoque2021cpath,205_vicory2015cpath,206_tosta2019cpath,207_dawood2023cpath}}                      & Domain generalization by specifically reducing the domain shift.                                     & Require large amounts of data to estimate the underlying distributions accurately.          \\
                                             &                                        & Learning domain-invariant feature representations, which can be useful for knowledge transfer.     & Some techniques require careful tuning of hyperparameters.                                    \\
                                             &                                        & Adaptable to different types of data and model architectures.                                      &                                                                                               \\ \midrule
\multirow{5}{2cm}{Data augmentation (\cref{sec:augment})}              & \multirow{5}{2cm}{\cite{21_tellez2019cpath,25_wu2019cpath,26_chen2022cpath,29_pohjonen2022cpath,32_li2022cpath,35_vasiljević2021cpath,56_chang2021cpath,57_shen2022cpath,58_akram2018cpath,59_alemi-koohbanani2020cpath,60_vanderwal2021cpath,61_wang2023cpath,62_lin2022cpath,63_jarkman2022cpath,85_zhang2022cpath,92_yamashita2021cpath,93_bándi2023cpath,97_zhu2021cpath,98_falahkheirkhah2023cpath,100_xue2021cpath,101_chen2022cpath,102_scalbert2022cpath,104_roy2021cpath,114_xu2022cpath,117_cho2021cpath,128_mahmood2020cpath,136_ozdemir2012cpath,140_tellez2018cpath,141_fan2022cpath,146_tsirikoglou2021cpath,148_marini2023cpath,150_holland2020cpath,161_zaffar2023cpath,162_yang2021cpath,163_faryna2021cpath,185_kondo2021cpath,186_nateghi2021cpath,187_chung2021cpath,193_haan2021cpath,196_jahanifar2022cpath,208_foote2022cpath,209_jahanifar2022cpath,210_shephard2022cpath}}                      & Introduce more variation of data.                                                                    & Excessive image transformations might cause label shifts.                                     \\
                                             &                                        & Increase robustness to common variation (stain, contrast, blur, etc).                              & Generative model may be hard to train in Cpath.                                             \\
                                             &                                        & Does not require domain labels and works good even with a single domain.                             & Do not explicitly look for domain-invariant features.                                         \\
                                             &                                        & Easy to implement and can be used with any model.                                                  &                                                                                               \\
                                             &                                        & Does not add computation overhead during inference.                                                  &                                                                                               \\ \midrule
\multirow{4}{2cm}{Domain separation (\cref{sec:separation})}         & \multirow{4}{2cm}{\cite{49_wagner2021cpath,107_chikontwe2022cpath}}                      & Improves   generalization by separating domain-specific and agnostic features.                       & Challenging to achieve true feature disentanglement.                                        \\
                                             &                                        & Easier adaptation to new domains with   domain-agnostic features.                                    & Needs   diverse multi-domain data for effective learning.                                     \\
                                             &                                        & Easier model behavior understanding through   feature separation.                                    & Risk   of performance drop if done imperfectly.                                               \\
                                             &                                        & Resists domain shifts by not overfitting to   domain specifics.                                      &                                                                                               \\ \midrule                                             
\multirow{4}{2cm}{Meta-learning (\cref{sec:meta})}             & \multirow{4}{2cm}{\cite{16_sikaroudi2021cpath,22_shakeri2022cpath,44_liu2021cpath,45_yuan2021cpath,46_han2022cpath,48_fagerblom2021cpath}}                      & Can be applied to any model architecture.                                                            & Can be computationally intensive.                                                             \\
                                             &                                        & Can scale to large and complex datasets.                                                             & Relies on several hyperparameters and therefore can be hard to train.                         \\
                                             &                                        & Learns to quickly adapt to new domains   during inference.      & Relies on having access to a diverse multiple (usually $>4$) source   domains during training. \\
                                             &                                        & Tend to work better than other   methods when domain shift is large.                                 &                                                                                               \\ \midrule
\multirow{3}{2cm}{Ensemble learning (\cref{sec:ensemble})}         & \multirow{3}{2cm}{\cite{50_sohail2021cpath,51_hägele2022cpath,52_yengec-tasdemir2023cpath,53_luz2021cpath,108_peyret2018cpath,139_linmans2023cpath,191_liang2021cpath,192_kotte2022cpath}}                      & Reduces the chance of overfitting, by combining the predictions of   different models                & Involves training multiple   models.                                                          \\
                                             &                                        & Reduce the impact of outliers/error   caused by any single model.                                    & Harder to interpret an   aggregation of multiple models.                                      \\
                                             &                                        & Can be applied to any model   architecture.                                                          &                                                                                               \\ \midrule
\multirow{3}{2cm}{Tailored model design (\cref{sec:design})}              & \multirow{3}{2cm}{\cite{23_anand2020cpath,36_silva2022cpath,37_graham2020cpath,38_lafarge2021cpath,39_razavi2022cpath,40_zhang2022cpath,41_xie2018cpath,42_javed2021cpath,43_wang2021cpath,96_chan2022cpath,115_rojas-moraleda2017cpath,119_graham2023cpath,120_dabass2022cpath,130_yu2023cpath,131_yaar2020cpath,145_tang2021cpath,151_chen2021cpath,155_li2023cpath,160_dooper2023cpath,182_vuong2021cpath}}                      & Tailored   designs for DG problem and learning shared representation can  improve performance.       & Sophisticated model architectures increase design complexity.                               \\
                                             &                                        & With MTL, one model for multiple tasks saves   computational resources especially in WSI processing. & Balancing   learning across multiple tasks can be difficult.                                  \\
                                             &                                        & Multi-tasking acts as a form of   regularization against overfitting.                                & MTL is hard to scale, where increased tasks can lead to model   size and complexity issues.   \\  \midrule              
\multirow{3}{2cm}{Pretraining (\cref{sec:pretrain})}               & \multirow{3}{2cm}{\cite{0_neto2022cpath,1_marini2021cpath,2_fashi2022cpath,4_qin2022cpath,5_yang2022cpath,6_li2022cpath,86_galdran2022cpath,87_bozorgtabar2021cpath,88_liu2022cpath,89_koohbanani2021cpath,91_abbet2022cpath,116_haq2022cpath,123_chikontwe2022cpath,142_tran2022cpath,144_fan2021cpath,154_sikaroudi2020cpath,156_huang2021cpath,157_li2021cpath,158_wang2022cpath,159_kang2023cpath,165_dercksen2019cpath,177_sikaroudi2023cpath,178_lazard2023cpath,180_vuong2023cpath,194_chen2022cpath,195_filiot2023cpath}}                      & Make use of abondance of unlabeled data in Cpath.                                                    & Hard to validated the quality of features for all applications.                               \\
                                             &                                        & Suitable for feature extraction for   weakly-supervised approaches (such as MIL).                    & Marginal to no improvement when labeled intermediate-size datasets are   available.          \\
                                             &                                        & Pretrained models are robust   to “nuisance factors” (exact location of objects, lighting, or color).& Expensive to train.                                                                           \\ \midrule                                             
\multirow{3}{2cm}{Regularization strategies (\cref{sec:regularize})} & \multirow{3}{2cm}{\cite{9_jiang2023cpath,10_minhas2021cpath,20_cai2021cpath,24_su2023cpath,95_su2023cpath,138_mehrtens2023cpath,153_liang2023cpath}}                      & Reduces the chance of overfitting   by reducing the influence of irrelevant features.                & Methods like batch norm and dropout can increase computational   complexity.                  \\
                                             &                                        & Can   help prevent the model from memorising training data in large models.                                          & Over-regularisation can lead to under-fitting.                                                \\
                                             &                                        & Can help reduce model   complexity.                                                                  & Sensitivity to   hyper-parameters.                                                            \\ \midrule
\bottomrule
\end{tabular}
\end{table*}

\section{Resources for domain generalization studies} \label{sec:resources}
\begin{figure}
    \centering
    \includegraphics[width=\columnwidth]{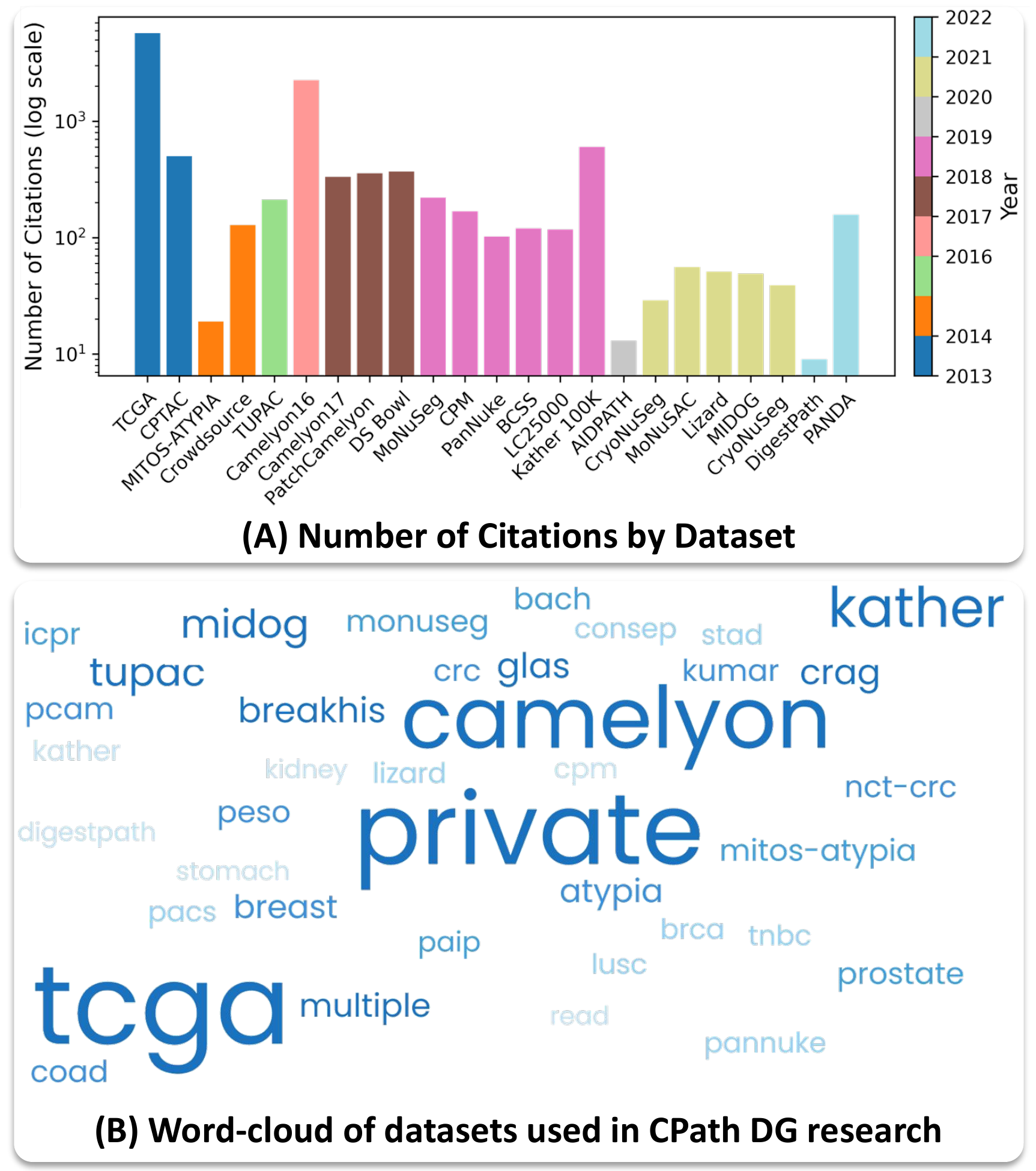}
    \caption{Dataset usage in CPath research studies. (A) Number of citations for major CPath datasets based on Google Scholar (updated in September 2023), (B) Word cloud of datasets used in DG studies in particular.}
    \label{fig:dataset}
\end{figure}

A significant catalyst for the recent advancements in DG studies within CPath has been the introduction of publicly accessible datasets and code repositories. Some of these resources originated from challenge competitions or were developed in response to specific challenges. Notably, the Camelyon dataset and associated challenges \cite{74_ehteshamibejnordi2017cpath,75_bandi2019cpath} have driven numerous DG investigations in CPath \cite{160_dooper2023cpath,75_bandi2019cpath,139_linmans2023cpath,146_tsirikoglou2021cpath,89_koohbanani2021cpath,130_yu2023cpath,92_yamashita2021cpath}. This trend is illustrated in \cref{fig:dataset}A, which depicts the citation counts for major DG-related datasets, and in \cref{fig:dataset}B, which presents the usage frequency of each dataset in DG studies through a word cloud visualization. Based on \cref{fig:dataset}B, TCGA dataset (different projects), Kather datasets \cite{173_kather2019cpath}, and MIDOG datasets \cite{64_aubreville2023cpath} are also among the most utilized datasets fueling DG in CPath. To support further DG research endeavors, we have included lists of not only pertinent CPath datasets but also toolboxes in the subsequent sections.

\subsection{Datasets and Challenges}
The following section provides a brief overview of some popular DG datasets that have been developed and utilized in the field of CPath. These datasets have been employed individually \cite{64_aubreville2023cpath, 74_ehteshamibejnordi2017cpath} or integrated with other datasets to create larger and more diverse collections \cite{72_graham2021cpath, midog22dataset}. This allowed for the curation of datasets that are able to capture a broader spectrum of domain variations and expand the representation of different tissue types, tumor types, scanners, staining techniques, etc. Moreover, these datasets have also been employed in combination with additional datasets from related domains or tasks to help evaluate the efficacy of different DG algorithms across multiple datasets \cite{91_abbet2022cpath, 180_vuong2023cpath, 22_shakeri2022cpath, 120_dabass2022cpath}. 


\begin{table*}[htbp]
\centering
\caption{Publicly available datasets for DG experiments in CPath.}
\label{table:datasets}
\begin{tabular}{p{3.2cm}p{7cm}p{0.8cm}p{4.5cm}}
\toprule \toprule
\textbf{Dataset} & \textbf{Application/Task} & \textbf{DS} & \textbf{Domains} \\
\midrule

\multicolumn{4}{l}{\phantom{This create space for task ~~}\textbf{Detection}}\\
\midrule
ATYPIA14 \cite{126_roux2014_cpath} & Mitosis detection in breast cancer &  1 & 2 scanners\\ 
Crowdsource \cite{71_irshad2014cpath} & Nuclei detection in renal cell carcinoma &  3 & 6 annotators \\
TUPAC-Aux \cite{124_veta2019cpath} & Mitosis detection in breast cancer &  1 & 3 centers \\
MIDOG21 \cite{64_aubreville2023cpath} & Mitosis detection in breast cancer & 1, 2 & 6 scanners \\ 
DigestPath \cite{125_da2022cpath} & Signet ring cell detection in colon cancer &  1 & 4 centers \\
MIDOG22 \cite{65_aubreville2022cpath, midog22dataset} & Mitosis detection in multiple cancer types &  1, 2, 3 & 6 tumor types, 2 species \\
TiGER-Cells \cite{170_rijthoven2022cpath} & TILs detection in breast cancer &  1 & 3 sources  \\
MIDOG++ \cite{166_aubreville2023cpath} & Mitosis detection in multiple cancer types &  1, 2, 3 & 7 tumor types, 2 species \\ 
\midrule

\multicolumn{4}{l}{\phantom{This create space for task ~~}\textbf{Classification}}\\
\midrule
TUPAC-Mitosis \cite{124_veta2019cpath} & BC proliferation scoring based on mitosis score &  1 & 3 centers \\
Camelyon16 \cite{74_ehteshamibejnordi2017cpath} & Lymph node WSI classification for BC metastasis &  1 & 2 centers \\
PatchCamelyon \cite{76_veeling2018cpath} & BC tumor classification based on Camelyon16 &  1 & 2 centers \\ 
Camelyon17 \cite{75_bandi2019cpath} & BC metastasis detection and pN-stage estimation &  1 & 5 centers  \\
LC25000 \cite{172_borkowski2019cpath} & Lung and colon tumor classification &  4 & 2 organs \\
Kather 100K \cite{173_kather2019cpath} & Colon cancer tissue phenotype classification &  1 & 3 centers  \\
WILDS-Camelyon \cite{137_koh2021cpath} &  BC tumor classification based on Camelyon17 &  1 & 5 centers \\
PANDA \cite{175_bulten2022cpath} &  ISUP and Gleason grading of prostate cancer &  1, 2, 3 & 2 centers \\
\midrule

\multicolumn{4}{l}{\phantom{This create space for task ~~}\textbf{Regression}}\\
\midrule
TUPAC-PAM50 \cite{124_veta2019cpath} & BC proliferation scoring based on PAM50 &  1 & 3 centers\\
LYSTO \cite{169_jiao2020cpath} & Lymphocyte assessment (counting) in IHC images &  1 & 3 cancer types, 9 centers \\
CoNIC (Lizard \cite{72_graham2021cpath}) & Cellular composition in colon cancer &  1, 3 & 6 sources \\
TiGER-TILs \cite{170_rijthoven2022cpath} & TIL score estimation in breast cancer &  1 & 3 sources  \\
\midrule

\multicolumn{4}{l}{\phantom{This create space for task ~~}\textbf{Segmentation}}\\
\midrule
Crowdsource \cite{71_irshad2014cpath} & Nuclear instance segmentation in renal cell carcinoma &  3 & 6 annotators \\
Camelyon \cite{74_ehteshamibejnordi2017cpath,75_bandi2019cpath} & BC metastasis segmentation in lymph node WSIs &  1 & 2 centers, 5 centers \\
DS Bowl 2018 \cite{70_caicedo2019cpath} & Nuclear instance segmentation  &  1, 4 & 31 sets, 22 cell types, 5 image types  \\
CPM \cite{69_vu2019cpath} & Nuclear instance segmentation &  1, 4 & 4 cancer types \\
BCSS \cite{174_amgad2019cpath} & Semantic tissue segmentation in BC (from TCGA) &  1 & 20 centers  \\
AIDPATH \cite{171_bueno2020cpath} & Glomeruli segmentation in Kidney biopsies &  1 & 3 centers  \\
PanNuke \cite{66_gamper2020cpath} & Nuclear instance segmentation and classification &  1, 2, 4 & 19 organs\\
MoNuSeg \cite{68_kumar2020cpath} & Nuclear instance segmentation in H\&E images &  1 & 9 organs, 18 centers\\
CryoNuSeg \cite{67_mahbod2021cpath} & Nuclear instance segmentation in cryosectioned H\&E &  1, 3 & 10 organs from 3 annotations\\
MoNuSAC \cite{73_verma2021cpath} & Nuclear instance segmentation and classification &  1, 2 & 37 centers, 4 organs\\
Lizard \cite{72_graham2021cpath} & Nuclear instance segmentation and classification &  1, 3 & 6 sources \\
MetaHistoSeg \cite{45_yuan2021cpath} & Multiple segmentation tasks through various cancers &   1 & 5 sources (different tasks)\\
PANDA \cite{175_bulten2022cpath} &  Tissue segmentation in prostate cancer &  1, 2 & 2 centers \\
TiGER-BCSS \cite{170_rijthoven2022cpath} & Semantic tissue segmentation in BC (BCSS extension) &  1 & 3 sources  \\
DigestPath \cite{125_da2022cpath} & Colon tissue segmentation &  1 & 4 centers \\
NuInsSeg \cite{176_mahbod2023cpath} & Nuclear instance segmentation pan-cancer and species &  1,4 & 2 centers, 31 organs, 2 species\\
\midrule

\multicolumn{4}{l}{\phantom{This create space for task ~~}\textbf{Survival and gene expression prediction}}\\
\midrule
TCGA \cite{168_weinstein2013cpath} & Pan-cancer survival and gene expression prediction &  1, 2, 4 & 33 cancer types (multiple sources)\\
CPTAC \cite{ellis2013connecting,167_edwards2015cpath} & Pan-cancer survival and gene expression prediction &  1, 2 & 10 cancer types (multiple sources)\\

\bottomrule \bottomrule
\end{tabular}
\end{table*}

\cref{table:datasets} presents a summary of some of the more popular DG datasets, providing information on the name, the specific task or application within CPath that the dataset is designed for, the type of domain shift (`Type 1: Covariate Shift', `Type 2: Prior Shift', `Type 3: Posterior Shift', and `Type 4: Class-Conditional Shift') and the number of distinct domains present in the dataset. More detailed descriptions of the datasets can be found in Appendix \ref{app:resources}.

\subsection{Toolboxes}
\label{sec:toolbox}
\begin{figure}
    \centering
    \includegraphics[width=\linewidth]{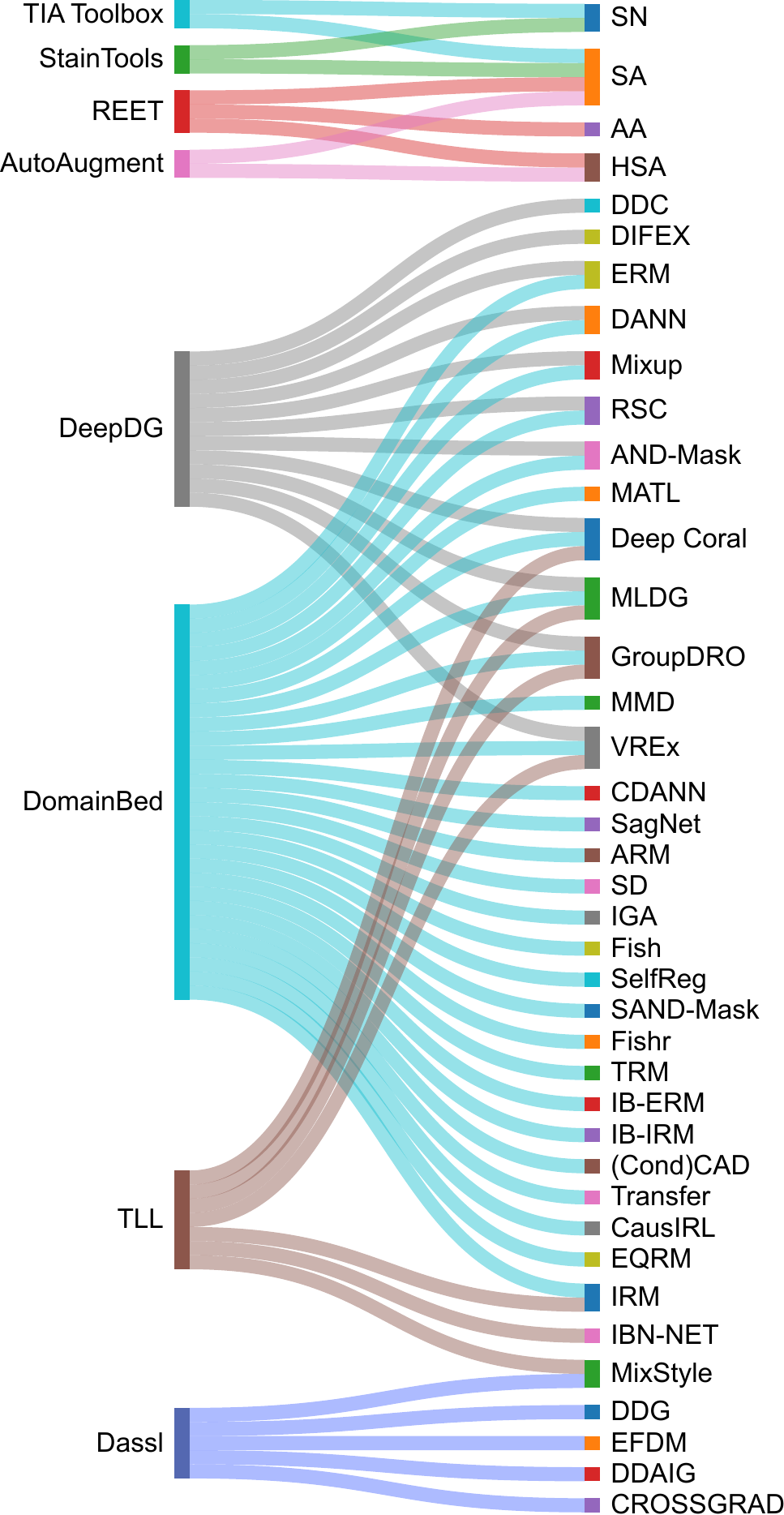}
    \caption{DG toolboxes and their respective algorithms introduced in \cref{sec:toolbox}. Other abbreviations are SN: Stain normalization, SA: Stain Augmentation, AA: Adversarial Augmentation, HSA: Histology-Specific Augmentations.}
    \label{fig:Toolboxes}
\end{figure}

Numerous toolboxes and frameworks are publicly available, offering a range of implemented algorithms and pretrained models for DG. Although not all of these toolboxes or algorithms were initially designed for CPath, they can be customized to suit the specific tasks in this field. In order to aid researchers in this area, we have curated comprehensive lists of the most widely used toolboxes and code bases (see Appendix \ref{app:resources}) for DG. A brief overview of the DG algorithms implemented in these toolboxes is also depicted in \cref{fig:Toolboxes}. Please note that we exclude well-known standard image augmentation toolboxes (such as Albumnetations \cite{info11020125}, ImgAug \cite{imgaug}, and RandAugment \cite{cubuk2020randaugment}) from our lists here.

\begin{enumerate}
    \item DomainBed \cite{34_gulrajani2021general} is a PyTorch suite implemented by the Facebook Research group. The toolkit currently supports 28 DG algorithms, 10 computer vision datasets, and 1 CPath Dataset - Camelyon17 WILDS \cite{137_koh2021cpath}. Explaining all of the investigated algorithms is beyond the extent of the current work but a list of the utilized algorithms and their reference is as follows: Empirical Risk Minimization (ERM) \cite{vapnik1999erm}, Invariant Risk Minimization (IRM) \cite{arjovsky2019irm}, Group Distributionally Robust Optimization (GroupDRO) \cite{sagawa2019groupdro}, Interdomain Mixup (Mixup) \cite{yan2020mixup}, Marginal Transfer Learning (MATL) \cite{blanchard2021mtl}, Meta Learning Domain generalization (MLDG) \cite{li2018mldg}, Maximum Mean Discrepancy (MMD) \cite{li2018mmd}, Deep CORAL (CORAL) \cite{sun2016deepcoral}, Domain Adversarial Neural Network (DANN) \cite{ganin2016dann}, Conditional Domain Adversarial Neural Network (CDANN) \cite{135_li2018general}, Style Agnostic Networks (SagNet) \cite{nam2021sagnet}, Adaptive Risk Minimization (ARM) \cite{zhang2021arm}, Variance Risk Extrapolation (VREx) \cite{krueger2021vrex}, Representation Self-Challenging (RSC) \cite{huang2020rsc}, Spectral Decoupling (SD) \cite{pezeshki2021sd}, Learning Explanations that are Hard to Vary (AND-Mask) \cite{parascandolo2020andmask}, Out-of-Distribution generalization with Maximal Invariant Predictor (IGA) \cite{koyama2020iga}, Self-supervised Contrastive Regularization (SelfReg) \cite{kim2021selfreg}, Smoothed-AND mask (SAND-mask) \cite{shahtalebi2021sand}, Learning Representations that Support Robust Transfer of Predictors (TRM) \cite{xu2021trm}, Invariance Principle Meets Information Bottleneck for Out-of-Distribution generalization (IB-ERM) \cite{ahuja2021ib}, Optimal Representations for Covariate Shift (CAD \& CondCAD) \cite{dubois2021cad}, Quantifying and Improving Transferability in Domain generalization (Transfer) \cite{zhang2021transfer}, Invariant Causal Mechanisms through Distribution Matching (CausIRL with CORAL or MMD) \cite{chevalley2022CausIRL}, and Empirical Quantile Risk Minimization (EQRM) \cite{eastwood2022eqrm}. To showcase the efficacy of the resources suggested in this section when adapted to CPath tasks, we have conducted benchmark experiments using DomainBed for mitosis vs. mimicker classification, further explained in Section \ref{sec:benchmark}.
    
    \item DeepDG: Deep Domain generalization Toolkit \cite{deepdg} is a PyTorch based toolkit currently supporting 6 CV datasets and 11 DG algorithms, namely, Empirical Risk Minimization (ERM) \cite{vapnik1999erm}, Deep Domain Confusion (DDC) \cite{tzeng2014deep}, Deep CORAL (CORAL) \cite{sun2016deepcoral}, Domain Adversarial Neural Network (DANN) \cite{ganin2016dann}, Meta-Learning Domain generalization (MLDG) \cite{li2018mldg}, Mixup \cite{zhang2017mixup}, Representation Self-Challenging (RSC) \cite{huang2020rsc}, Group Distributionally Robust Optimization (GroupDRO) \cite{sagawa2019groupdro}, Learning Explanations that are Hard to Vary (AND-Mask) \cite{parascandolo2020andmask}, Variance Risk Extrapolation (VREx) \cite{krueger2021vrex} and Domain-Invariant Feature EXploration (DIFEX) \cite{lu2022domain}. 
    
    \item Domain Adaptation and Semi-Supervised Learning (Dassl) \cite{zhou2021ensemble}, \cite{110_zhou2023general} is a Pytorch toolkit that implements algorithms for domain generalization, single source and multi-source domain adaptation and semi-supervised learning. The framework currently supports 10 datasets (including Camelyon17 WILDS \cite{137_koh2021cpath}) and 5 methods for DG including Dynamic Domain generalization (DDG) \cite{sun2022dynamic}, Exact Feature Distribution Matching (EFDM) \cite{zhang2022exact}, MixStyle \cite{zhou2021mixstyle}, Deep Domain-Adversarial Image Generation (DDAIG) \cite{zhou2020deep} and CROSSGRAD \cite{shankar2018generalizing}.
    
    \item The Transfer Learning Library \cite{jiang2022transferability} (TLL), \cite{tllib} is a PyTorch-based framework that supports algorithms for domain adaptation, task adaptation (fine-tuning), pre-trained model selection, semi-supervised learning, and domain generalization. The 7 algorithms included in the aforementioned toolkit for DG are as follows: IBN-Net \cite{pan2018two}, MixStyle \cite{zhou2021mixstyle}, Meta-Learning Domain generalization (MLDG) \cite{li2018mldg}, Invariant Risk Minimization (IRM) \cite{arjovsky2019irm}, Variance Risk Extrapolation (VREx) \cite{krueger2021vrex}, Group Distributionally Robust Optimization (GroupDRO) \cite{sagawa2019groupdro}, and Deep CORAL (CORAL) \cite{sun2016deepcoral}.
\end{enumerate}

For the sake of completeness, it is perhaps essential to include popular CPath toolboxes that support dedicated methods for stain normalization (SN) and stain augmentation (SA). These methods are commonly used as pre-processing steps or in conjunction with other DG algorithms to help facilitate the development of models that can generalize well across different staining patterns. 

\begin{enumerate}
    \item TIAtoolbox \cite{pocock2022tiatoolbox} is a Python toolkit developed by the TIA center. It supports implementations of stain normalization methods including Ruifrok  \cite{198_ruifrok2001cpath}, Macenko  \cite{199_macenko2009cpath}, Reinhard  \cite{200_reinhard2001cpath}, and Vahadane \cite{201_vahadane2016cpath}. Additionally, the toolbox also provides utilities for stain augmentation by using either the Macenko  \cite{199_macenko2009cpath} or Vahadane \cite{201_vahadane2016cpath} methods for extracting the stain matrices and concentrations.
    \item The StainTools \cite{Byﬁeld_2019} toolkit, implemented in Python, contains methods for stain normalization and extraction using the Macenko  \cite{199_macenko2009cpath} and Vahadane \cite{201_vahadane2016cpath} methods. However, the repository has now been archived and is not likely to be updated.
    \item Robustness Evaluation and Enhancement Toolbox \cite{208_foote2022cpath}, implemented in Python and contains methods for measuring the robustness of classification and segmentation algorithms to the perturbations in CPath-specific factors such stain, resolution, brightness, compression, focus, and blurring, etc. It also supports strategies for efficient adversarial training such as adversarial stain augmentation that can be adapted for DG.
    \item  Automated data augmentation for H\&E (AutoAugment) \cite{163_faryna2021cpath}, implemented in Python  and introduces automated data augmentation policy selection for histopathological slides, enhancing the RandAugment framework \cite{cubuk2020randaugment} with domain-specific modifications, leading to improved generalizability on histology images.
\end{enumerate}

\section{Benchmark domain generalization algorithms} \label{sec:benchmark}
During the review of DG papers in CPath, we realized that there are many novel DG algorithms proposed in the CV or ML communities that have never been investigated for CPath applications. To close this gap and to showcase the functionality of one of the resources we reviewed in the previous section, we try 28 DG algorithms implemented in the `DomainBed' toolbox \cite{34_gulrajani2021general} on the mitosis classification task. To do this, we leverage MIDOG22 dataset \cite{midog22dataset} where we extract small patches of size $128\time128$ pixels around mitotic figures and mimickers (based on the annotations provided) and form a dataset of 20,552 samples coming from 5 different domains: Canine Lung Cancer, Human Breast Cancer, Canine Lymphoma, Human neuroendocrine tumor, Human neuroendocrine tumor. As seen, with respect to the mitosis detection task, this dataset covers a large variation of covariate shift as images come from different centers, scanners, tissue types, and species. For more information on the dataset refer to Appendix \ref{app:benchmark}.

The benchmarking experiments in this work have been done similarly to \cite{34_gulrajani2021general}, however, we only considered the model selection method in which the validation data is extracted from the training domains (20\% of the training data). We have also considered a modality-specific domain generalization technique which is stain augmentation \cite{21_tellez2019cpath} during the training of the model using the ERM method and we call it `StainAug'. Cross-validation experiments have been done by considering each domain as the unseen test set once (hence leave-one-domain-out cross-validation in 5 folds). Furthermore, for each method, cross-validation experiments have 3 different random selections of hyperparameters as well as 3 different independent runs (to reduce the chance of using optimal seed). Therefore, in total for MIDOG22 classification benchmarking a total of $3\times3\times5\times28=1260$ classification experiments were run on an NVidia DGX-2 machine (details of experiments can be found in Appendix \ref{app:benchmark}).

The performance of various algorithms was compared using the MIDOG22 dataset \cite{midog22dataset} and reported in \cref{tab:benchmark}, with the ERM algorithm serving as the baseline. The F1 Score is especially significant given the unbalanced nature of the data. In terms of F1 Score, the StainAug algorithm emerged as the superior method, achieving an impressive F1 score of 76.0\%. This is a notable improvement compared to the baseline ERM algorithm which had an F1 score of 74.7\%.
The RSC and EQRM algorithms closely follow StainAug, with respective F1 scores of 75.7\% and 75.2\%. It is worth mentioning that despite EQRM's slightly lower F1 score, it delivered the highest accuracy of 80.2\%.
On the other hand, the IGA algorithm performed notably poorly in terms of both accuracy and F1 Score, yielding the lowest scores among the evaluated algorithms. Detailed results over different domains can be found in Appendix \ref{app:benchmark}.

These results suggest that if there is a well-labeled dataset that covers enough variation (such MIDOG22 dataset), using a model with a large enough capacity (such as ResNet50), and a carefully designed ERM training paradigm one can achieve good enough results in CPath classification algorithms. This is in line with what authors found in \cite{34_gulrajani2021general} when experimenting with natural images. However, it is clear that adding simple stain augmentation techniques during the training of the model can considerably improve the performance. This is expected because one of the main reasons for covariate shifts in CPath is changes in the color palette of the images (stain variation). The effectiveness of using stain augmentation technique in various CPath tasks has also been shown before \cite{209_jahanifar2022cpath,196_jahanifar2022cpath,21_tellez2019cpath,163_faryna2021cpath,208_foote2022cpath}.

It is pertinent to acknowledge that these results and conclusions are only valid for this dataset and alternative approaches exist for conducting cross-validation experiments to assess DG methods. For instance, one can explore methodologies involving the utilization of a solitary domain for training, or incorporate multiple domains for both training and testing. Furthermore, the examination of DG methods across diverse dataset size scenarios holds significance, as outcomes may vary due to certain DG techniques demonstrating superior efficacy under conditions of limited data availability. However, it is imperative to recognize that delving into more intricate explorations of such experiments falls beyond the scope of the current study.

\begin{table}[]
\centering
\caption{Benchmarking of different DG algorithms for mitosis vs. mimicker classification task using MIDOG22 dataset. The results in each column are colored based on their values, green and red for better and worse results, respectively.}
\begin{tabular}{lll}
\toprule \hline
\textbf{Algorithm} & \textbf{Accuracy} & \textbf{F1 Score} \\
\hline
\textbf{StainAug \cite{21_tellez2019cpath}} & \cellcolor[HTML]{81C77D}79.9 $\pm$ 0.3                                               & \cellcolor[HTML]{63BE7B}\textbf{76.0 $\pm$ 0.4}                                      \\
RSC   \cite{huang2020rsc}                   & \cellcolor[HTML]{ECE683}78.8 $\pm$ 0.4                                               & \cellcolor[HTML]{7AC57D}75.7 $\pm$ 0.2                                               \\
EQRM   \cite{eastwood2022eqrm}              & \cellcolor[HTML]{63BE7B}\textbf{80.2 $\pm$ 0.1}                                      & \cellcolor[HTML]{9FD07F}75.2 $\pm$ 0.1                                               \\
CausIRL-CORAL   \cite{chevalley2022CausIRL} & \cellcolor[HTML]{E2E383}78.9 $\pm$ 0.3                                               & \cellcolor[HTML]{B5D680}74.9 $\pm$ 0.5                                               \\
CORAL   \cite{sun2016deepcoral}             & \cellcolor[HTML]{C5DB81}79.2 $\pm$ 0.4                                               & \cellcolor[HTML]{B5D680}74.9 $\pm$ 0.2                                               \\
SagNet   \cite{nam2021sagnet}               & \cellcolor[HTML]{F6E984}78.7 $\pm$ 0.2                                               & \cellcolor[HTML]{BDD881}74.8 $\pm$ 0.3                                               \\
Mixup   \cite{yan2020mixup}                 & \cellcolor[HTML]{B1D580}79.4 $\pm$ 0.3                                               & \cellcolor[HTML]{C4DA81}74.7 $\pm$ 0.3                                               \\ \hline
\textit{ERM \cite{vapnik1999erm}}           & \cellcolor[HTML]{CFDD82}79.1 $\pm$ 0.2                                               & \cellcolor[HTML]{C4DA81}74.7 $\pm$ 0.2                                               \\ \hline
GroupDRO   \cite{sagawa2019groupdro}        & \cellcolor[HTML]{ECE683}78.8 $\pm$ 0.1                                               & \cellcolor[HTML]{D3DF82}74.5 $\pm$ 0.4                                               \\
CDANN   \cite{135_li2018general}                  & \cellcolor[HTML]{ECE683}78.8 $\pm$ 0.6                                               & \cellcolor[HTML]{E2E383}74.3 $\pm$ 0.3                                               \\
VREx   \cite{krueger2021vrex}               & \cellcolor[HTML]{BBD881}79.3 $\pm$ 0.2                                               & \cellcolor[HTML]{E2E383}74.3 $\pm$ 0.2                                               \\
MLDG   \cite{li2018mldg}                    & \cellcolor[HTML]{E2E383}78.9 $\pm$ 0.3                                               & \cellcolor[HTML]{E9E583}74.2 $\pm$ 0.4                                               \\
DANN   \cite{ganin2016dann}                 & \cellcolor[HTML]{C5DB81}79.2 $\pm$ 0.3                                               & \cellcolor[HTML]{E9E583}74.2 $\pm$ 0.4                                               \\
MTL   \cite{blanchard2021mtl}               & \cellcolor[HTML]{E2E383}78.9 $\pm$ 0.4                                               & \cellcolor[HTML]{FFEB84}73.9 $\pm$ 0.7                                               \\
SD   \cite{pezeshki2021sd}                  & \cellcolor[HTML]{FEEA83}78.5 $\pm$ 0.6                                               & \cellcolor[HTML]{FFEB84}73.9 $\pm$ 0.3                                               \\
ARM   \cite{zhang2021arm}                   & \cellcolor[HTML]{FEEA83}78.5 $\pm$ 0.3                                               & \cellcolor[HTML]{FEE883}73.4 $\pm$ 0.2                                               \\
IRM   \cite{arjovsky2019irm}                & \cellcolor[HTML]{FEE883}78.1 $\pm$ 0.5                                               & \cellcolor[HTML]{FEE582}72.9 $\pm$ 0.5                                               \\
MMD   \cite{li2018mmd}                      & \cellcolor[HTML]{FEDC81}75.3 $\pm$ 1.8                                               & \cellcolor[HTML]{FDD07E}69.0 $\pm$ 2.7                                               \\
TRM   \cite{xu2021trm}                      & \cellcolor[HTML]{FDD880}74.2 $\pm$ 1.6                                               & \cellcolor[HTML]{FDCA7D}68.0 $\pm$ 2.9                                               \\
IB-ERM   \cite{ahuja2021ib}                 & \cellcolor[HTML]{FEDB80}74.9 $\pm$ 0.8                                               & \cellcolor[HTML]{FDCA7D}67.9 $\pm$ 1.5                                               \\
CausIRL-MMD   \cite{chevalley2022CausIRL}   & \cellcolor[HTML]{FCC37C}69.3 $\pm$ 4.3                                               & \cellcolor[HTML]{FCC57C}67.1 $\pm$ 1.8                                               \\
SelfReg   \cite{kim2021selfreg}             & \cellcolor[HTML]{FDD07E}72.4 $\pm$ 0.3                                               & \cellcolor[HTML]{FCBD7B}65.6 $\pm$ 1.8                                               \\
SANDMask   \cite{shahtalebi2021sand}        & \cellcolor[HTML]{FDD27F}72.8 $\pm$ 0.9                                               & \cellcolor[HTML]{FCB97A}64.8 $\pm$ 1.6                                               \\
Transfer   \cite{zhang2021transfer}         & \cellcolor[HTML]{FBB078}65.0 $\pm$ 4.9                                               & \cellcolor[HTML]{FCB77A}64.6 $\pm$ 2.8                                               \\
ANDMask   \cite{parascandolo2020andmask}    & \cellcolor[HTML]{FDD57F}73.5 $\pm$ 0.8                                               & \cellcolor[HTML]{FCB479}64.0 $\pm$ 2.0                                               \\
IGA   \cite{koyama2020iga}                  & \cellcolor[HTML]{F8696B}48.2 $\pm$ 2.5                                               & \cellcolor[HTML]{FBA376}60.9 $\pm$ 0.0                                               \\
CondCAD   \cite{dubois2021cad}              & \cellcolor[HTML]{FA9473}58.3 $\pm$ 6.3                                               & \cellcolor[HTML]{F98770}55.7 $\pm$ 4.9                                               \\
CAD   \cite{dubois2021cad}                  & \cellcolor[HTML]{FA9273}58.0 $\pm$ 6.4                                               & \cellcolor[HTML]{F8696B}50.2 $\pm$ 7.1                \\
\hline \bottomrule
\end{tabular}
\label{tab:benchmark}
\end{table}

\section{Guidelines for domain generalization}
\label{sec:guidelines}

\subsection{Experiment design and model selection}
\label{sec:experiment}

In the realm of DG studies, robust experimental design and proper model validation stand as initial critical steps. Before grappling with DS issues or any scientific inquiry about DG, the groundwork of problem experiment design and model validation must be examined. To ascertain the validity of their endeavor concerning datasets and objectives, researchers should assess whether their problem and experiment designs align. Sometimes, perceived DS concerns may actually stem from flawed experiment design, rendering attempts at resolution impractical.

For example, revisiting the scenario presented in \cref{sec:ds3}, when disparities exist in treatment between source and target groups, the resulting survival outcomes can diverge, even if the initially observed covariates align (owing to shifts in the distribution of events $P(Y|X)$). In line with the specific objective, the experiment design can be scrutinized. For instance, predicting the effectiveness of a third treatment for patients by stratifying them into high- and low-risk groups demands caution. Relying on a model trained on source domain outcomes might not yield generalizable or valid results for the target domain, where different initial treatments were applied. Here, reevaluation of the experiment design and its application becomes pivotal. However, in alternate scenarios, assessing the prognostic potential of a biomarker across source and target domains, irrespective of treatment variations, can remain a valid pursuit.
Consider another instance, focusing on a classification task aiming to predict tumors from non-tumor image tiles. A pitfall emerges when training data comprises tumor-negative samples from FFPE slides and tumor samples from frozen sections. In such a case, the trained model might inadvertently distinguish FFPE from Frozen sections, overshadowing the tumor classification. Consequently, applying this model to a target domain containing tumor and non-tumor samples from FFPE slides might lack generalizability, unless substantial adjustments are made. However, exploiting the same dataset to train a model capable of consistently segmenting nuclear instances across both domains remains feasible.

These examples underline the pivotal role of problem experiment design within the realm of CPath, particularly in DG investigations. Furthermore, averting domain shifts (within practical limits) emerges as a potent strategy to address DS concerns. For instance, favoring objective labeling techniques over subjective ones can mitigate prior shifts.

Lastly, conscientious experimentation and rigorous model evaluation hold immense significance in DG studies. Various cross-validation approaches, like leave-one-domain-out or k-fold, can be employed \cite{birba2020splits}. These entail selecting the best-performing model through training-validation splits, followed by testing on the held-out test set. As detailed in \cite{34_gulrajani2021general}, the choice of model selection approach significantly influences performance on unseen domains. In their study, Gulrajani and Lopez-Paz \cite{34_gulrajani2021general} explored three alternatives: 1) validation splits drawn from all training domains, 2) a hold-out domain from training domains, and 3) a validation set from the test domain (oracle). Notably, the best approach for model selection emerged as using a validation split from all training domains. However, several considerations are pivotal when forming the validation set \cite{94_kleppe2021cpath,asif2023unleashing}. It should be sufficiently representative, encompassing random samples that capture population variations \cite{bishop2006pattern}. Additionally, stringent precautions must be taken to prevent data leakage within the validation subset \cite{shabtai2012survey}. For instance, when multiple samples pertain to a single case, case-wise validation sampling prevents using images from the same case in both training and validation, thereby averting overestimation of performance and erroneous model selection.

\subsection{Identifying domain shifts}
\label{sec:guidelines-indentify}
Below, we have explained the most common approaches to detect domain shifts in a CPath problem. Please note that these are general methods and the suitability and effectiveness of each method may vary depending on the specific datasets, the application, and the task at hand.

\subsubsection{Detecting covariate shift}
Although it may be possible to detect covariate shifts by comparing the distribution of raw pixel values over the different domains \cite{gonzalez2009digital,quinonero2008dataset,sugiyama2007covariate,197_bejnordi2014cpath} or by simply measuring the drop in the performance of the model on unseen domains \cite{190_aubreville2021cpath}, it is highly recommended to detect covariate shifts by comparing the feature distributions of the source and target domains \cite{17_nisar2022cpath,207_dawood2023cpath,de2019stain,13_stacke2021cpath}. 
Some methods used simple statistics of feature representations (such means and medians) or a direct importance estimation method that does not require density estimation \cite{sugiyama2008direct} while others suggested investigating the differences in the distribution of features \cite{ganin2016dann,nair2019covariate,pathak2022new}.
A common method is to train a binary classifier to distinguish between the data from the source and target domains. If the classifier can accurately differentiate the source and target data, it indicates a significant covariate shift \cite{gretton2012kernel}. This method is sometimes referred to as a two-sample test, and tools like Maximum Mean Discrepancy can be used to measure the statistical distance between the source and target distributions \cite{gretton2012kernel,207_dawood2023cpath}.

In the context of CPath, Stacke \etal \cite{13_stacke2021cpath} introduced a new measure called the "representation shift", which uses internal neural network representations to quantify model-specific domain shift which is presented target set and may cause poor model generalization. Experimental results show that training data processing and network architecture have a significant impact on learned representations, and the proposed metric can serve as an initial test to evaluate how well a trained model handles new, similar data without requiring annotations. They also showed that the proposed values of the "representation shift" metric inversely correlate with classification accuracy \cite{13_stacke2021cpath}. 
Furthermore, Nisar \etal \cite{17_nisar2022cpath} showed that by using PixelCNN and a domain shift metric based on feature space distances \cite{de2019stain}, domain shift in digital histopathology can be detected and quantified, and this measurement correlates strongly with the generalization performance of DNNs.
Alternatively, visualization methods such as t-SNE \cite{van2008tsne}, UMAP \cite{mcinnes2018umap}, or PCA \cite{gewers2021principal} can provide intuitive insight into distributional differences and have been utilized intensively to illustrate covariate shift in CPath \cite{82_vu2022cpath}.

\subsubsection{Detecting prior shift}
Prior shifts can be detected by examining the class balance in the source and target domains. One way to identify prior shifts is by comparing label distribution histograms for the source and target datasets in classification applications (example in \cref{fig:shifts}F bottom plot). If the distributions differ significantly, it may indicate a prior shift. Statistical tests, such as the Chi-Squared test or Kolmogorov-Smirnov test \cite{mitchell1971comparison}, can be used to assess the statistical significance of the observed differences in distribution. For regression tasks, the same approach can be utilized by quantizing the target values and plotting their histogram over source and target domains. For detection and segmentation tasks, the distribution of the number of objects belonging to each category can be investigated. In survival analysis applications, one can plot the survivor curves for the source and target domain data on the same axis and compare them to check if they have significantly different distributions of events visually (similar to \cref{fig:shifts}F top plot). With survival analysis, average hazard rates of source and target domain data can also be informative of possible prior shifts \cite{kleinbaum1996survival}.

\subsubsection{Detecting posterior shift}
Detecting posterior shifts is more challenging due to its dependency on accurate label information. If multiple annotators are available, intra- and inter-annotator variability can be used to detect potential posterior shifts. This involves comparing the variability in labeling for the same data between different annotators or comparing the variability in labeling for the same annotator over different periods of time. If the annotator's labels are inconsistent, it may indicate a posterior shift as it has been illustrated in the example of \cref{fig:shifts}G. Alternatively, Cohen's Kappa or Fleiss' Kappa can be used to measure the level of agreement between annotators for categorical data \cite{de2011measurement} or a more robust chance-corrected statistic \cite{gwet2002kappa}. These are only simple and generic guidelines to detect posterior shifts but there is a vast literature on inter-rater agreement that can be helpful in this area \cite{chaturvedi2015evaluation,hallgren2012computing}.

\subsubsection{Detecting class-conditional shift}
Detecting class-conditional shifts can be achieved by analyzing the data distribution within each class \cite{moreno2012unifying} over different domains. One method is to use a class-conditional version of the two-sample test for covariate shift \cite{sugiyama2007covariate}. This involves comparing the distributions of data within each class between the source and target domains. If significant differences are detected, it suggests a class-conditional shift. This can be assessed using statistical distance measures or through visualization techniques, similar to covariate shift detection \cite{quinonero2008dataset}.
One simple way to detect class-conditional shifts is by employing a clustering mechanism within each class. The underlying intuition here is that if the data from the source and target domain for a specific class have diverged or shifted, the clustering algorithm would tend to form distinct clusters for data from each domain. Similar to covariate shifts, visualization techniques provide an intuitive understanding of high-dimensional data by projecting it into a lower-dimensional space (using tools such as UMAP, t-SNE, and PCA). When detecting class-conditional shifts, these methods can offer insights into how the distributions of features vary between source and target domains for different classes, similar to the example shown in \cref{fig:shifts}H.

\subsection{Handling domain shifts}
\label{sec:guidelines-handle}
Once the predominant domain shift type has been determined using the techniques outlined in the preceding section, individuals can then adhere to the DG directives tailored to that specific DS category. It is important to note that these recommendations are of a general nature and should not be treated as universally applicable remedies. Instead, we advise exploring diverse methodologies, making use of a range of resources (such as DomainBed), and conducting rigorous cross-validation experiments to pinpoint the optimal approach for your particular scenario or application.
\begin{figure*}
    \centering
    \includegraphics[width=\textwidth]{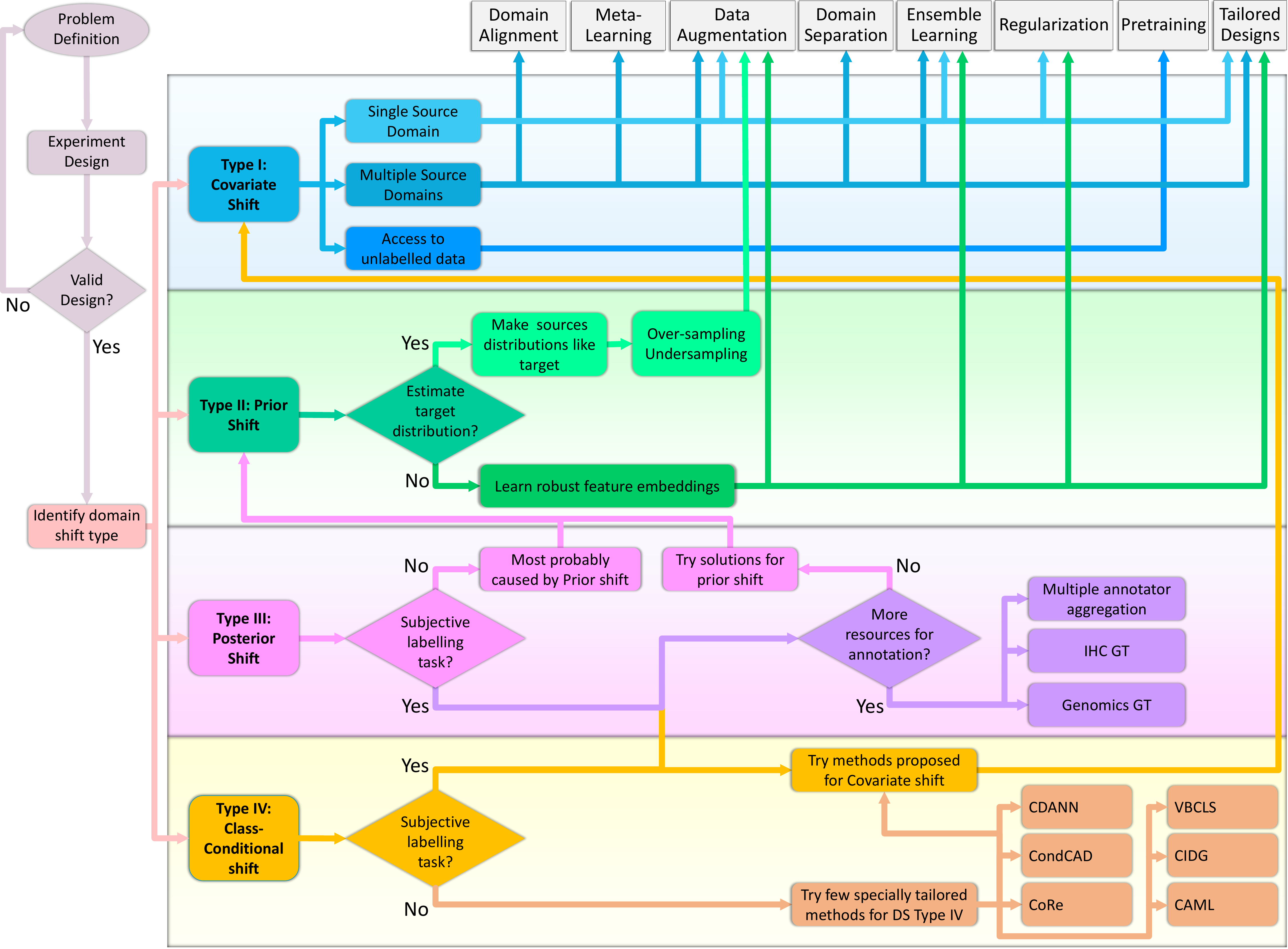}
    \caption{Guidelines for handling different types of domain shift in computation pathology. First, type of prominent domain shift should be identified using guidelines in \cref{sec:guidelines-indentify} and then proper solutions can be selected according to \cref{sec:guidelines-handle}.}
    \label{fig:guidline}
\end{figure*}

\subsubsection{Handling covariate shift}
As mentioned before, this type of domain shift is the most investigated problem for DG \cite{nair2019covariate,pathak2022new}. Therefore, most of the proposed DG methods are dedicated to this DS \cite{110_zhou2023general,34_gulrajani2021general,112_wang2021general} (as discussed in \cref{sec:future}). We propose general guidelines on selecting the best category of DG methods based on your problem definition and an extra condition on the properties of the data that you have access to during the training, as shown in the top part of \cref{fig:guidline}. 

When dealing with covariate shifts and having access to only one source domain, your choice of methods becomes limited. Data augmentation techniques (\cref{sec:augment}) have proven to be very effective in this situation \cite{29_pohjonen2022cpath,208_foote2022cpath,21_tellez2019cpath}. Based on the problem at hand, one can utilize on-the-fly stain augmentation techniques \cite{21_tellez2019cpath,56_chang2021cpath,163_faryna2021cpath,pocock2022tiatoolbox} or GAN-based image generators \cite{104_roy2021cpath,187_chung2021cpath,31_boyd2022cpath,35_vasiljević2021cpath,62_lin2022cpath} to increase the diversity of data seen during the training process. It has been shown before that the more diverse the training distributions are, the more generalizable to OOD data the model will be \cite{xu2020neural,29_pohjonen2022cpath}. Furthermore, very few methodologies that utilized reinforcement learning \cite{tang2021probeable,zhao2023rlogist,yao2022improving,qaiser2019learning} can also benefit from data augmentation to become more robust against common covariate shifts \cite{laskin2020reinforcement,yarats2020image}.
Depending on the problem and the kind of labels available within the single source domain, one may be able to incorporate novel model designs that are robust to specific data variations (see \cref{sec:design}). Also, MTL can lead to the training of robust feature encoders \cite{graham2019hover}. Using regularization techniques (such as different types of dropout \cite{srivastava2014dropout,labach2019survey,105_liu2019mri}) is always possible and recommended. Finally, it has been shown in several research papers and challenge leaderboards that ensemble learning can gain an extra bit of generalizability \cite{eisenmann2023winner,102_scalbert2022cpath,53_luz2021cpath,50_sohail2021cpath} and is worth including if test-time computational cost is not a concern.

Moving on to cases where there is access to more than one source domain, the number of options in DG methods increases. Apart from data augmentation, ensemble learning, regularization, and model design techniques, one can also consider using domain alignment techniques (\cref{sec:alignment}), domain separation (learning disentangled representations) (\cref{sec:separation}), and meta-learning (\cref{sec:meta}). Meta-learning allows for the inclusion of other DG alignment methods \cite{44_liu2021cpath,46_han2022cpath} and show promising results in various problems, however, they would require more data from more source domains (usually >3) and they are harder to train \cite{finn2017maml}. Although stain normalization has been proposed as one of the earliest domain adaptation techniques in CPath \cite{198_ruifrok2001cpath}, it has been shown several times that in H\&E image analysis stain normalization does not have the practical stability to be used on real-world applications \cite{82_vu2022cpath}, do not remove domain-specific data \cite{207_dawood2023cpath}, and will add another step of computation during inference \cite{21_tellez2019cpath}.

Finally, when there is an abundance of unlabeled data from different domains, self/unsupervised methods can be leveraged to pretrain DL models so they can generate more robust feature representations and would require a smaller number of labeled images from the training domains to be able to generalize well to unseen domains (\cref{sec:pretrain}). In particular, the pretrained model can be used to extract more relevant histology features from input tile images of WSIs to be used in MIL \cite{bilal2021development,bilal2023aggregation,lu2021data} or GNN \cite{levy2020topological,zheng2022graph,li2018graph} frameworks. Many diverse publicly available datasets (labeled and unlabeled, such as TCGA, CPTAC, PAIP, etc) and code bases (see \cref{app:resources-code} for an extensive list) can be leveraged to pretrain a reliable feature extractor for CPath. Alternatively, already existing pretrained feature extractors such as Cerberus \cite{119_graham2023cpath}, HistoNet \cite{hoefling2021histonet}, HistoEncoder \cite{pohjonen2023histoencoder}, SSLDP \cite{159_kang2023cpath}, CTransPath \cite{158_wang2022cpath}, and SelfPath \cite{89_koohbanani2021cpath} are publicly available.

\subsubsection{Handling prior shift}
Recommendations for handling this type of domain shift are broadly similar to those in other modalities and machine learning problems with some slight tailoring to the specific nature of CPath-based problems.

When the label distribution of the target domain is available or can be inferred, class-balancing strategies like over-sampling and under-sampling can be employed. Studies have demonstrated that oversampling minority classes enhances model generalization compared to undersampling \cite{Johnson2019SurveyOD}. Hence, we advocate for oversampling the minority class or adopting weighted sampling based on the frequency of each class in the training dataset. Beyond these, synthetic data generation methods, such as SMOTE \cite{Chawla2002SMOTESM}, offer additional avenues. A variety of data augmentation techniques in CPath are discussed in \cref{sec:augment}, which can be harnessed to generate new data points. For a comprehensive understanding of synthetic data generation in machine learning, readers should refer to \cite{chlap2021review,shorten2019survey,Lu2023MachineLF}. Notably, merging oversampling with synthetic data generation, with an emphasis on minority classes, has been observed to boost performance in the target domain.

In situations where the label distribution of the target domain remains elusive or cannot be determined, the aforementioned techniques retain their relevance. Yet, they might necessitate supplementary, sophisticated methods. One approach involves ensemble learning where the identical model is trained on diverse subsets of the source domain, each with notably varied label distributions. This raises the likelihood that a subset of the trained models might align closely with the target domain's prior distribution, thereby bolstering ensemble prediction efficacy. Analogous to the challenges posed by covariate shift, strategies such as pertaining on large-scale datasets, which foster the development of robust and transferable features, can also be instrumental in mitigating prior shifts, especially when the target domain remains concealed.

Model design and selecting appropriate loss functions for training and evaluation metrics are also pivotal. For instance, in segmentation tasks, the Dice and Jaccard loss functions have shown resilience against label imbalances in datasets \cite{hirling2023segmentation,jahanifar2018segmentation} which can alleviate prior shifts among different domains. Similarly, employing the F1 Score as a performance metric for classification models is recommended, given its capacity to account for both sensitivity and precision, making it resistant to distributional shifts \cite{reinke2022metrics}.


\subsubsection{Handling posterior shift}
Managing posterior shifts is notably more challenging than dealing with prior and covariate shifts. When discussing this type of domain shift, it is crucial to distinguish between two scenarios: dataset creation and intervention on a pre-existing dataset.

In the context of CPath, dataset creation often reveals pronounced posterior shifts, primarily due to differing opinions among pathologists. To mitigate this from the outset, adopting robust annotation protocols and guidelines can be instrumental \cite{wahab2022semantic,ibrahim2022assessment}. Alternatively, consolidating inputs from multiple annotators to assign a final label to a specific instance can be effective \cite{wilm2021influence,175_bulten2022cpath, 72_graham2021cpath,bertram2019large}. If there is access to labels from multiple observers, the Latent Doctor Model \cite{212_linmans2023cpath} can be used which trains on the full label distribution to predict expert panel uncertainty and the most likely ground-truth label. This has been shown to outperform models trained on majority vote labels. Ideally, when feasible, sourcing annotations from a molecular baseline, such as PHH3 IHC staining for mitosis annotation, is recommended \cite{140_tellez2018cpath,ibrahim2023improving}.

When dealing with a predefined dataset, addressing posterior shift becomes considerably intricate. For non-subjective tasks like survival analysis prediction, the presence of a posterior shift is likely indicative of an underlying prior shift. In such cases, the strategies mentioned for the prior shift can be adopted. However, if a posterior shift emerges in subjective tasks — for instance, if the source domain of Gleason grading is annotated by one pathologist and the target domain grades by another — then techniques like label smoothing can be beneficial \cite{goodfellow2016deep}. Label smoothing has been documented to counteract label noise effects, especially in the knowledge distillation context when implemented on the teacher model \cite{Lukasik2020DoesLS}. Introducing controlled noise into the source data serves as a regularization strategy during model training, potentially enhancing its adaptability to the target data, albeit possibly compromising source domain performance \cite{xie2020self,song2022learning}. Nonetheless, when feasible, we advise pursuing label refinement, especially for validation and test datasets, using the approaches outlined earlier.

\subsubsection{Handling class-conditional shift}
As highlighted in \cref{sec:ds4}, there exists an intersection between class-conditional shift and covariate shift. Additionally, it is important to note that inter-observer variability can contribute to class-conditional shifts by assigning differing image features to the same class label. Consequently, a primary step in addressing class-conditional shifts involves examining whether the labels stem from a subjective task. If affirmative, it is advisable to explore techniques recommended for addressing posterior shifts to mitigate this issue. Furthermore, the array of proposed approaches for managing covariate shifts remains applicable.
However, more sophisticated DG methods tailored to tackle class-conditional shifts are also available. A subset of these methods is introduced below.

Li \etal \cite{135_li2018general} proposed class-conditional domain adversarial neural networks (CDANN) which involved training a neural network with a novel loss function that focused on matching the conditional distribution of labels given data, rather than just the marginal distribution. In another work, Li \etal \cite{132_li2018general} proposed a method called CIDG for addressing class-conditional shift by enforcing joint distribution invariance across domains, considering situations where both marginal distribution and class-conditional distribution change; this method used regularization terms to ensure that the class-conditional distributions of learned features remained consistent across different domains. 
Liu \etal proposed \cite{134_liu2021general} a Variational Bayesian Conditional and Label Shift (VBCLS) framework that employed variational inference and posterior alignment to address class-conditional shift by modeling both conditional and label distributions, effectively reducing the effects of shift in DG tasks. 
Heinze‑Deml and Meinshausen \cite{heinze2021conditional} proposed a method called Conditional variance Regularization (CoRe) that addressed class-conditional shift by categorizing latent features into `conditionally invariant' core features and `orthogonal' style features, and then penalized the conditional variance of predictions based on core features and class labels. This regularization yielded an estimator that remained invariant under changes in the distribution of style features, protecting against domain shifts and was shown to be robust using a causal framework. Jiang \etal \cite{133_jiang2018general} proposed a meta-learning framework called conditional Class-Aware Meta-Learning (CAML) that used structured class information to modulate representations in few-shot learning tasks. CAML consists of four components: an embedding function, a base learner, an adaptation function, and a meta-learner. The goal was to learn a function that maps inputs to a metric space where semantic distances between instances follow Euclidean geometry and use the class structure to inform the model to reshape the representation landscape.
Lastly, the "CondCAD" method, proposed in \cite{dubois2021cad}, addressed class-conditional domain shift by introducing a domain bottleneck and utilizing a contrastive objective to learn representations that disentangled class and domain variations. Among the reviewed methods in this category, CDANN, CondCAD, and CoRe are publicly available to try (the first two are included in DomainBed framework \cite{34_gulrajani2021general}).

\section{Discussions}
\label{sec:discussion}
The implications of our survey on DG methods in CPath are far-reaching, both in terms of research and practical application. As this field evolves, the issues of DS become increasingly important to address in order to ensure robust, accurate, and generalizable diagnostic and prognostic models. Although some of the current DG methods often require substantial computational resources and expertise to implement, carefully designed DG and evaluation methods should be incorporated into the model development process to make them robust enough to be embedded into routine clinical workflows.  It is also important to note that while various DG methods exist, there is currently no one-size-fits-all solution. The effectiveness of a given method is highly dependent on the specific context and data, making it a challenging task to select the most suitable method for a particular problem.

\subsection{Implications of the study}
The analysis of DS types underscores the complexity of image analysis in CPath. Various medical contexts, patient populations, imaging technologies, and data sources inherently involve different data distributions. Understanding and quantifying these shifts provide a path toward building models that can perform well across domains, an essential goal in CPath.

The surveyed methods in domain alignment, data augmentation, domain separation, meta-learning, ensemble learning, model design, pretraining, and regularization strategies showcase a wide range of approaches that researchers and practitioners can utilize to address DS. Each method offers unique strengths and considerations, inviting the opportunity to select or even combine techniques based on specific use cases. For instance, domain alignment methods can be particularly effective in minimizing the distribution discrepancy between the source and target domains. Meanwhile, meta-learning offers a way to learn how to learn across domains, paving the way for flexible and adaptive models.

Through the exploration of available resources, we highlight the importance of diverse, challenging datasets and dedicated toolboxes in advancing the field. Furthermore, by benchmarking DG algorithms through these resources, we try to foster progress and innovation.

In the guidelines section, we provided a comprehensive guide for addressing DS in CPath. By integrating identification and handling strategies for each DS type, along with model validation and selection, these guidelines offer a roadmap for implementing DG in practice. The benefits could range from more consistent performance across different medical contexts to improved patient outcomes as a result of more accurate and reliable predictions.

In light of these implications, the potential of DG in CPath becomes apparent. As we continue to navigate and unravel this rich field, the goal remains clear: to develop robust, generalizable models that can truly adapt to the complexity and variability inherent in histology images.

\begin{figure*}
    \centering
    \includegraphics[width=\textwidth]{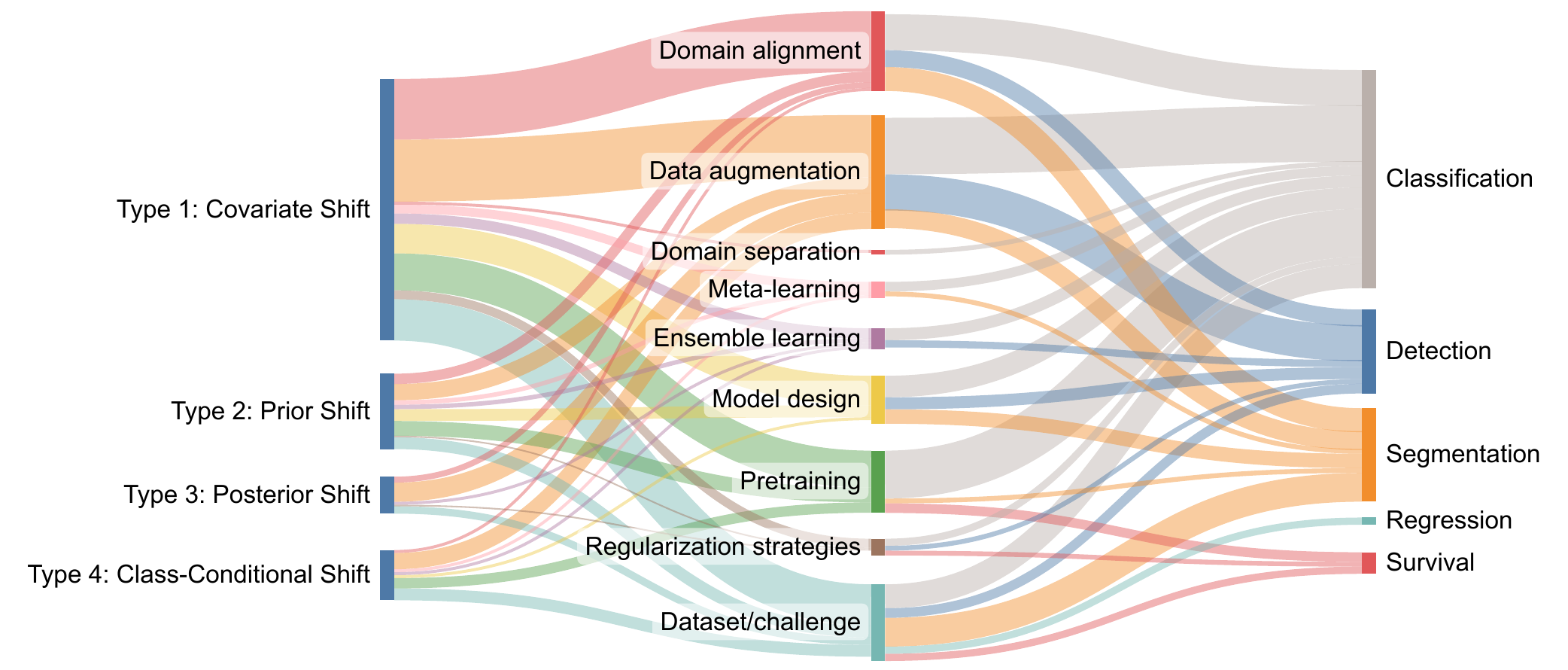}
    \caption{Sankey diagram showing the flow from domain shift types through method categories to applications. Line thickness indicates the frequency of each path in the literature.}
    \label{fig:snakey}
\end{figure*}

\subsection{Possible future research directions}
\label{sec:future}
\cref{fig:snakey} illustrates the relationships between DS types, method categories, and application details of DG-related papers reviewed in this work. The thickness of the links represents the number of occurrences of each transition. On the left side, four types of DS are shown. In the middle, the various method categories are displayed alongside the `Dataset/Challenges' that are DG-enabled. On the right, the specific applications are represented. This visual representation allows for an easy understanding of the most common paths from DS types, through methods, to applications, and reveals the prevalence of specific combinations in the literature.

As observed from \cref{fig:snakey}, mostly covariate shift has been investigated in CPath literature. Few papers focused on prior shift \cite{40_zhang2022cpath,45_yuan2021cpath,8_mahapatra2021multi,23_anand2020cpath}, and posterior shift \cite{140_tellez2018cpath,128_mahmood2020cpath,67_mahbod2021cpath,10_minhas2021cpath}, and class-conditional shift \cite{6_li2022cpath,28_sikaroudi2022cpath,108_peyret2018cpath,141_fan2022cpath}, while these are very common problems in CPath due to the use of the tile-trained network on WSIs, inter-observer variability of most annotation tasks in CPath, and heterogeneity of specific classes in different cancer type, respectively. We believe one of the main reasons for these shortcomings is that no specific dataset or challenge contest has been designed to directly investigate prior or posterior shifts. However, some datasets can be still leveraged to research these overlooked domain shift types in CPath. For example, MIDOG datasets  \cite{166_aubreville2023cpath,65_aubreville2022cpath} contain all kinds of DS and can foster research on all of them (see \cref{app:benchmark}). Similarly, in MetaHistoSeg dataset \cite{45_yuan2021cpath}, which is a collection of different datasets, not only images are coming from different sources and cancers (covariate shift), but also, the target entity in each domain is different from the others and therefore the distribution of labeled pixels will be different (prior shift). Furthermore, for some domains (and collectively), there exists a subjectivity in the labeling which can cause a posterior shift. The CryoNuSeg dataset \cite{67_mahbod2021cpath} provides three manual mark-ups to measure intra-observer and inter-observer variability but, as far as we know, no research paper has investigated their proposed DG method against this property of the CryoNuSeg dataset. We agree that curating new datasets and challenges to specifically address prior and posterior DS is very time-consuming and expensive, but we also believe that researchers can leverage several different datasets in the field to form new datasets (similar to what MetaHistoSeg  \cite{45_yuan2021cpath} has done) suitable for investigating these domain shifts.

In the context of DG, we can also talk about Homogeneous and Heterogeneous settings. Homogeneous Domain generalization (HDG) assumes that all domains (both source and target) share the same feature space and task space. In other words, the same set of features is measured across all domains, and the same prediction task is performed \cite{geng2020recent}. Heterogeneous Domain generalization (HeDG), on the other hand, assumes that the feature space and/or task space can differ across domains \cite{geng2020recent,bendale2015towards}. For instance, in CPath, one domain might be H\&E images of breast cancer, while another domain could be IHC images of lung cancer. Similarly, the tasks can vary across domains: one task might involve the classification of breast cancer subtypes based on H\&E images, while another task might entail predicting the survival rate for lung cancer patients from IHC images. This variation in both data types and tasks in HeDG mirrors the complexity and diversity of real-world scenarios, like those in CPath, where data types and tasks often differ across domains. Such problems have not been investigated in the CPath domain but the emerging of new methods \cite{li2018mldg,shu2021open} and datasets like  MetaHistoSeg  \cite{45_yuan2021cpath} make it possible to research this kind of problems more thoroughly. 

On the right side of \cref{fig:snakey} we can get an idea of how DG methods are distributed over different CPath applications in the reviewed papers. It is obvious that survival analysis studies have the least share of DG In other words, among many survival studies in the field of CPath \cite{wen2023deep,fuchs2011computational,atallah2023deciphering}, very few have particularly investigated DS problems or DG in the context of survival analysis \cite{10_minhas2021cpath,144_fan2021cpath,88_liu2022cpath,156_huang2021cpath,9_jiang2023cpath}. This can also be attributed to the scarcity of multi-centric datasets with well-curated patient outcome data. However, we believe the existing large-scale data repositories such as TCGA \cite{168_weinstein2013cpath} and CPTAC datasets and their combination can enable many DG studies in survival analysis, especially in pan-cancer studies. Another important takeaway from \cref{fig:snakey} is that no work with the application of regression has investigated DG specifically, although there is a plethora of methods dedicated to regression applications in CPath \cite{dawood2021albrt,peikari2017automatic,petrick2021spie,rakhlin2019breast,vo2022mf,li2022ultra,song2022deep,laves2020well}. We believe that datasets like CoNIC \cite{graham2021conic,graham2023conic}, that includes cell composition regression task over a multi-source dataset, and LYSTO \cite{169_jiao2020cpath}, that poses lymphocyte assessment in multiple cancer types as a regression task, can fuel more dedicated research in the field of DG for regression applications. 

Lastly, focusing on the DG methods depicted in the central segment of the \cref{fig:snakey} diagram, three key techniques—data augmentation, domain alignments, and pretraining—have garnered substantial attention in the literature. This prominence can be attributed to their practicality, adaptability, and straightforward implementation, particularly in the case of data augmentation.
However, our examination has revealed that meta-learning holds significant promise for enhancing DG. Its potential lies in its compatibility with a variety of other DG methods, as evidenced by its incorporation with various techniques in studies such as \cite{28_sikaroudi2022cpath,44_liu2021cpath,129_li2022multi}. Consequently, further investigation into meta-learning is warranted, although it may necessitate access to a more extensive set of training domains.
Likewise, specialized model designs have exhibited substantial effectiveness across various tasks within CPath, as indicated by references such as \cite{151_chen2021cpath,42_javed2021cpath,115_rojas-moraleda2017cpath,36_silva2022cpath}. Particularly in the context of multi-task learning, this approach not only enhances generalizability \cite{120_dabass2022cpath,40_zhang2022cpath}, but also bolsters processing efficiency for large WSIs by enabling concurrent execution of multiple prediction tasks \cite{119_graham2023cpath}.

\subsection{Emerging approaches}
\label{sec:emerging}
Although we have covered many methodologies in the review section (\cref{sec:dg_survey}) and many different methods have been investigated in the realm of CPath that were not researched in the general ML community before (such as special model designs and pretraining algorithms), there are still many methodologies that have been proposed and investigated in general ML community but not yet in CPath. For some examples, we can mention  Task-Adversarial Gradients \cite{volpi2018generalizing}, Random Augmentation Networks \cite{xu2020robust}, Learnable Augmentation Networks \cite{zhou2020learning}, and Domain-Specific Batch normalization \cite{seo2020learning}. However, here we want to focus on categories of methodologies that are emerging in recent years and show potential in improving DG in medical image analysis.

\subsubsection{Causal domain generalization}
\label{sec:causal}

Most \emph{non-causal} domain generalization approaches (\eg, regularization or classical data augmentation) work by minimizing the worst-case error expected over examples in a source domain as well as in its \emph{neighborhood} but typically without any explicit modeling of causal or non-causal attributes. For example, domain adversarial training or data augmentation approaches minimize risk over given training examples as well as their perturbed variants. From a causal perspective, domains can be thought of as explicit interventions over non-causal features of a prediction problem, \ie, domain shifts can only change features that are not causal \cite{pearl2009causality,scholkopf2012causal,scholkopf2021toward,heinze2021conditional}. For example, staining variations across different centers can be considered as non-causal attributes in a prediction problem aimed at detecting cancer in WSIs. This is because such variations will not change the underlying disease status. On the other hand, morphological characteristics (\eg, size and shape) of cells are causally linked to the underlying disease status. This implies that if we can identify causal features, either through domain knowledge, interventions, or exploiting statistical asymmetries, then we can learn to disregard or be invariant to such non-causal variations. This is the fundamental principle underlying causal DG approaches which works by minimizing the worst-case error over examples in the source domain as well as over examples obtained by explicit interventions over non-causal features of a prediction problem. Given that the non-causal impact of stain variations is known from background domain knowledge, using stain augmentation during training can be classified as a causal mechanism for DG. However, causal DG approaches go beyond this and can be broadly classified into three categories as in the recent review on this subject by Sheth \etal \cite{sheth2022domain}:
 
\paragraph{Causal Data Augmentation} methods work by synthetically generating counter-factual examples in which non-causal features are perturbed and then an ML model is trained to obtain the lowest error over both the original and perturbed data. Examples of such methods include using causal GANs, generative interventions or gradient-based causal augmentation \cite{zhang2021learning,kocaoglu2017causalgan,bai2021decaug}.

\paragraph{Causal Representation Learning} methods which aim to learn representations that are invariant to causal factors that are different across domains. An example of such a method is causal matching which finds pairs of examples from different training domains that are likely to be produced by the same underlying factors based on similarity in the space of causal factors and then trains a predictive model to learn a representation that is invariant to such causal factors \cite{mahajan2021domain,lv2022causality}. Gradient reversal can also be regarded as a causal representation learning method as it can be generalized to disregard factors related to differences across domains \cite{ganin2016dann,135_li2018general}.

\paragraph{Causal Invariance Mechanisms} aim to learn invariances across environments by finding a data representation such that an optimal classifier based on these representations matches across all environments. An example of such an approach is Invariant Risk Minimization (IRM) \cite{arjovsky2019irm} and causal-IRM \cite{chevalley2022CausIRL}. Another example is the "Diversify and Disambiguate" approach which learns the most diverse set of classifiers each giving a sufficiently low error and then picking the classifier that gives consistently good performance across domains \cite{lee2022diversify}.

In medical image analysis and CPath, causal learning can be beneficial in identifying the causal relationships between observed biomedical factors and certain medical conditions \cite{magliacane2018domain,prosperi2020causal}. However, the use of causal learning in the domain generalization context within medical image analysis is still largely untapped. Future research could focus on developing algorithms that infer causal structures from medical images across different domains, like varying imaging techniques or demographic groups. This could potentially provide better generalizable models that perform well even when the data distribution changes to ensure robustness and fairness.


\subsubsection{Continual learning}
Continual learning, sometimes referred to as `lifelong learning' or `learning without forgetting' \cite{li2017learning}, is a learning paradigm where the model continually learns from a stream of data, ideally retaining previously learned information while adapting to new tasks \cite{parisi2019continual}. In DG, continual learning has been proposed as a solution to improve models' ability to generalize across unseen domains by sequentially learning from multiple source domains \cite{li2017meta}.

In the medical imaging field, continual learning can be used to continually adapt to new datasets, imaging modalities, patient demographics, or disease patterns, without forgetting prior learned knowledge \cite{93_bándi2023cpath}. For example, in a recent work, Sadafi \etal \cite{211_sadafi2023cpath} developed an uncertainty-aware continual learning framework that uses exemplar samples from previous stages of learning and a distillation loss to preserve the information learned in the previous stages. The authors evaluated their incremental learning method against different scenarios and reported consistent performance at all stages and overall improved generalization. However, the use of continual learning for DG in CPath remains relatively under-explored. Future research can explore methods to reduce catastrophic forgetting when dealing with multiple CPath tasks or domains \cite{de2021continual}. In addition, designing strategies to balance the trade-off between stability and plasticity during continual learning can also be a valuable research direction in CPath.

\subsubsection{Multiple Instance Learning (MIL)}
MIL has seen burgeoning popularity for WSI processing in CPath \cite{bilal2021development,lu2021data,campanella2018terabyte,ilse2018attention,gadermayr2022multiple,bilal2023aggregation}. Although many works have investigated generalizability of their MIL-based method to unseen domains in various problems \cite{130_yu2023cpath,131_yaar2020cpath,160_dooper2023cpath}, there are very few studies like \cite{shephard2023fully} that crafted special designs (apart from stain normalization) in their MIL framework to address DS challenges.
Given the potential of MIL methods for leveraging weak labels in CPath, the exploration of innovative DG methods within MIL frameworks becomes imperative.

\subsubsection{Foundation models}
The utilization of pretrained transformer models in CPath has recently gained significant momentum, especially Vision Transformer (ViT)-based models \cite{dosovitskiy2020image}. These ViT models have demonstrated their superiority over conventional CNN-based models across various histology tasks in CPath \cite{xu2023vision}. Notably, the hierarchical image pyramid transformer \cite{chen2022scaling} has excelled in tasks such as cancer subtyping and survival prediction within WSIs. This success is attributed to its adept utilization of the multi-scale and hierarchical characteristics inherent in WSIs during pretraining. Additionally, researchers have ventured into creating custom transformer architectures tailored specifically to histology tasks \cite{vu2023handcrafted,158_wang2022cpath}.

The triumph of recent large language models like GPT-3 \cite{GPT3} and LLaMA \cite{touvron2023llama} has spurred the development of large vision-language models like the Segment Anything Model (SAM) \cite{SAM}. SAM and similar models have emerged as potential foundation models that provide a robust starting point for further advancements. However, evaluating these models in CPath often occurs in zero-shot or few-shot scenarios, revealing that models like SAM, originally trained on natural images, may initially struggle with medical images \cite{Zero-shotSAMPath,SAM_Med}. Nonetheless, these models exhibit a high adaptability potential, and through in-domain pretraining, they can acquire pan-cancer representations beneficial for various CPath tasks \cite{filiot2023scaling}.
Pretrained large transformers hold promise for advancing CPath due to their enhanced generalization capabilities. While some efforts have been made to introduce foundation models in CPath \cite{pohjonen2023histoencoder,PLIP}, there remains ample room for further research in this direction to establish a comprehensive vision-language foundation model for CPath. Achieving this goal will likely require access to extensive datasets and significant computational resources.

\subsubsection{Multi-modal learning}
Adding more data and using more diverse source domains has always been shown to improve the generalizability of trained models \cite{zhang2021understanding,taleb2019much,kandel2020effect}. This is not limited to the imaging data. Alongside histology images, researchers have been using additional data extracted from different modalities (such as other imaging modalities, genomic data, patient history/metadata, and lab reports) and reported improved generalizability of their models on various tasks \cite{lipkova2022artificial,miao2018genomic,mobadersany2018predicting,gillies2016radiomics,zuo2022identify,chen2020pathomic,chen2022pan}. However, training DL models with multi-modal data can be a challenging task due to data availability, cost, and the potential for missing data \cite{ravi2016deep,baltruvsaitis2018multimodal}. Several research works have investigated these issues in the past \cite{baltruvsaitis2018multimodal, tran2017missing,radu2018multimodal} but there is still room for investigation in the context of multi-modality and DG in CPath.

Nevertheless, additional data can also be scrapped from the existing resources. Sources like TCGA and CPTAC can provide useful additional information in conjunction with WSIs. Furthermore, researchers can use novel approaches to extract extra information from WSIs (such as nuclei distribution descriptors \cite{zamanitajeddin2021cells}) to be used in conjunction with deep features.

\subsection{Limitations of the current study}

While our survey provides a comprehensive overview of DG methods in CPath, there are several limitations to our study which we acknowledge.
First, due to the rapidly evolving nature of this field, there may be recent advancements and methodologies that were not covered in our survey. While we aimed to provide a broad and inclusive review, it is possible that emerging strategies, especially those published after our review period, have been overlooked. Furthermore, our focus was on peer-reviewed published works and highly cited preprints. Therefore, some of the preprint manuscripts that did not include novelty or relevant ideas were also excluded to maintain a review of high-quality works.
 
Second, our survey predominantly focused on quantitative methods for DG. Other complementary approaches such as studies that reported generalizability to (internal) test sets, where no domain shift exists, have not been discussed in depth.

Third, while we have made an effort to consider various forms of domain shifts and dissect them separately, in real-life applications these domain shifts may not be isolated. We acknowledge this limitation although we have explained why this approach to classifying DS can be beneficial \cref{sec:ds}. 

Fourth, our study did not provide a detailed performance comparison of different DG methods on other applications rather than classification, due to the lack of time, computational power, and experimental setups across the field. This could be addressed in future benchmark studies or the reader can refer to similar works that have reported the performance of different methods over various tasks \cite{graham2023conic,64_aubreville2023cpath,124_veta2019cpath,74_ehteshamibejnordi2017cpath,75_bandi2019cpath}.

Finally, the ethical, legal, and social implications of DG, such as issues related to privacy, fairness, bias, and data use, have not been fully addressed in this study. Readers can gain more insights into these topics by referring to \cite{chen2023algorithmic, nazer2023bias}.

These limitations provide opportunities for future updates to this survey, as well as areas for new research in the field of DG in CPath and medical image analysis.

\section{Conclusion}

In this survey, we have provided a comprehensive review of DG methods in CPath, offering insights into a wide range of strategies for handling domain shifts and a detailed discussion on their implementation. Despite certain limitations, this survey highlights the significant potential of DG in improving the robustness and generalizability of models in CPath. As we continue to navigate this rich field, the need for the development of more effective and adaptable DG techniques becomes increasingly apparent. Through continued research and innovation, we anticipate further progress in this field, ultimately contributing to more accurate, reliable, and universally applicable diagnostic and prognostic models in pathology.

\section*{Acknowledgment}
MJ, NZ, TV, JTK, and NR report financial support provided by the Medical Research Council (MRC) UK and South Korea biomedical and health researcher exchange scheme grant No. MC/PC/210-14. 
JTK and TV report financial support provided by the National Research Foundation of Korea (NRF) grant funded by the Korean government (MSIP) (No. 2021K1A3A1A88100920 and No. 2021R1A2C2-014557).
MJ, SEAR, and NR report financial support provided by the BigPicture project which is funded by the European Commission.
SEAR, FM, and NR report financial support provided by UK Research and Innovation (UKRI), outside of this work.
MR and NR report financial support from AstraZeneca, outside of this work.
NR, RJ, and FM report financial support from GlaxoSmithKline, outside of this work.
AS reports financial support from Cancer Research UK (CRUK), outside of this work.
NR is a co-founder of Histofy Ltd.
\appendices

\setcounter{table}{0}
\renewcommand{\thetable}{A\arabic{table}}
\setcounter{figure}{0}
\renewcommand{\thefigure}{A\arabic{figure}}
\section{Full results of benchmarking experiments}
\label{app:benchmark}

Benchmarking 28 deep learning algorithms for distinguishing between mitosis and mimicker classes in the context of the MIDOG22 training set \cite{65_aubreville2022cpath} was executed using the DomainBed framework \cite{34_gulrajani2021general} (introduced in Section \ref{sec:toolbox}). The chosen model for this experimentation is ResNet-50 \cite{he2016deep}, renowned for its ample capacity to generalize, and extensively employed within the CPath and computer vision communities. As previously indicated, the leave-one-domain-out cross-validation strategy was employed, and model selection for each experiment was based on a 20\% in-domain validation set, evaluated using the F1 Score metric.

\begin{figure}
    \centering
    \includegraphics[width=\columnwidth]{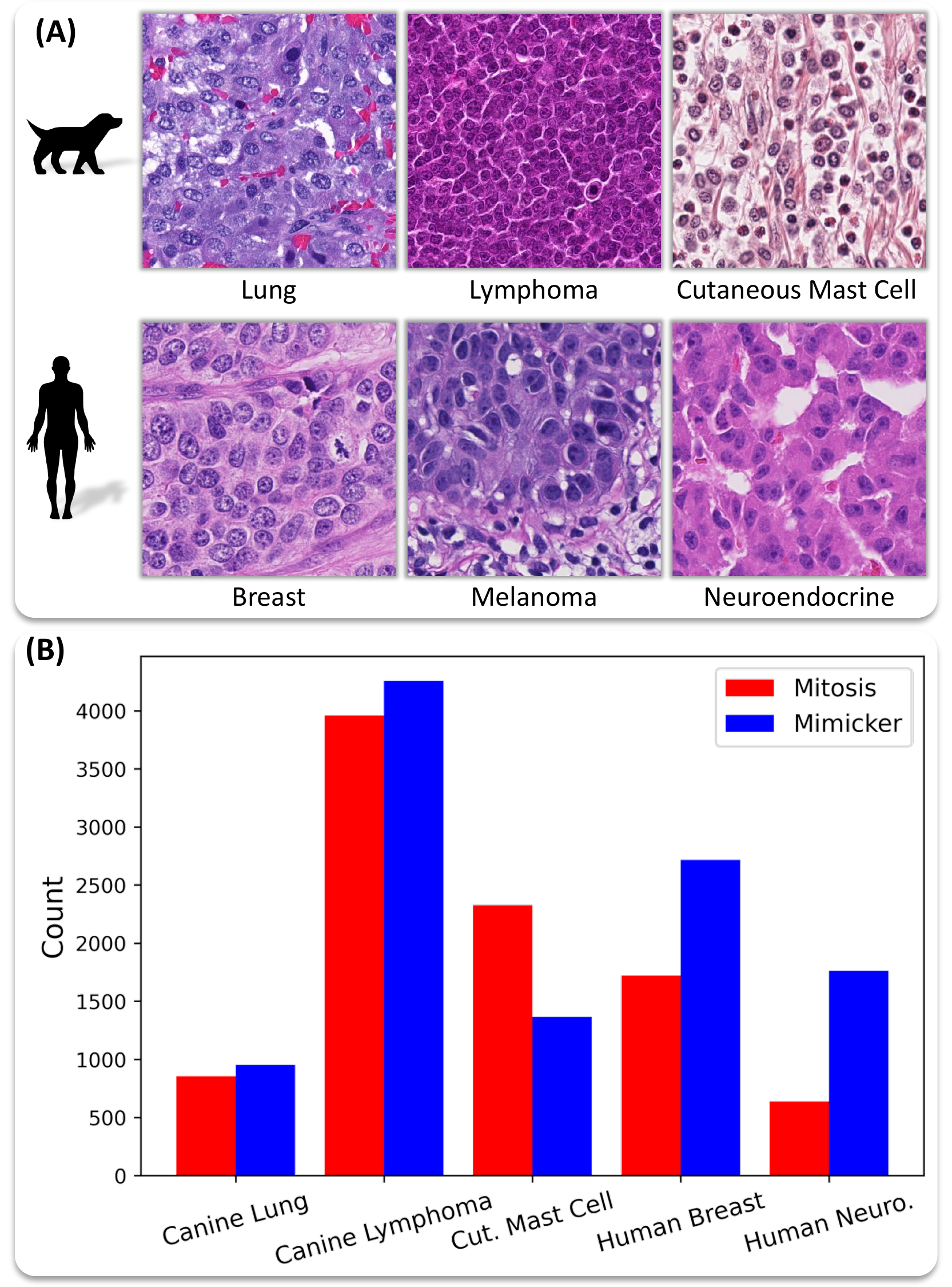}
    \caption{Mitosis vs. Mimicker population in different domains in MIDOG22 dataset.}
    \label{fig:midog}
\end{figure}

The MIDOG22 training set \cite{65_aubreville2022cpath} comprises 354 annotated images, encompassing canine lung cancer, human breast cancer, canine lymphoma, human neuroendocrine tumor, and canine cutaneous mast cell tumor domains. Each image is accompanied by annotations designating 9501 mitotic instances (positive class) and 11051 mimickers (negative class, representing challenging non-mitotic entities). Employing these annotations, patches of size $128\times128$ were extracted around each candidate, at a magnification of $40\times$ (equating to a resolution of 0.25 microns-per-pixel). The distribution of mitotic and mimicker instances across all domains is depicted in Figure \ref{fig:midog}(B), revealing a discernible prior shift particularly between the 'canine cutaneous mast cell tumor' domain and the others. Additionally, it is evident that label distributions for 'canine lung' and 'canine lymphoma' are closely aligned, though distinct from those of 'human breast' and 'human neuroendocrine'. Given the diverse sources of the samples within MIDOG22 (captured using different scanners), the occurrence of covariate shifts among the domains is inevitable. Considering the subjective nature of mitosis annotation \cite{veta2016mitosis,alkhasawneh2015interobserver,saldanha2020global}, it is reasonable to anticipate posterior shifts across the domains, even extending to images acquired from distinct scanners \cite{64_aubreville2023cpath}. Moreover, although there might be visual similarities among mitotic figures across various cancers and species, the inherent disparities in tissue structure and cell morphology across diverse species and tumor types (evident in Figure \ref{fig:midog}(A)) can lead to class-conditional shifts within the mimicker class. In summary, achieving domain generalization across the MIDOG22 dataset presents a formidable challenge due to the potential prevalence of all four types of domain shift, rendering it an ideal candidate for benchmarking diverse algorithms. However, evaluating model generalization against each of these shifts individually is not feasible; therefore, performance assessment is conducted on unseen domains, which collectively encapsulate these shifts.

Comprehensive results, encompassing F1 Score and Accuracy metrics for each domain within the benchmarking experiments on MIDOG22, are detailed in Tables \ref{tab:f1} and \ref{tab:acc} respectively. Evidently, the most proficient algorithm, StainAug, consistently performs well across diverse domains. The least favorable F1 Score performance occurs in the Human Neuroendocrine domain, possibly attributable to a pronounced prior shift; notably, the dataset contains a considerably higher count of mimickers in comparison to mitoses, unlike other domains, potentially impacting algorithm precision. Leveraging this analysis and the insights elaborated upon in Section \ref{sec:guidelines-handle}, it is feasible to adopt a strategic approach to enhance performance further.

\begin{table*}[]
\centering
\caption{Benchmarking F1 score of different DG algorithms for mitosis vs. mimicker classification task using MIDOG22 dataset over 5 different domains. Result cells for different methods on each domain are colored, green and red indicating better and worse results, respectively.}
\begin{tabular}{lcccccc}
\toprule \hline
\textbf{Algorithm} & \textbf{Canine Lung} & \textbf{Canine Lymph} & \textbf{Cut. Mast Cell} & \textbf{Human Breast} &  \textbf{Human Neuro.} & \textbf{Average} \\ \hline
\textbf{StainAug} \cite{21_tellez2019cpath}                         & \cellcolor[HTML]{63BE7B}\textbf{77.6±0.6}   & \cellcolor[HTML]{63BE7B}\textbf{74.3±0.8}    & \cellcolor[HTML]{FEEA83}80.4±1.2                    & \cellcolor[HTML]{63BE7B}\textbf{78.3±0.4}    & \cellcolor[HTML]{63BE7B}\textbf{69.5 ± 0.3}          & \cellcolor[HTML]{63BE7B}\textbf{76.0}     \\
EQRM \cite{eastwood2022eqrm}                                     & \cellcolor[HTML]{8CCA7E}76.8±0.6            & \cellcolor[HTML]{9FD07F}70.7±2.2             & \cellcolor[HTML]{A7D27F}83.3±0.8                    & \cellcolor[HTML]{7AC57D}78.0±0.9             & \cellcolor[HTML]{6BC17C}69.3±1.0                   & \cellcolor[HTML]{77C47D}75.6            \\
SD \cite{pezeshki2021sd}                                        & \cellcolor[HTML]{ABD380}76.2±1.0            & \cellcolor[HTML]{C7DB81}68.3±1.9             & \cellcolor[HTML]{63BE7B}85.3±0.4                    & \cellcolor[HTML]{B0D480}77.3±0.4             & \cellcolor[HTML]{87C97E}68.6±1.0                     & \cellcolor[HTML]{8BCA7E}75.2            \\
RSC \cite{huang2020rsc}                                        & \cellcolor[HTML]{9CCF7F}76.5±1.0            & \cellcolor[HTML]{A9D380}70.1±1.2             & \cellcolor[HTML]{8CCA7E}84.1±0.4                    & \cellcolor[HTML]{B0D480}77.3±0.7             & \cellcolor[HTML]{C2DA81}67.1±0.6                     & \cellcolor[HTML]{95CD7E}75.0              \\
ARM \cite{zhang2021arm}                                       & \cellcolor[HTML]{FEE783}73.2±1.3            & \cellcolor[HTML]{8BCA7E}71.9±0.1             & \cellcolor[HTML]{99CE7F}83.7±1.9                    & \cellcolor[HTML]{B7D780}77.2±0.8             & \cellcolor[HTML]{C6DB81}67.0±1.0                     & \cellcolor[HTML]{A9D380}74.6            \\
GroupDRO \cite{sagawa2019groupdro}                                 & \cellcolor[HTML]{78C47D}77.2±0.2            & \cellcolor[HTML]{B3D580}69.5±2.3             & \cellcolor[HTML]{BBD881}82.7±2.6                    & \cellcolor[HTML]{91CC7E}77.7±0.8             & \cellcolor[HTML]{FEEA83}65.5±0.4                     & \cellcolor[HTML]{AED480}74.5            \\
MTL \cite{blanchard2021mtl}                                      & \cellcolor[HTML]{63BE7B}77.6±0.8            & \cellcolor[HTML]{E4E483}66.5±0.9             & \cellcolor[HTML]{99CE7F}83.7±0.7                    & \cellcolor[HTML]{B0D480}77.3±0.6             & \cellcolor[HTML]{E2E383}66.3±0.3                     & \cellcolor[HTML]{B8D780}74.3            \\
VREx \cite{krueger2021vrex}                                     & \cellcolor[HTML]{87C97E}76.9±0.8            & \cellcolor[HTML]{D7E082}67.3±2.5             & \cellcolor[HTML]{D6DF82}81.9±1.9                    & \cellcolor[HTML]{BFD981}77.1±0.7             & \cellcolor[HTML]{E2E383}66.3±2.2                     & \cellcolor[HTML]{CBDC81}73.9            \\
SagNet \cite{nam2021sagnet}                                   & \cellcolor[HTML]{82C77D}77.0±1.0            & \cellcolor[HTML]{D2DE82}67.6±1.7             & \cellcolor[HTML]{FEEA83}80.4±1.4                    & \cellcolor[HTML]{B7D780}77.2±0.2             & \cellcolor[HTML]{DAE182}66.5±0.2                     & \cellcolor[HTML]{D0DE82}73.8            \\
CausIRL-CORAL \cite{chevalley2022CausIRL}                             & \cellcolor[HTML]{FEEA83}74.5±0.8            & \cellcolor[HTML]{C0D981}68.7±2.5             & \cellcolor[HTML]{B4D680}82.9±0.9                    & \cellcolor[HTML]{A8D27F}77.4±0.5             & \cellcolor[HTML]{FEE983}65.3±0.1                     & \cellcolor[HTML]{D0DE82}73.8            \\
CORAL \cite{sun2016deepcoral}                                    & \cellcolor[HTML]{A6D27F}76.3±0.5            & \cellcolor[HTML]{C7DB81}68.3±1.1             & \cellcolor[HTML]{FEEA83}80.6±1.2                    & \cellcolor[HTML]{FEEA83}76.1±0.6             & \cellcolor[HTML]{BAD881}67.3±1.3                     & \cellcolor[HTML]{D5DF82}73.7            \\
\hline
\textit{ERM} \cite{vapnik1999erm}                               & \cellcolor[HTML]{ABD380}76.2±0.5            & \cellcolor[HTML]{EEE683}65.9±0.4             & \cellcolor[HTML]{8FCB7E}84.0±0.6                    & \cellcolor[HTML]{ECE683}76.5±1.3             & \cellcolor[HTML]{FEEB84}65.6±1.4                     & \cellcolor[HTML]{DAE182}\textit{73.6}   \\
\hline
MLDG \cite{li2018mldg}                                     & \cellcolor[HTML]{FDEB84}74.6±0.8            & \cellcolor[HTML]{FEEA83}64.8±2.3             & \cellcolor[HTML]{E0E283}81.6±1.7                    & \cellcolor[HTML]{D6DF82}76.8±0.5             & \cellcolor[HTML]{DAE182}66.5±0.6                     & \cellcolor[HTML]{FDEB84}72.9            \\
DANN \cite{ganin2016dann}                                     & \cellcolor[HTML]{82C77D}77.0±0.8            & \cellcolor[HTML]{FEE983}64.4±0.2             & \cellcolor[HTML]{FEE983}79.9±2.5                    & \cellcolor[HTML]{F4E884}76.4±0.2             & \cellcolor[HTML]{D6E082}66.6±1.5                     & \cellcolor[HTML]{FDEB84}72.9            \\
Mixup \cite{yan2020mixup}                                       & \cellcolor[HTML]{FEE883}73.6±1.6            & \cellcolor[HTML]{FEE883}64.0±3.7             & \cellcolor[HTML]{FEEB84}80.7±1.3                    & \cellcolor[HTML]{6BC17C}78.2±0.7             & \cellcolor[HTML]{AFD480}67.6±0.3                     & \cellcolor[HTML]{FEEA83}72.8            \\
CDANN \cite{135_li2018general}                                    & \cellcolor[HTML]{BAD881}75.9±0.9            & \cellcolor[HTML]{FEDB80}59.5±10.5            & \cellcolor[HTML]{6AC07C}85.1±0.9                    & \cellcolor[HTML]{FEE683}74.4±3.1             & \cellcolor[HTML]{D6E082}66.6±0.6                     & \cellcolor[HTML]{FEE983}72.3            \\
IRM \cite{arjovsky2019irm}                                       & \cellcolor[HTML]{FEE783}73.3±3.0            & \cellcolor[HTML]{FEE883}64.0±0.3             & \cellcolor[HTML]{FEE482}78.0±1.0                    & \cellcolor[HTML]{FEDB80}69.9±5.5             & \cellcolor[HTML]{FEDE81}62.6±2.6                     & \cellcolor[HTML]{FEE182}69.5            \\
TRM \cite{xu2021trm}                                      & \cellcolor[HTML]{FEDE81}69.3±2.8            & \cellcolor[HTML]{FDCB7D}54.2±7.9             & \cellcolor[HTML]{F0E784}81.1±2.1                    & \cellcolor[HTML]{FEE983}75.8±0.3             & \cellcolor[HTML]{FEDE81}62.5±1.5                     & \cellcolor[HTML]{FEDE81}68.6            \\
IB-ERM \cite{ahuja2021ib}                                   & \cellcolor[HTML]{FEE082}70.4±1.2            & \cellcolor[HTML]{FDD880}58.6±1.3             & \cellcolor[HTML]{FDD880}72.6±4.8                    & \cellcolor[HTML]{FEDA80}69.5±0.5             & \cellcolor[HTML]{FEE583}64.3±0.2                     & \cellcolor[HTML]{FEDA80}67.1            \\
SANDMask \cite{shahtalebi2021sand}                                 & \cellcolor[HTML]{FCBD7B}56.1±10.0           & \cellcolor[HTML]{FDD880}58.5±1.8             & \cellcolor[HTML]{FEDC81}74.5±1.5                    & \cellcolor[HTML]{FEDD81}70.7±3.0             & \cellcolor[HTML]{FEE783}64.7±0.9                     & \cellcolor[HTML]{FDD47F}64.9            \\
ANDMask \cite{parascandolo2020andmask}                                   & \cellcolor[HTML]{FDD07E}63.9±3.2            & \cellcolor[HTML]{FCB479}46.4±3.9             & \cellcolor[HTML]{FEE582}78.2±1.4                    & \cellcolor[HTML]{FEE382}73.4±1.3             & \cellcolor[HTML]{FEDB81}61.8±0.8                     & \cellcolor[HTML]{FDD37F}64.7            \\
SelfReg \cite{kim2021selfreg}                                  & \cellcolor[HTML]{FDD47F}65.6±2.6            & \cellcolor[HTML]{FDD07E}56.0±3.8             & \cellcolor[HTML]{FDD680}72.1±3.0                    & \cellcolor[HTML]{FEDA80}69.7±2.8             & \cellcolor[HTML]{FDD37F}59.8±2.3                     & \cellcolor[HTML]{FDD37F}64.6            \\
CausIRL-MMD \cite{chevalley2022CausIRL}                               & \cellcolor[HTML]{FEE382}71.4±1.7            & \cellcolor[HTML]{FEE082}61.4±3.4             & \cellcolor[HTML]{FBA175}49.4±20.2                   & \cellcolor[HTML]{FDD680}68.2±7.3             & \cellcolor[HTML]{FCC17C}55.5±5.6                     & \cellcolor[HTML]{FDC97D}61.2            \\
Transfer \cite{zhang2021transfer}                                 & \cellcolor[HTML]{FEE583}72.5±3.9            & \cellcolor[HTML]{EDE683}66.0±1.0             & \cellcolor[HTML]{FEE382}77.6±4.3                    & \cellcolor[HTML]{FA9C74}44.8±12.6            & \cellcolor[HTML]{FA9373}44.3±0.9                     & \cellcolor[HTML]{FDC97D}61.1            \\
MMD \cite{li2018mmd}                                      & \cellcolor[HTML]{ABD380}76.2±1.1            & \cellcolor[HTML]{FEE082}61.4±2.6             & \cellcolor[HTML]{A7D27F}83.3±1.0                    & \cellcolor[HTML]{F86C6B}25.5±20.8            & \cellcolor[HTML]{FA9573}44.7±18.3                    & \cellcolor[HTML]{FCC17B}58.2            \\
IGA \cite{koyama2020iga}                                      & \cellcolor[HTML]{FBA376}45.5±15.2           & \cellcolor[HTML]{FFEB84}64.9±0.1             & \cellcolor[HTML]{FEE282}77.2±0.2                    & \cellcolor[HTML]{FCB87A}56.0±0.2             & \cellcolor[HTML]{F8696B}33.8±6.7                     & \cellcolor[HTML]{FCB97A}55.5            \\
CondCAD \cite{dubois2021cad}                                  & \cellcolor[HTML]{F8696B}21.8±17.2           & \cellcolor[HTML]{F8696B}21.2±16.3            & \cellcolor[HTML]{FBA777}52.0±21.3                   & \cellcolor[HTML]{F86C6B}25.5±20.8            & \cellcolor[HTML]{F8716C}35.9±13.8                    & \cellcolor[HTML]{F8746D}31.3            \\
CAD \cite{dubois2021cad}                                      & \cellcolor[HTML]{F8706C}25.0±14.3           & \cellcolor[HTML]{F8716C}24.0±16.6            & \cellcolor[HTML]{F8696B}25.1±20.5                   & \cellcolor[HTML]{F8696B}24.2±16.1            & \cellcolor[HTML]{F8786E}37.7±13.6                    & \cellcolor[HTML]{F8696B}27.2            \\ \hline \bottomrule
\end{tabular}
\label{tab:f1}
\end{table*}

\begin{table*}[]
\centering
\caption{Benchmarking accuracy of different DG algorithms for mitosis vs. mimicker classification task using MIDOG22 dataset over 5 different domains. Result cells for different methods on each domain are colored, green and red indicating better and worse results, respectively.}
\begin{tabular}{lcccccc}
\toprule \hline
\textbf{Algorithm} & \textbf{Canine Lung} & \textbf{Canine Lymph} & \textbf{Cut. Mast Cell} & \textbf{Human Breast} &  \textbf{Human Neuro.} & \textbf{Average} \\ \hline
\textbf{EQRM} \cite{eastwood2022eqrm}                          & \cellcolor[HTML]{6BC17C}79.0±0.3 & \cellcolor[HTML]{9FD07F}75.0±0.9 & \cellcolor[HTML]{93CC7E}80.6±0.1  & \cellcolor[HTML]{63BE7B}\textbf{83.9}±0.2 & \cellcolor[HTML]{A6D27F}82.4±0.1  & \cellcolor[HTML]{63BE7B}\textbf{80.2} \\
StainAug \cite{21_tellez2019cpath}                         & \cellcolor[HTML]{DDE283}77.5±0.6 & \cellcolor[HTML]{63BE7B}\textbf{76.2}±0.9 & \cellcolor[HTML]{FEE883}77.6±0.8  & \cellcolor[HTML]{6FC27C}83.8±0.4 & \cellcolor[HTML]{63BE7B}\textbf{84.4}±0.9  & \cellcolor[HTML]{81C77D}79.9 \\
Mixup \cite{yan2020mixup}                         & \cellcolor[HTML]{B7D780}78.0±0.6 & \cellcolor[HTML]{D5DF82}73.9±0.3 & \cellcolor[HTML]{EFE784}78.7±1.1  & \cellcolor[HTML]{C8DB81}83.0±0.8 & \cellcolor[HTML]{8BCA7E}83.2±0.9  & \cellcolor[HTML]{B1D580}79.4 \\
VREx \cite{krueger2021vrex}                          & \cellcolor[HTML]{F4E884}77.2±0.6 & \cellcolor[HTML]{BDD881}74.4±1.3 & \cellcolor[HTML]{B5D680}79.9±0.7  & \cellcolor[HTML]{7AC57D}83.7±0.3 & \cellcolor[HTML]{D1DE82}81.1±0.8  & \cellcolor[HTML]{BBD881}79.3 \\
CORAL \cite{sun2016deepcoral}                        & \cellcolor[HTML]{FEE883}76.5±0.8 & \cellcolor[HTML]{E4E483}73.6±0.7 & \cellcolor[HTML]{7BC57D}81.1±0.7  & \cellcolor[HTML]{6FC27C}83.8±0.4 & \cellcolor[HTML]{D8E082}80.9±0.2  & \cellcolor[HTML]{C5DB81}79.2 \\
DANN  \cite{ganin2016dann}                        & \cellcolor[HTML]{BFD981}77.9±0.2 & \cellcolor[HTML]{F3E884}73.3±0.5 & \cellcolor[HTML]{63BE7B}\textbf{81.6}±0.4  & \cellcolor[HTML]{DEE283}82.8±0.6 & \cellcolor[HTML]{E8E583}80.4±0.4  & \cellcolor[HTML]{C5DB81}79.2 \\
\hline
\textit{ERM}  \cite{vapnik1999erm}                          & \cellcolor[HTML]{B0D480}78.1±0.2 & \cellcolor[HTML]{FEE983}72.8±1.6 & \cellcolor[HTML]{EFE784}78.7±1.5  & \cellcolor[HTML]{90CB7E}83.5±0.2 & \cellcolor[HTML]{A9D380}82.3±0.5  & \cellcolor[HTML]{CFDD82}\textit{79.1} \\
\hline
MLDG \cite{li2018mldg}                          & \cellcolor[HTML]{99CE7F}78.4±0.5 & \cellcolor[HTML]{D5DF82}73.9±1.2 & \cellcolor[HTML]{B0D580}80.0±0.4  & \cellcolor[HTML]{B1D580}83.2±0.2 & \cellcolor[HTML]{FEE983}79.1±0.4  & \cellcolor[HTML]{E2E383}78.9 \\
MTL \cite{blanchard2021mtl}                           & \cellcolor[HTML]{D6DF82}77.6±0.2 & \cellcolor[HTML]{FEE783}72.5±1.2 & \cellcolor[HTML]{E5E483}78.9±0.2  & \cellcolor[HTML]{90CB7E}83.5±0.4 & \cellcolor[HTML]{B0D580}82.1±1.1  & \cellcolor[HTML]{E2E383}78.9 \\
CausIRL-CORAL \cite{chevalley2022CausIRL}                 & \cellcolor[HTML]{D6DF82}77.6±0.4 & \cellcolor[HTML]{95CD7E}75.2±1.5 & \cellcolor[HTML]{FEE783}77.5±0.8  & \cellcolor[HTML]{DEE283}82.8±0.6 & \cellcolor[HTML]{BDD881}81.7±1.5  & \cellcolor[HTML]{E2E383}78.9 \\
GroupDRO  \cite{sagawa2019groupdro}                    & \cellcolor[HTML]{63BE7B}\textbf{79.1}±0.1 & \cellcolor[HTML]{F8E984}73.2±0.4 & \cellcolor[HTML]{BFD981}79.7±0.9  & \cellcolor[HTML]{FFEB84}82.5±0.4 & \cellcolor[HTML]{FEE983}79.2±1.0  & \cellcolor[HTML]{ECE683}78.8 \\
CDANN \cite{135_li2018general}                        & \cellcolor[HTML]{DDE283}77.5±0.1 & \cellcolor[HTML]{E4E483}73.6±0.7 & \cellcolor[HTML]{A2D07F}80.3±1.3  & \cellcolor[HTML]{C8DB81}83.0±0.4 & \cellcolor[HTML]{FCEB84}79.8±1.3  & \cellcolor[HTML]{ECE683}78.8 \\
RSC  \cite{huang2020rsc}                         & \cellcolor[HTML]{FEE883}76.6±0.4 & \cellcolor[HTML]{C2DA81}74.3±0.7 & \cellcolor[HTML]{8ACA7E}80.8±0.3  & \cellcolor[HTML]{FEEA83}82.4±0.7 & \cellcolor[HTML]{F6E984}80.0±0.7  & \cellcolor[HTML]{ECE683}78.8 \\
SagNet \cite{nam2021sagnet}                       & \cellcolor[HTML]{B7D780}78.0±0.4 & \cellcolor[HTML]{EEE784}73.4±1.6 & \cellcolor[HTML]{E0E283}79.0±0.7  & \cellcolor[HTML]{B1D580}83.2±0.4 & \cellcolor[HTML]{FFEB84}79.7±0.2  & \cellcolor[HTML]{F6E984}78.7 \\
ARM  \cite{zhang2021arm}                         & \cellcolor[HTML]{FEEA83}77.0±0.4 & \cellcolor[HTML]{F8E984}73.2±0.5 & \cellcolor[HTML]{C4DA81}79.6±1.3  & \cellcolor[HTML]{FEE883}81.9±0.5 & \cellcolor[HTML]{DBE182}80.8±1.0  & \cellcolor[HTML]{FEEA83}78.5 \\
SD  \cite{pezeshki2021sd}                          & \cellcolor[HTML]{99CE7F}78.4±1.1 & \cellcolor[HTML]{DFE283}73.7±1.2 & \cellcolor[HTML]{FEE983}77.9±1.9  & \cellcolor[HTML]{D3DF82}82.9±0.3 & \cellcolor[HTML]{FEEA83}79.5±0.7  & \cellcolor[HTML]{FEEA83}78.5 \\
IRM \cite{arjovsky2019irm}                           & \cellcolor[HTML]{FCEA84}77.1±0.7 & \cellcolor[HTML]{FEEA83}72.9±0.5 & \cellcolor[HTML]{FEE983}78.0±0.7  & \cellcolor[HTML]{FFEB84}82.5±0.5 & \cellcolor[HTML]{F2E884}80.1±1.3  & \cellcolor[HTML]{FEE883}78.1 \\
MMD  \cite{li2018mmd}                          & \cellcolor[HTML]{FEE983}76.7±0.6 & \cellcolor[HTML]{FEE983}72.8±1.6 & \cellcolor[HTML]{8ACA7E}80.8±0.1  & \cellcolor[HTML]{FBA977}68.4±6.1 & \cellcolor[HTML]{FEE582}77.8±1.7  & \cellcolor[HTML]{FEDC81}75.3 \\
IB-ERM \cite{ahuja2021ib}                        & \cellcolor[HTML]{FDD47F}72.8±0.4 & \cellcolor[HTML]{FDD880}69.9±0.5 & \cellcolor[HTML]{FDD57F}72.8±3.5  & \cellcolor[HTML]{FEDB81}79.2±0.4 & \cellcolor[HTML]{FFEB84}79.7±0.7  & \cellcolor[HTML]{FEDB80}74.9 \\
TRM \cite{xu2021trm}                            & \cellcolor[HTML]{FEE182}75.2±1.9 & \cellcolor[HTML]{FDC77D}67.1±2.6 & \cellcolor[HTML]{FDD07E}71.5±3.8  & \cellcolor[HTML]{FEE582}81.3±0.4 & \cellcolor[HTML]{FEDE81}75.7±2.2  & \cellcolor[HTML]{FDD880}74.2 \\
ANDMask \cite{parascandolo2020andmask}                      & \cellcolor[HTML]{FDC77D}70.2±0.8 & \cellcolor[HTML]{FCC37C}66.4±3.5 & \cellcolor[HTML]{FEDA80}73.9±2.6  & \cellcolor[HTML]{FEDF81}80.1±0.4 & \cellcolor[HTML]{FEE182}76.6±2.2  & \cellcolor[HTML]{FDD57F}73.5 \\
SANDMask \cite{shahtalebi2021sand}                     & \cellcolor[HTML]{FCC07B}68.9±3.4 & \cellcolor[HTML]{FCC07B}66.0±1.4 & \cellcolor[HTML]{FDD27F}72.0±0.6  & \cellcolor[HTML]{FEDE81}79.8±0.9 & \cellcolor[HTML]{FEE382}77.4±0.7  & \cellcolor[HTML]{FDD27F}72.8 \\
SelfReg \cite{kim2021selfreg}                      & \cellcolor[HTML]{FDD07E}71.9±1.0 & \cellcolor[HTML]{FDD07E}68.6±1.6 & \cellcolor[HTML]{FDCF7E}71.1±1.9  & \cellcolor[HTML]{FEDC81}79.5±1.0 & \cellcolor[HTML]{FDCE7E}70.7±1.0  & \cellcolor[HTML]{FDD07E}72.4 \\
CausIRL-MMD \cite{chevalley2022CausIRL}                  & \cellcolor[HTML]{FEDE81}74.6±1.6 & \cellcolor[HTML]{FBB179}63.5±6.5 & \cellcolor[HTML]{FBAC77}62.0±8.2  & \cellcolor[HTML]{FCB77A}71.5±8.2 & \cellcolor[HTML]{FEDB81}74.9±1.4  & \cellcolor[HTML]{FCC37C}69.3 \\
Transfer \cite{zhang2021transfer}                     & \cellcolor[HTML]{FCBD7B}68.3±6.4 & \cellcolor[HTML]{FA9072}57.9±5.9 & \cellcolor[HTML]{FCBC7B}66.3±11.7 & \cellcolor[HTML]{FBA576}67.6±5.4 & \cellcolor[HTML]{FCBB7A}64.7±5.8  & \cellcolor[HTML]{FBB078}65.0 \\
CondCAD  \cite{dubois2021cad}                     & \cellcolor[HTML]{F98A71}58.6±5.1 & \cellcolor[HTML]{F98A71}56.9±4.8 & \cellcolor[HTML]{FCB97A}65.3±8.4  & \cellcolor[HTML]{FA9B74}65.4±6.8 & \cellcolor[HTML]{F97D6F}45.1±11.6 & \cellcolor[HTML]{FA9473}58.3 \\
CAD \cite{dubois2021cad}                          & \cellcolor[HTML]{F98770}57.9±4.9 & \cellcolor[HTML]{F98971}56.8±3.9 & \cellcolor[HTML]{F97B6E}49.2±9.7  & \cellcolor[HTML]{FAA075}66.5±4.6 & \cellcolor[HTML]{FBAB77}59.6±12.5 & \cellcolor[HTML]{FA9273}58.0 \\
IGA \cite{koyama2020iga}                            & \cellcolor[HTML]{F8696B}52.1±0.9 & \cellcolor[HTML]{F8696B}51.3±0.6 & \cellcolor[HTML]{F8696B}44.3±5.8  & \cellcolor[HTML]{F8696B}54.7±0.8 & \cellcolor[HTML]{F8696B}38.5±6.7  & \cellcolor[HTML]{F8696B}48.2           \\ \hline \bottomrule
\end{tabular}
\label{tab:acc}
\end{table*}

\section{Full list of resources} \label{app:resources}
\subsection{Datasets} \label{app:resources-data}
This section provides a comprehensive overview of the domain generalization datasets previously mentioned in  \cref{table:datasets}. A detailed description of each dataset is provided below, highlighting key characteristics such as the number of images, dataset sources, classes or categories and the tissue types represented. 

\begin{enumerate} 

\item Mitosis Domain Generalization Challenge 2021 (MIDOG21) \cite{64_aubreville2023cpath}:
The training set contained 200 cases obtained using four different whole slide image scanners (50 cases each) contributing to the Covariate Shift:  
\begin{enumerate} 
\item Hamamatsu XR  nanozoomer 2.0  
\item Hamamatsu S360 (0.5 NA)  
\item Aperio ScanScope  CS2  
\item Leica GT450  
\end{enumerate} 
The test set contained 80 cases from different tumor cases, with two scanners that were part of the training set and two (undisclosed) scanners, with 20 slides per scanner.

\item Mitosis Domain Generalization Challenge 2022 (MIDOG22) \cite{65_aubreville2022cpath} \cite{midog22dataset}: This particular dataset encompasses a large variation of covariate shift as images are sourced from different centres and consequently different scanners as well as various tissue types, and species. Additionally, as the distribution of labelled mitosis differs across the various cancer type this dataset also contains a Prior Shift. The Posterior shift is a consequence of the subjectivity involved in the data labeling process. The training dataset comprised of the following: 
\begin{enumerate} 
     \item Human Breast Cancer (150 cases from the MIDOG21 dataset  \cite{64_aubreville2023cpath}) 
     \item Human neuroendocrine tumor (55 cases scanned with Hamamatsu NanoZoomer XR) 
     \item Human melanoma (49 cases scanned using Hamamatsu NanoZoomer XR) 
     \item Canine Lung Cancer (44 cases scanned with 3DHistech Pannoramic Scan II) 
     \item Canine Lymphoma (55 cases scanned with 3DHistech Pannoramic Scan II) 
     \item Canine Cutaneous Mast Cell Tumor (50 cases scanned using Aperio ScanScope CS2) 
\end{enumerate}  
The test dataset comprised of 100 cases, from 10 different tumor types. 

\item A comprehensive multi-domain dataset for mitotic figure detection (MIDOG++): The MIDOG ++ dataset is an extension of the previous MIDOG21 and MIDOG22 datasets \cite{64_aubreville2023cpath, 65_aubreville2022cpath}): The dataset consists of regions of interest (RoIs) from 503 distinct cases across seven different tumor types from both canine and humor tissue:
\begin{enumerate}
    \item breast carcinoma
    \item lung carcinoma
    \item lymphosarcoma
    \item neuroendocrine tumor
    \item cutaneous mast cell tumor 
    \item cutaneous melanoma
    \item (sub)cutaneous soft tissue sarcoma.
\end{enumerate}
The dataset contained labels for 11,937 mitotic figur. All images were scanned magnification of 40 using one of the following scanners:
\begin{enumerate}
    \item Hamamatsu XR
    \item Hamamatsu S360
    \item Leica ScanScope CS2
    \item 3DHistech Pannoramic Scan II 
    \item Aperio ScanScope CS2
\end{enumerate}

\item MITOS-ATYPIA-14 \cite{126_roux2014_cpath}: The dataset consisted of cases scanned using two different scanners : 
\begin{enumerate} 
     \item Aperio Scanscope XT  
     \item Hamamatsu Nanozoomer 2.0-HT. 
\end{enumerate} 

\item TUmor Proliferation Assessment Challenge 2016 (TUPAC16) \cite{124_veta2019cpath}: The training dataset comprised of 500 cases sourced from The Cancer Genome Atlas (TCGA) \cite{168_weinstein2013cpath} . Additionally, two auxiliary datasets were provided : 
\begin{enumerate} 
    \item A Region of Interest (ROI) dataset consisting of 148 randomly selected cases  
    \item A Mitosis dataset consisting of 73 cases sourced from three different centers  
\end{enumerate} 
The test dataset consisted of 321 cases of breast cancer and the mitosis detection test set comprised of 34 breast cancer cases sourced from two different pathology labs.

\item AIDPATH Kidney Database \cite{171_bueno2020cpath}: This dataset comprises of 47 WSIs (at $20 \times$ magnification) obtained from 5 different datasets of human kidney tissue cohorts from three medical institutions: 
\begin{enumerate}
    \item Castilla-La Mancha’s Healthcare services
    \item The Andalusian Health Service 
    \item The Vilnius University Hospital Santaros Klinikos
\end{enumerate}

\item The Tumor InfiltratinG lymphocytes in
breast cancER (TiGER) Challenge \cite{170_rijthoven2022cpath}: This dataset contained manual annotatated images of Her2 positive (HER2+) and Triple Negative (TNBC) breast cancer WSIs sourced from the Radboud University Medical Center (RUMC), the Jules Bordet Institut (JB) and the TCGA-BRCA archive \cite{168_weinstein2013cpath}. The training data was released as three different datasets:
\begin{enumerate}
    \item Whole-slide images with manual annotations in regions of interest (WSIROIS) consisted of 195 WSIs manually annotated with the following class labels: invasive tumor, tumor-associated stroma, in-situ tumor, healthy glands, necrosis not in-situ, inflamed stroma and rest. 
    \item Whole-slide images with coarse manual annotation of the tumor bulk (WSIBULK) containing 93 WSIs from RUMC and JB. For each WSI, coarse (manual) annotations of one or more regions that contain invasive tumor cells, "tumor bulk", are provided.
    \item Whole-slide images with visual estimation of the TILs at slide level (WSITILS) consisting of 82 WSI sourced from RUMC and JB with one TIL value per slide provided. 
\end{enumerate}
Four test sets, (two for each of the two leaderboards) were also provided, including one experimental test set and a final test set. 

\item Camelyon16 \cite{74_ehteshamibejnordi2017cpath}: The training dataset consisted of 270 whole slide images (WSIs) comprising of 110 WSIs with and 160 WSIs without nodal metastases, sourced from two centres using two different scanners, namely, Radboud University Medical Center (RUMC) (Pannoramic 250 Flash II) and University Medical Center Utrecht (UMCU) (NanoZoomer-XR Digital slide scanner C12000-01). The test dataset consisted of 129 WSIs, 49 containing nodal metastases and 80 without nodal metastases, collected from both Universities. 

\item Camelyon17 \cite{75_bandi2019cpath}: The training data consisted of 899 WSIs collected  from the lymph
node resections of 100 patients from five different medical institutions as follows: \begin{enumerate}
\item slides from 130 resections from the Radboud University Medical Center (RUMC)
\item 144 from Canisius-Wilhelmina
Hospital (CWZ)
\item 129 from the University Medical
Center Utrecht (UMCU) 
\item 168 from Rijnstate Hospital (RST)
\item 140 from the Laboratory of Pathology
East-Netherlands (LPON)
\end{enumerate}
The test set consisted of 500 slides across 100 patients with 5 slides per patient. 

\item PatchCamelyon benchmark (PCam) \cite{76_veeling2018cpath}: The PCam dataset is a patch-level dataset curated from the aforementioned Camelyon16 \cite{74_ehteshamibejnordi2017cpath} dataset. The dataset consists of 327,680 patches of size $96 \times 96$ pixels at a magnification of $10 \times$ with 75\% of the patches allocated as the training set, 12.5\% as the validation set and 12.5\% as the test set. 

\item Camelyon17-WILDS \cite{137_koh2021cpath}: 
 The dataset was curated from the aforementioned  Camelyon17 \cite{75_bandi2019cpath} dataset but unlike the original dataset the Camelyon17-WILDS dataset takes distribution shifts into consideration and splits the dataset according to the domain. Domain in this context, refers to the medical institutions the WSIs were sourced from. Camelyon17-WILDS consists of a total of 450,000 patches extracted from 50 WSIs sourced from  Camelyon17 \cite{75_bandi2019cpath} with 10 WSIs sourced from each of the 5 hospitals (domains). The training and testing splits are as follows: 
\begin{enumerate}
    \item Training: The training set consisted of 302,436 patches sourced from three different hospitals.
    \item In-distribution (ID) Validation: This validation set comprised of 33,560 patches extracted from the same 30 WSIs as in the training set, hence the term "in-distribution".
    \item Out-of-distribution (OOD) Validation: This validations set contained 34,904 patches extracted from 10 WSIs obtained from a 4th hospital, hence the term "out-of-distribution".
    \item Out-of-distribution (OOD) Test: Like the OOD Validation set the OOD Test set consisted of 85,054 patches taken from 10 WSIs from the 5th hospital. 
\end{enumerate}

\item Lung and Colon Cancer Histopathological Image Dataset (LC25000) \cite{172_borkowski2019cpath}: This dataset contains 25,000 images of size $768 \times 768$ pixels for two different organs, lung and colon. The images are equally divided into five distinct categories :
\begin{enumerate}
    \item colon adenocarcinoma
    \item benign colonic tissue
    \item lung adenocarcinoma
    \item lung squamous cell carcinoma
    \item benign lung tissue
\end{enumerate}

\item Kather 19 \cite{173_kather2019cpath}: The training dataset (NCT-CRC-HE100K) contains 100,000 patches from 86 H\&E slides sourced from the National Center for Tumor diseases (NCT) and University Medical Center Mannheim (UMM). 

The test dataset (CRC-VAL-HE-7K) contains 7180 image patches acquired from 25 H\&E slides of CRC tissue obtained from the DACHS study \cite{brenner2011long}, \cite{hoffmeister2015statin} in the NCT biobank.

All images were of size $224 \times 224$ pixels and obtained at $20 \times$. They belonged to 9 different classes: Adipose, background, debris, lymphocytes, mucus, smooth muscle, normal colon mucosa, cancer-associated stroma, colorectal adenocarcinoma epithelium. 

\item Prostate Cancer Grade Assessment (PANDA) Challenge \cite{175_bulten2022cpath}: The dataset consisted of 12,625 WSIs of prostate biopsies. 10,616 biopsies were used as the development set, 393 biopsies as the tuning set during the competition phase and 545 biopsies were used for validation in the post-competition phase. All these biospies were soured from the Radboud University Medical Center and Karolinska Institutet. Additionally, 1,071 biopsies were sourced from the Karolinska University Hospital and two medical laboratories and a tertiary teaching hospital in the US and used as an external validation set. 

\item The Lymphocyte Assessment Hackathon and Benchmark Dataset (LYSTO) \cite{169_jiao2020cpath}: The dataset consisted of a total of 83 WSIs with of 28 colon, 33 breast and 22 prostate slides. The images were curated from 9 different medical centers. The WSIs were split into a training set consisting of 43 slides derived from 2 centers and a test set containing 40 slides sourced from eight centers. The training and test datasets had an overlap of one centre. 

\item Multi-Organ Nucleus Segmentation Challenge (MoNuSeg) \cite{68_kumar2020cpath}: The training set for this challenge consisted of 30 tissue images of size $1000 \times 1000$, each extracted from a different WSI (at a magnification of $40 \times$) belonging to an individual patient. Th tissue images were downloaded from the TCGA \cite{168_weinstein2013cpath} and belonged to 18 different hospitals and encompassed 7 different organs, namely,  breast, liver, kidney, prostate, bladder, colon and stomach. The dataset contained 21, 623 hand-annotated nuclear boundaries and included tissue samples from both benign and diseased tissue.

Similarly, test set also consisted of 14 tissue images of size $1000 \times 1000$ and approximately 7,223 hand-annotated nuclei. The test set too contained images from 7 different organs, kidney, lung,
colon, breast, bladder, prostate, brain out of which two (lung and brain) were not previously included in the training set. 

\item A Dataset for Nuclei Segmentation of Cryosectioned H\&E-stained Histological Images (CryoNuSeg) \cite{67_mahbod2021cpath}: In order to create this particular dataset 30 WSIs (at a magnification of $40 \times$) from The Cancer Genome Atlas (TCGA) dataset \cite{168_weinstein2013cpath}. The dataset encompassed 10 different organs (with three WSIs for each organ): the adrenal gland, larynx, lymph node, mediastinum, pancreas, pleura, skin, testis, thymus, and thyroid gland. One image patch of size $512 \times 512$ pixels was extracted from each WSI. Additionally, the nuclei in the dataset were annotated by two experts, one of which relabelled the entire dataset. This provided three sets of manual mark-ups for the dataset. 

\item A Multi-Organ Nuclei Segmentation and Classification Challenge (MoNuSAC2020) \cite{73_verma2021cpath}: The dataset contained four organs (Lung, Prostate, Kidney, and Breast) and four nucleus types, epithelial cells, lymphocytes, neutrophils, and macrophages. The training set comprised of WSIs (at a magnification of $40 \times$) sourced from 46 patients and 32 hospitals with 31,000 nuclear boundary annotations from The Cancer Genome Atlas (TCGA) \cite{168_weinstein2013cpath} data portal. The test data was collected from 19 hospitals out of which 14 were shared with the training dataset and 25 patients that did not overlap with the patients present in training set. The testing data also contained annotations for ambiguous regions which was not present in the training data resulting in a type 2 domain shift.

\item The 2018 Data Science Bowl \cite{70_caicedo2019cpath}: The dataset contained 37,333 manually annotated nuclei across 841 images split into five broad image categories based on visual differences, namely, purple tissue images, pink tissue images, gray-scale tissue images, small fluorescent images and large fluorescent images. These images were collected across 31 different biological experiments and represented 22 different cell types as well as 15 different image resolutions.

\item Crowdsource 2015 \cite{71_irshad2014cpath}: The Crowdsource dataset comprises of 810 images of size $400 \times 400$ pixels sourced from 10 WSIs (at a magnification of $40 \times$) of kidney renal clear cell carcinoma (KIRC) downloaded from the TCGA portal \cite{168_weinstein2013cpath}. The dataset also contained annotations from four types of annotators:
\begin{enumerate}
    \item crowdsourced annotations across a range of contributor skill levels (1, 2, or 3)
    \item published state-of-the-art automated nucleui detection and segmentation algorithms
    \item research fellows
    \item expert pathologists (Ground truth)
\end{enumerate}

\item CPM-17 \cite{69_vu2019cpath}: The CPM-17 dataset contains tissue images of four types of cancers, non-small cell lung cancer (NSCLC), head and neck squamous cell carcinoma (HNSCC), glioblastoma multiforme (GBM), and lower grade glioma (LGG) tumors. It consisted of a total of 64 images captured at $20 \times$ and $40 \times$ magnification.

\item PanNuke: An Open Pan-Cancer Histology Dataset for Nuclei Instance Segmentation and Classification \cite{66_gamper2020cpath}: The PanNuke dataset consists of 189,744 labeled nuclei from 2000 visual fields sampled from more than 20,000 WSIs obtained from the TCGA data portal \cite{168_weinstein2013cpath} and University Hospitals Coventry and Warwickshire (UHCW). The WSIs encompassed a wide range of tissue types, totaling 19 distinct organs: adrenal, bile duct, bladder, breast, colon, cervix, esophagus, head\&neck, kidney, liver, lung, ovarian, pancreatic, prostate, skin, testis, stomach, thyroid and uterus. The nuclei were classified into five categories, inflammatory, neoplastic, dead, connective and non-neoplastic
epithelial, the distributions of which may vary from tissue to tissue. 

\item Lizard: A large-scale dataset for colonic nuclear instance segmentation and classification \cite{72_graham2021cpath}: The lizard dataset was curated from six different sources:
\begin{enumerate}
    \item Gland Segmentation Challenge Contest (GlaS) challenge contest \cite{sirinukunwattana2017gland}
    \item Colorectal Adenocarcinoma Gland (CRAG) \cite{graham2019mild}
    \item Colorectal Nuclear Segmentation and Phenotypes (CoNSeP) \cite{graham2019hover}
    \item Digestive-System Pathological Detection and Segmentation Challenge (DigestPath) 2019 \cite{125_da2022cpath}
    \item PanNuke: An Open Pan-Cancer Histology Dataset for Nuclei Instance Segmentation and Classification \cite{66_gamper2020cpath}
    \item The Cancer Genome Atlas (TCGA) \cite{168_weinstein2013cpath}
\end{enumerate}
In total, 291 image regions were extracted with an average size of $1016 \times 917$ pixels at $20 \times$ magnification.
Nuclei class labels were provided for epithelial cells, connective tissue cells, lymphocytes, plasma cells, neutrophils and eosinophils resulting in nearly half a million labelled nuclei.

\item Digestive-System Pathological Detection and Segmentation Challenge (DigestPath) 2019 \cite{125_da2022cpath}: The DigestPath challenge released two datasets:
\begin{enumerate}
    \item The signet ring cell detection dataset: This dataset consisted of 682 images of size $2000 \times 2000$ pixels from WSIs (scanned at $40 \times$ belonging to a total of 155 patients. The dataset encompassed two organs namely, gastric mucosa and intestine and consisted of 104 positive images (with signet ring cells) and 583 negative images and 14,859 annotated cells (noisey dataset- not fully annotated).
    \item The colonoscopy tissue segmentation and classification dataset: This dataset consisted of 872 tissue images with an average size of $5000 \times 5000$ pixels, acquired from 476 patients and re-scaled to $20 \times$. 
\end{enumerate}
The images for both datasets were sourced from four medical institutions as follows:
\begin{enumerate}
    \item Ruijin Hospital
    \item Xijing Hospital
    \item Shanghai Songjiang District Central Hospital 
    \item Histo Pathology Diagnostic Center
\end{enumerate}

\item Breast Cancer Semantic Segmentation (BCSS) \cite{174_amgad2019cpath} : This dataset contains over 20,000 segmentation annotations for regions of interest in breast cancer tissue images sourced from 20 different centers from TCGA \cite{168_weinstein2013cpath}. 

\item MetaHistoSeg: A Python Framework for Meta Learning in Histopathology Image Segmentation \cite{45_yuan2021cpath}: The dataset was curated from five different datasets:
\begin{enumerate}
    \item Automatic Gleason grading of prostate cancer in digital histopathology (Gleason2019) \cite{nir2018automatic, karimi2019deep}
    \item BreastPathQ: Cancer Cellularity Challenge 2019 \cite{peikari2017automatic}
    \item Multi-Organ Nucleus Segmentation Challenge (MoNuSeg) \cite{68_kumar2020cpath}
    \item Gland Segmentation Challenge Contest (GlaS) challenge contest \cite{sirinukunwattana2017gland}
    \item Digestive-System Pathological Detection and Segmentation Challenge (DigestPath) 2019 \cite{125_da2022cpath}
\end{enumerate}

\item A Fully Annotated Dataset for Nuclei
Instance Segmentation in H\&E-Stained Histological
Images (NuInsSeg) \cite{176_mahbod2023cpath}: The NuInsSeg dataset contains 665 image tiles sourced from the Medical University of Vienna (MUV) with 30,698 manually segmented nuclei from 31 human and mouse organs. 

\item The Clinical Proteomic Tumor Analysis Consortium (CPTAC) \cite{ellis2013connecting,167_edwards2015cpath}: The dataset is a comprehensive resource encompassing diverse proteogenomics data from various cancer types. It includes data from thousands of primary cancer and matched normal samples, spanning 10 cancer types, and sourced from 11 collaborating institutions.

\item The Cancer Genome Atlas (TCGA) \cite{168_weinstein2013cpath}: The dataset contains data from 20,000 primary cancer and matched normal samples spanning 33 cancer types collected from 11,000 patients with 20 collaborating institutions.
\end{enumerate}

\subsection{Code bases}
\label{app:resources-code}
In this section, we present a concise summary table, Table. \ref{table:codebases}, which highlights the papers covered in this study on domain generalization in computational pathology, which have publicly available codebases. The table includes information about the domain generation (DG) method type, the paper title, and author name/reference, categorized according to the method category. 

\onecolumn
\setlength{\tabcolsep}{8pt}
\begin{center}
\begin{longtable}{p{3cm}p{5cm}p{7cm}}
\caption{Domain Generalization Publicly Available Code Bases}
\label{table:codebases} \\
\toprule \hline
 \textbf{Reference} & \textbf{DG method} & \textbf{Title} \\
 \hline 
 \endfirsthead
 \multicolumn{3}{c}%
{{\tablename\ \thetable{} -- continued from previous page}} \\
\hline
\textbf{Reference} & \textbf{DG method} & \textbf{Title} \\
\hline
\endhead

\hline \multicolumn{3}{r}{{Continued on next page}} \\ \hline
\endfoot

\hline
\endlastfoot

\multicolumn{3}{c}{\textbf{Pretraining}}\\
\hline
Yang \emph{et al}. \cite{5_yang2022cpath} & Minimizing Contrastive Loss & CS-CO: A Hybrid Self-Supervised Visual Representation Learning Method for H\&E-stained Histopathological Images \\

Li \emph{et al}. \cite{6_li2022cpath} & Minimizing Contrastive Loss & Lesion-Aware Contrastive Representation Learning For Histopathology Whole Slide Images Analysis \\

Galdran \emph{et al}. \cite{86_galdran2022cpath} & Unsupervised or Self-supervised learning  & Test Time Transform Prediction for Open Set Histopathological Image Recognition  \\

Bozorgtabar \emph{et al}. \cite{87_bozorgtabar2021cpath} & Unsupervised or Self-supervised learning & SOoD: Self-Supervised Out-of-Distribution Detection Under Domain Shift for Multi-Class Colorectal Cancer Tissue Types \\

Koohbanani \emph{et al}. \cite{89_koohbanani2021cpath} & Multiple Pretext Tasks & Self Path: Self Supervision for Classification of Histology Images with Limited Budget of Annotation \\

Abbet \emph{et al}. \cite{91_abbet2022cpath} & Unsupervised or Self-supervised learning & Self-rule to multi-adapt: Generalized multi-source feature learning using unsupervised domain adaptation for colorectal cancer tissue detection \\

Cho\emph{et al}. \cite{117_cho2021cpath} & Unsupervised or Self-supervised learning & Cell Detection in Domain Shift Problem Using Pseudo-Cell-Position Heatmap \\

Chikontwe \emph{et al}. \cite{123_chikontwe2022cpath} & Unsupervised or Self-supervised learning & Weakly supervised segmentation on neural compressed histopathology with self-equivariant regularization \\

Tran \emph{et al}. \cite{142_tran2022cpath} & Minimizing Contrastive Loss & S5CL: Unifying Fully-Supervised, Self-Supervised, and Semi-Supervised Learning Through Hierarchical Contrastive Learning \\

Sikaroudi \emph{et al}. \cite{154_sikaroudi2020cpath} & Unsupervised or Self-supervised learning & Supervision and Source Domain Impact on Representation Learning: A Histopathology Case Study \\

Wang \emph{et al}. \cite{158_wang2022cpath} & Unsupervised or Self-supervised learning & Transformer-based unsupervised contrastive learning for histopathological image classification \\

Kang \emph{et al}. \cite{159_kang2023cpath} & Unsupervised or Self-supervised learning & Benchmarking Self-Supervised Learning on Diverse Pathology Datasets \\

Lazard \emph{et al}. \cite{178_lazard2023cpath} & Contrastive Learning & Giga-SSL: Self-Supervised Learning for Gigapixel Images \\

Vuong \emph{et al}. \cite{180_vuong2023cpath} & Contrastive Learning & IMPaSh: A Novel Domain-Shift Resistant Representation for Colorectal Cancer Tissue Classification \\

Chen \emph{et al}. \cite{194_chen2022cpath} & Unsupervised or Self-supervised learning & Fast and scalable search of whole-slide images via self-supervised deep learning \\

\hline

\multicolumn{3}{c}{\textbf{Meta-Learning}}\\
\hline
Sikaroudi \emph{et al}. \cite{16_sikaroudi2021cpath} & Meta-learning & Magnification Generalization For Histopathology Image Embedding \\

Yuan \emph{et al}. \cite{45_yuan2021cpath}  & Meta-learning & MetaHistoSeg: A Python Framework for Meta Learning in Histopathology Image Segmentation \\
\hline

\multicolumn{3}{c}{\textbf{Domain Alignment}}\\
\hline
Sharma \emph{et al}. \cite{19_sharma2022cpath} & Mutual Information & MaNi: Maximizing Mutual Information for Nuclei Cross-Domain Unsupervised Segmentation \\

Boyd \emph{et al}. \cite{31_boyd2022cpath} & Generative Models & Region-guided CycleGANs for Stain Transfer in Whole Slide Images \\

Kather \emph{et al}. \cite{79_kather2019cpath} & Stain Normalization & Deep learning can predict microsatellite instability directly from histology in gastrointestinal cancer \\

Zheng \emph{et al}. \cite{203_zheng2019cpath} & Stain Normalization & Adaptive color deconvolution for histological WSI normalization \\

Sebai \emph{et al}. \cite{80_sebai2020cpath} & Stain Normalization & MaskMitosis: a deep learning framework for fully supervised, weakly supervised, and unsupervised mitosis detection in histopathology images \\

Zhang \emph{et al}. \cite{90_zhang2022cpath} & Minimizing Contrastive Loss & Stain Based Contrastive Co-training for Histopathological Image Analysis \\

Shahban \emph{et al}. \cite{103_shaban2019cpath} & Generative Models & Staingan: Stain Style Transfer for Digital Histological Images \\

Wagner \emph{et al}. \cite{113_wagner2022cpath} & Generative Models & Federated Stain Normalization for Computational Pathology\\

Quiros \emph{et al}. \cite{143_quiros2021cpath} & Domain Adversarial Learning & Adversarial learning of cancer tissue representations \\

Salehi \emph{et al}. \cite{147_salehi2022cpath} & Minimizing the KL Divergence & Unsupervised Cross-Domain Feature Extraction for Single Blood Cell Image Classification \\

Wilm \emph{et al}. \cite{188_wilm2021cpath} & Domain-Adversarial Learning & Domain adversarial retinanet as a reference algorithm for the mitosis domain generalization (midog) challenge \\

Haan \emph{et al}. \cite{193_haan2021cpath} & Generative models & Deep learning-based transformation of H\&E stained tissues into special stains \\

Dawood \emph{et al}. \cite{207_dawood2023cpath} & Stain Normalization & Do Tissue Source Sites leave identifiable Signatures in Whole Slide Images beyond staining? \\ 

\hline

\multicolumn{3}{c}{\textbf{Data Augmentation}}\\
\hline
Pohjonen \emph{et al}. \cite{29_pohjonen2022cpath} & Data augmentation & Augment like there’s no tomorrow: Consistently performing neural networks for medical imaging  \\

Chang \emph{et al}. \cite{56_chang2021cpath} & Stain Augmentation & Stain Mix-up: Unsupervised Domain Generalization for Histopathology Images \\

Shen \emph{et al}. \cite{57_shen2022cpath} & Stain Augmentation & RandStainNA: Learning Stain-Agnostic Features from Histology Slides by Bridging Stain Augmentation and Normalization \\

Koohbanani \emph{et al}. \cite{59_alemi-koohbanani2020cpath} & Data augmentation & NuClick: A deep learning framework for interactive segmentation of microscopic images  \\

Wang \emph{et al}. \cite{61_wang2023cpath}  & Data augmentation &  A generalizable and robust deep learning algorithm for mitosis detection in multicenter breast histopathological images \\ 

Lin \emph{et al}. \cite{62_lin2022cpath} & Generative Models & InsMix: Towards Realistic Generative Data Augmentation for Nuclei Instance Segmentation \\

Zhang \emph{et al}. \cite{85_zhang2022cpath} & Data augmentation & Benchmarking the Robustness of Deep Neural Networks to Common Corruptions in Digital Pathology \\

Yamashita \emph{et al}. \cite{92_yamashita2021cpath} & Style Transfer Models & Learning domain-agnostic visual representation for computational pathology using medically-irrelevant style transfer augmentation \\

Falahkheirkhah \emph{et al}. \cite{98_falahkheirkhah2023cpath} & Generative Models & Deepfake Histologic Images for Enhancing Digital Pathology \\

Scalbert \emph{et al}. \cite{102_scalbert2022cpath} & Generative Models & Test-time image-to-image translation ensembling improves out-of-distribution generalization in histopathology \\

Mahmood \emph{et al}. \cite{128_mahmood2020cpath} & Generative Models & Deep Adversarial Training for Multi-Organ Nuclei Segmentation in Histopathology Images \\

Fan \emph{et al}. \cite{141_fan2022cpath} & Generative Models & Fast FF-to-FFPE Whole Slide Image Translation via Laplacian Pyramid and Contrastive Learning \\

Marini \emph{et al}. \cite{148_marini2023cpath} & Stain Augmentation & Data-driven color augmentation for H\&E stained images in computational pathology \\

Faryna \emph{et al}. \cite{163_faryna2021cpath} & RandAugment for Histology & Tailoring automated data augmentation to H\&E-stained histopathology \\

\hline

\multicolumn{3}{c}{\textbf{Model Design}}\\
\hline
Graham \emph{et al}. \cite{37_graham2020cpath} & Model design & Dense Steerable Filter CNNs for Exploiting Rotational Symmetry in Histology Images \\

Lafarge \emph{et al}. \cite{38_lafarge2021cpath} & Model design & Roto-translation equivariant convolutional networks: Application to histopathology image analysis \\

Zhang \emph{et al}. \cite{40_zhang2022cpath} & Model design & DDTNet: A dense dual-task network for tumor-infiltrating lymphocyte detection and segmentation in histopathological images of breast cancer \\

Graham \emph{et al}. \cite{119_graham2023cpath} & Model Design & One model is all you need: Multi-task learning enables simultaneous histology image segmentation and classification \\ 

Yu \emph{et al}. \cite{130_yu2023cpath} & Model Design &  Prototypical multiple instance learning for predicting lymph node metastasis of breast cancer from whole-slide pathological images\\

Yaar \emph{et al}. \cite{131_yaar2020cpath} & Model Design & Cross-Domain Knowledge Transfer for Prediction of Chemosensitivity in Ovarian Cancer Patients \\

Tang \emph{et al}. \cite{145_tang2021cpath} & Model Design & Probeable DARTS with Application to Computational Pathology \\

Vuong \emph{et al}. \cite{182_vuong2021cpath} & Model Design & Joint categorical and ordinal learning for cancer grading in pathology images \\

\hline

\multicolumn{3}{c}{\textbf{Learning Disentangled Representations}}\\
\hline
Wagner \emph{et al}. \cite{49_wagner2021cpath} & Generative Models & HistAuGAN : Structure-Preserving Multi-Domain Stain Color Augmentation using Style-Transfer with Disentangled Representations  \\

Chikontwe \emph{et al}. \cite{107_chikontwe2022cpath} & Learning disentangled representations & Feature Re-calibration based Multiple Instance Learning for Whole Slide Image Classification \\

\hline
\multicolumn{3}{c}{\textbf{Ensemble Learning}}\\
\hline
Sohail \emph{et al}. \cite{50_sohail2021cpath} & Ensemble learning & Mitotic nuclei analysis in breast cancer histopathology images using deep ensemble classifier \\
\hline

\multicolumn{3}{c}{\textbf{Regularization Strategies}}\\
\hline
Mehrtens \emph{et al}. \cite{138_mehrtens2023cpath} & Regularization Strategies & Benchmarking common uncertainty estimation methods with histopathological images under domain shift and label noise \\

\hline
\multicolumn{3}{c}{\textbf{Other}}\\
\hline
Lu \emph{et al}.\cite{33_lu2022cpath} & Other & Federated learning for computational pathology on gigapixel whole slide images \\

Aubreville \emph{et al}. \cite{190_aubreville2021cpath} & Other & Quantifying the Scanner-Induced Domain Gap in Mitosis Detection \\ 

Sadafi \emph{et al.} \cite{211_sadafi2023cpath} & Other & A Continual Learning Approach for Cross-Domain White Blood Cell Classification \\
\hline \bottomrule

\end{longtable}
\end{center}

\twocolumn

\ifCLASSOPTIONcaptionsoff
  \newpage
\fi

\bibliographystyle{IEEEtran}
\bibliography{refs}

\end{document}